\documentclass[aps,pre,showpacs,amsmath,amssymb,amsfonts,lengthcheck,twocolumn,longbibliography,superscriptaddress, nofootinbib]{revtex4-2}

\usepackage{changes} %to highlight revisions for peer review

\usepackage{graphicx}
\usepackage{subfigure}
\usepackage{amsthm}
\usepackage{amsmath}
\usepackage{verbatim}
\usepackage{dcolumn}% Align table columns on decimal point
\usepackage{bm}% bold math
\usepackage{color}
\usepackage[colorlinks=true,citecolor=blue,linkcolor=blue,urlcolor=blue]{hyperref}%
\usepackage{xcolor}
\usepackage{dsfont}
\usepackage{longtable}
\usepackage{tabularx}
\allowdisplaybreaks
\newcommand{\ket}[1]{\left|#1\right\rangle}

\newcommand{\bla}{bla\\bla\\bla\\bla\\bla}

\begin{document}

\title{Quantum simulation costs for Suzuki-Trotter decomposition of quantum many-body lattice models}
\date{\today}
\author{Nathan M. Myers}
\email{myersn1@vt.edu}
\affiliation{Department of Physics, Virginia Tech, Blacksburg, VA 24061, USA}
\author{Ryan Scott}
\affiliation {Department of Physics, Virginia Tech, Blacksburg, VA 24061, USA}
\author{Kwon Park}
\affiliation {Quantum Universe Center, Korea Institute for Advanced Study, Seoul 02455, Korea}
\affiliation {School of Physics, Korea Institute for Advanced Study, Seoul 02455, Korea} 
\author{Vito W. Scarola}
\affiliation {Department of Physics, Virginia Tech, Blacksburg, VA 24061, USA}

\begin{abstract}
Quantum computers offer the potential to efficiently simulate the dynamics of quantum systems, a task whose difficulty scales exponentially with system size on classical devices. To assess the potential for near-term quantum computers to simulate many-body systems we develop a formalism to straightforwardly compute bounds on the number of Trotter steps needed to accurately simulate the time evolution of fermionic lattice models based on the first-order commutator scaling. We apply this formalism to two closely related many-body models prominent in condensed matter physics, the Hubbard and t-J models. We find that, while a naive comparison of the Trotter depth first seems to favor the Hubbard model, careful consideration of the model parameters and the allowable error for accurate simulation leads to a substantial advantage in favor of the t-J model. These results and formalism set the stage for significant improvements in quantum simulation costs.
\end{abstract}

\maketitle

\section{Introduction} 
\label{sec:1}

Despite rapid growth in accessible computational resources,  simulating quantum systems on classical computers has remained an elusive challenge. This difficulty arises from the exponential scaling of complexity with system size necessary to accurately model quantum systems. Quantum computers, at least in theory, present a solution to this problem. By relying on hardware that is itself quantum in nature, quantum computers can be used to efficiently simulate the dynamics of other quantum systems~\cite{Feynman1982,Lloyd1996, Abrams1997}. Quantum simulation is poised to be a potent tool, with potential applications in a diverse array of areas, including high-energy physics and biology~\cite{Georgescu2014}. In particular, the quantum simulation of electronic structure promises to have particular significance for quantum chemistry~\cite{Kassal2011, Cao2019, McArdle2020} and materials science~\cite{Babbush2018}. 

A critical step in any quantum simulation problem is to translate the Hamiltonian that generates the dynamics for the simulated system into a qubit Hamiltonian that can be implemented on a quantum computer. For the case of electronic structure, this translation can be accomplished using the celebrated Jordan-Wigner transformation~\cite{Jordan1928}, which maps the fermionic creation and annihilation operators to the Pauli operators. The cost of this mapping in terms of circuit design can be quantified by the ``Pauli depth," the maximum length of Pauli operators needed to implement any given term of the qubit-mapped Hamiltonian. A notable challenge in maintaining a low Pauli depth for fermionic Hamiltonains is the need to account for the underlying anticommutivity of the fermionic operators. To do so, the parity for each orbital must be stored using a string of Pauli $Z$ operators~\cite{Whitfield2011}. A consequence of the parity information being stored nonlocally in the ``Jordan-Wigner string" is that each fermionic operator is, in the most general case, mapped to $\mathcal{O}(N)$ Pauli operators, where $N$ is the total qubit number~\cite{McArdle2020} (unless non-local gates are used~\cite{Hastings2015}). This non-locality is an inevitable consequence of the fermion sign problem, which is responsible for the classification of such fermionic lattice models as NP-hard~\cite{Troyer2005}. With this in mind, in this work we focus on the Jordan-Wigner mapping as our primary tool for encoding fermionic Hamiltonians in qubit form. 

Along with encoding the Hamiltonian in qubit form, quantum simulation also requires a method of implementing the time-evolution of the system. This can be done by decomposing the time evolution operator into a series of local gate operations by means of the Suzuki–Trotter expansion~\cite{Trotter1959, Suzuki1976}, a method known as ``Trotterization"~\cite{Nielsen2010}. As the expansion is only exact in the limit of an infinite number of expansion steps, practical applications require truncating at a finite number of steps, and consequently introducing a truncation error, $\epsilon$. This error, along with the desired evolution time, $\tau$, are the model independent parameters that determine the required number of Trotter steps for the simulation, $r$. Thus, $\epsilon$, $\tau$, and $r$ are the three key quantities for benchmarking Trotterized quantum simulation. In general, the choice of $\epsilon$ and $\tau$ will depend on the simulation observable of interest and the model energy scale. For example, an accurate quantum simulation to determine the energy gap~\cite{Lee2022} of a particular many-body model will require an $\epsilon$ significantly smaller than the typical energy spacing. We therefore use the dimensionless quantity $r\epsilon/\tau^2$ as a model independent measure of the Trotter cost.             

Together, the Pauli depth and maximum number of Trotter steps, $r$, provide two significant benchmarks for the computational cost of quantum simulation. This, in turn, has led to significant study in how to optimize these costs, such as by circuit modifications that lead to the cancellation of Jordan-Wigner strings and reordering the terms in the Suzuki–Trotter expansion to reduce the error in each step~\cite{Hastings2015}.  

In this work, rather than seeking to optimize these costs for a particular model, we instead compare the costs between two prominent models for quantum condensed matter, namely the Hubbard and t-J models. We avoid performing detailed gate counts or circuit depth estimations, as has been previously explored~\cite{Wecker2014, Hastings2015, Babbush2018PRX2, Nam2019}. Our goal is to first answer the question of which condensed matter models are most amenable to simulation independent of any specific implementation or choice of simulation parameters, before proceeding with any detailed minimization of computational costs. In this way we can first demonstrate that any advantage either model shows in computational cost is inherent to the structure of the model itself. 

To date, the Hubbard model has served as one of the foremost models targeted for quantum simulation. Numerous approaches, have been pursued, including experimental analog simulations using using atoms confined in optical lattices~\cite{Gross2017, Schneider2008, Jordens2008,Jordens2010,Mazurenko2017, Drewes2017} as well as quantum dots~\cite{Hensgens2017} and NMR systems~\cite{Melo2021}. The Hubbard model has also received focus as test-bed model for simulation on near-term quantum computers~\cite{Cade2020, Choquette2021, Stanisic2022, Suchsland2022}. With this in mind, it is important to identify whether or not the Hubbard model is truly the most efficient model, in terms of simulation resources, that captures the physics of interest.

The t-J model, which arises out of a perturbative treatment of the Hubbard model in the limit where the on-site interaction dominates over the hopping energy~\cite{Takahashi1977, Hirsch1985, Gros1987PRB, MacDonald1988}, is of particular interest due to its potential to model high-temperature superconductivity~\cite{Gros1987, MacDonald1988, Zhang1988, Spalek1988, Lee2006}. An important feature of the t-J model Hamiltonian is the presence of Gutzwiller projection operators that eliminate the possibility of any site being doubly-occupied~\cite{Chao1977, Spalek1988, Fazekas1999, Auerbach1998}. This significantly reduces the Hilbert space of the t-J model in comparison to the more general Hubbard model, which motivates the supposition that the t-J model should be more tractable to simulate on near-term quantum computers. To confirm whether this hypothesis bears out, we apply the Jordan-Wigner transformation along with optimized bounds on the Trotter error scaling~\cite{Childs2021} to benchmark the simulation costs for two-dimensional Hubbard and t-J models on a square lattice in terms of the Pauli and Trotter depths. 

There remains debate about the presence of superconducting behavior in the ground state of the doped t-J model. Unbiased classical simulation using exact diagonlization techniques to probe this issue have been limited to 20 sites for the case of the 2D t-J model~\cite{Kwon2022}. Quantum simulation could help settle the debate. With the growing, but still very limited, qubit resources available to existing quantum computers, determining the model that makes the most efficient use of those resources while maintaining the potential to reveal impactful new physics is of significant importance to both the quantum computing and condensed matter communities. To address this question, we develop a formalism for applying the optimized commutator bound~\cite{Childs2021} to fermionic lattice models that yields analytical expressions for the Trotter depth in terms of the model parameters. We then apply this formalism to demonstrate that the bound on the Trotter depth is significantly lower for the t-J model in comparison to the Hubbard model in the parameter regime of validity. These results suggest that the t-J model should be a prominent candidate for simulation on near-term quantum computers.

In Section~\ref{sec:2} we begin by reformulating the upper bound on Trotter depth given in Ref.~\cite{Childs2021} for the case of square lattice models with both open and periodic boundary conditions. Then, in Section~\ref{sec:3} we review the 2D Hubbard and t-J models including a mapping of the spinful models onto bipartite spinless lattices, before we apply the Jordan-Wigner transformation in Section~\ref{sec:4}. From the Jordan-Wigner transformed Hamiltonians we can immediately evaluate the minimum Pauli depth for each model. In Section~\ref{sec:5} we apply our expanded form of the Trotter bound in order to compare the maximum Trotter depth for the t-J and Hubbard models in both one and two dimensions. Notably, we show that the Trotter depth for both models scales linearly with the system size in both 1D and 2D. Furthermore, we find that, within the parameter regimes where a valid comparison can be made, the t-J model is significantly less costly to simulate. Finally, in Section~\ref{sec:6} we summarize our results and offer some perspectives on their significance for near-term quantum simulation.  

\section{Bounding Trotter Depth}
\label{sec:2}

For a Hamiltonian that can be decomposed as $H = \sum_{\gamma}^{\Gamma} H_{\gamma}$ it is well established that the bound on the Trotter depth scales with $\sum_{\gamma}^{\Gamma} ||H_{\gamma}||$~\cite{Suzuki1985, Hadfield2018}, where $||A||$ denotes the spectral norm of operator $A$. However, this bound is typically a significant overestimate, as it neglects to account for any commutativity within the terms of the Hamiltonian. In Ref.~\cite{Childs2021} it was shown that this bound can be improved to,  
\begin{equation}
    \label{eq:ComBound}
     r_{\mathrm{com}} = \mathcal{O} \left(\frac{\tau^2}{\epsilon} \sum_{\gamma_1,\gamma_2}^{\Gamma} || \left[H_{\gamma_1},H_{\gamma_2}\right]||\right).
\end{equation}
This bound is a powerful improvement and can be orders of magnitude smaller than bounds arising from just the norm of the Hamiltonian terms. Note that Eq.~\eqref{eq:ComBound} is specific to a first order product formula. The bound derived in Ref.~\cite{Childs2021} is further generalized to any order, but for the purposes of this manuscript we consider only the first order result, and leave generalizations to higher orders to be explored in future work.    

Before applying this bound to any specific models, let us first consider it in more detail. First, we  note that the bound is primarily dependent on the spectral norms of commutators of the individual Hamiltonian terms. For Jordan-Wigner transformed fermionic Hamiltonians, the Hamiltonian will always consist of a sum over products of Pauli matrices, along with some prefactors. The spectral norm of any product of Pauli matrices is always one, as demonstrated in Appendix~\ref{Appendix A}. Thus, if each $H_{\gamma}$ consists of only products of Pauli matrices, $||\left[H_{\gamma_1}, H_{\gamma_2}\right]||$ will simply be given by the absolute value of the product of the respective prefactors of $H_{\gamma_1}$ and $H_{\gamma_2}$.

Fermionic lattice Hamiltonians are typically expressed as a sum over lattice sites, $H = \sum_{i,j}H_{i,j}$. In the following we assume translationally invariant lattices, up to boundary terms. As stated above, after applying the Jordan-Wigner transformation each single site term, $H_{i,j}$, can be decomposed into $D$ subterms,
\begin{equation}
    H_{i,j} = \sum_{\delta}^{D} H_{i,j}^{\delta}
\end{equation}
where each $H_{i,j}^{\delta}$ consists only of products of Pauli matrices. Note that each $||[H_{i,j}^{\delta},H_{i,j}^{\delta}]||$ is guaranteed to be zero, while the same is not true for $||[H_{i,j},H_{k,l}]||$ as each $H_{i,j}^{\delta}$ may not commute with all the others. Thus, to ensure we capture all non-commutation in the Hamiltonian, we decompose it as $H = \sum_{ij}^{N_y,N_x} \sum_{\delta}^{D} H_{i,j}^{\delta}$ for the purposes of evaluating Eq.~\eqref{eq:ComBound}. We note that $H_{i,j}$ and $H_{k,l}$ must share at least one pair of Pauli operators with the same indices for any $[H_{i,j}^{\delta},H_{k,l}^{\delta'}]$ to be nonzero.  

To simplify notation we define the operators,
\begin{equation}
    \label{eq:Adef}
    A^{i_1,j_1}_{i_2,j_2} \equiv \sum_{\delta_1, \delta_2 = 1}^D || \left[H_{i_1, j_1}^{\delta_1},H_{i_2, j_2}^{\delta_2}\right]||. 
\end{equation}
Let us consider some general properties of each $A^{i_1,j_1}_{i_2,j_2}$.
\begin{enumerate}
    \item $A^{i_1,j_1}_{i_2,j_2} = A^{i_2,j_2}_{i_1,j_1}$
    \item $\sum_{i_1,i_2} A^{i_1,j_1}_{i_2,j_2} = \sum_{i_1,i_2} A^{i_1,j_2}_{i_2,j_1}$
\end{enumerate}
The first property follows from
\begin{equation}
     ||\left[B,C\right]|| = ||\left[C,B\right]||.
\end{equation}
We can verify the second is true by expanding the summations as, 
\begin{equation}
    \label{eq:SumProof1}
    \sum_{i_1,i_2 = 1}^{N_y} A^{i_1,j_1}_{i_2,j_2} = A^{1,j_1}_{1,j_2} + A^{1,j_1}_{2,j_2} + ... + A^{1,j_1}_{N_y,j_2} + A^{2,j_1}_{1,j_2} + ... + A^{N_y,j_1}_{N_y,j_2},
\end{equation}
and,
\begin{equation}
    \label{eq:SumProof2}
    \sum_{i_1,i_2 = 1}^{N_y} A^{i_1,j_2}_{i_2,j_1} = A^{1,j_2}_{1,j_1} + A^{1,j_2}_{2,j_1} + ... + A^{1,j_2}_{N_y,j_1} + A^{2,j_2}_{1,j_1} + ... + A^{N_y,j_2}_{N_y,j_1}.
\end{equation}
We see that for each term $A^{i_1,j_1}_{i_2,j_2}$ in Eq.~\eqref{eq:SumProof1} there will be a corresponding $A^{i_2,j_2}_{i_1,j_1}$ term in Eq.~\eqref{eq:SumProof2}.

Finally, we also note that, since each single-site term $H_{i,j}$ in a lattice Hamiltonian is identical to each other single-site term, up to a change in indices, we have $A^{i,j}_{i + a,j + b} = A^{k,l}_{k + a,l + b}$. This property is in essence a statement that the rectangular lattice is invariant under horizontal and vertical translations, up to boundary terms. Using these properties we can expand Eq.~\eqref{eq:ComBound} as,    
\begin{equation}
    \label{eq:ComExpanded}
    \begin{split}
    r_{\mathrm{com}}& = \frac{\tau^2}{\epsilon} \Bigg\{N_x N_y A^{1,1}_{1,1} + 2 N_y \sum_{p = 1}^{N_x - 1}\left[\left(N_x-p\right)A^{1,1}_{1,1+p}\right] \\ &+ 2 N_x \sum_{q = 1}^{N_y - 1} \left[\left(N_y-q\right)A^{1,1}_{1+q,1}\right]
    \\&+ 2 \sum_{p = 1}^{N_x - 1} \sum_{q = 1}^{N_y - 1} \left[\left(N_x-p\right)\left(N_y-q\right)A^{1,1}_{1+q,1+p}\right] \\&+ 2 \sum_{p = 1}^{N_x - 1} \sum_{q = 1}^{N_y - 1} \left[\left(N_x-p\right)\left(N_y-q\right)A^{1+q,1}_{1,1+p}\right] \Bigg\}.
    \end{split}
\end{equation}
This expression constitutes the centerpiece of our formalism.
We pause here for a clarification of notation. Using the condition $A^{i,j}_{i + a,j + b} = A^{k,l}_{k + a,l + b}$ we see that all $A_{i,j}^{i,j}$ are equal, $A_{1,1}^{1,1} = A_{1,2}^{1,2} = A_{2,1}^{2,1} = ...$. Thus we have replaced the summation that would appear in Eq.~\eqref{eq:ComExpanded} with a product over the total number of $A_{i,j}^{i,j}$ terms, i.e. $\sum_{ij} A_{i,j}^{i,j} = N_x N_y A_{1,1}^{1,1}$. Similar statements hold for all $A^{i,j}_{i,j+p}$, $A^{i,j}_{i+q,j}$, $A^{i,j}_{i+q,j+p}$, and $A^{i+q,j}_{i,j+p}$. For simplicity's sake, we have chosen $i = j = 1$, but we stress that Eq.~\eqref{eq:ComExpanded} is true for \textit{any} choice of $i$ and $j$. However, to account for the lattice boundaries, for any choice other than $i=j=1$, all $j$ indices should be considered modulo $N_x$ and all $i$ indices should be considered modulo $N_y$. We emphasize further that Eq.~\eqref{eq:ComExpanded} is completely general in regards to hopping, including all-to-all interactions. As we will demonstrate in subsequent sections, model-dependent rules and specification of boundary conditions can considerably simplify this expression.     

To ensure that we have expanded Eq.~\eqref{eq:ComBound} properly, we can check to make sure Eq.~\eqref{eq:ComExpanded} has the same number of terms. We see that Eq.~\eqref{eq:ComBound} has $N_x^2 N_y^2$ total terms. Using the summation identity,
\begin{equation}
    \sum_{i = 1}^{N-1}(N-i) = \frac{N(N-1)}{2},
\end{equation}
we find that Eq.~\eqref{eq:ComExpanded} has,
\begin{equation}
    \begin{split}
    &N_x N_y + N_y N_x(N_x-1)+ N_x N_y(N_y-1) \\&+N_y(N_y-1)N_x(N_x-1) \\
    &= N_x^2 N_y^2
    \end{split}
\end{equation}
total terms. Therefore, as expected, Eqs.~\eqref{eq:ComBound} and~\eqref{eq:ComExpanded} have an equal number of terms.

\section{The Hubbard and \lowercase{t}-J models}
\label{sec:3}

\begin{figure}
    \centering
    \includegraphics[width=0.3\textwidth]{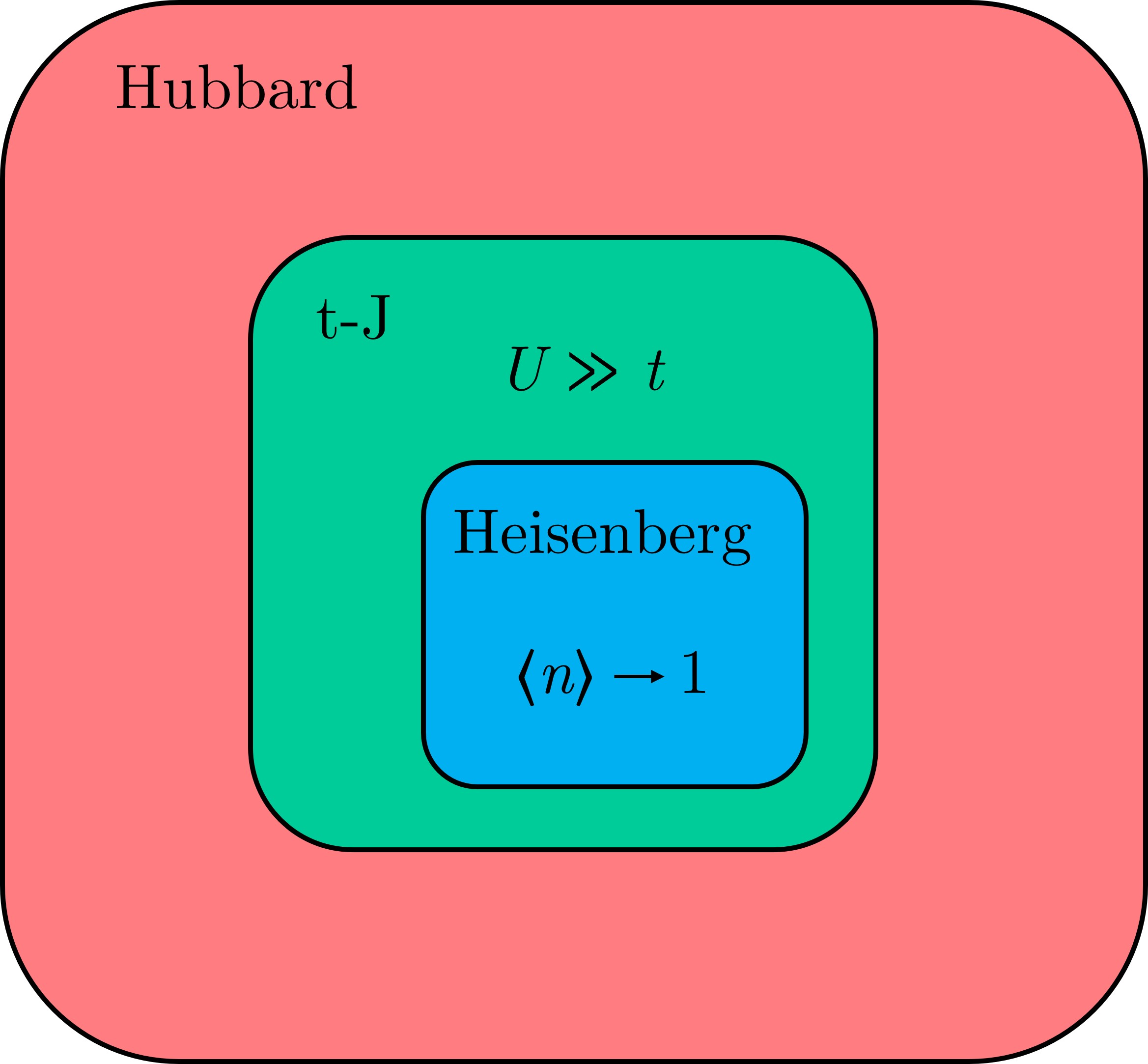}
    \caption{Schematic representation of the parameter space for common fermionic lattice models near half-filling. The t-J model emerges from the Hubbard model near half filling under the constraint that $U \gg t$. The Heisenberg model then emerges from the t-J model under the additional restriction of half-filling.}
    \label{fig:Regimes}
\end{figure}

\begin{figure}
    \centering
    \vspace{0.75 cm}
    \includegraphics[width=0.3\textwidth]{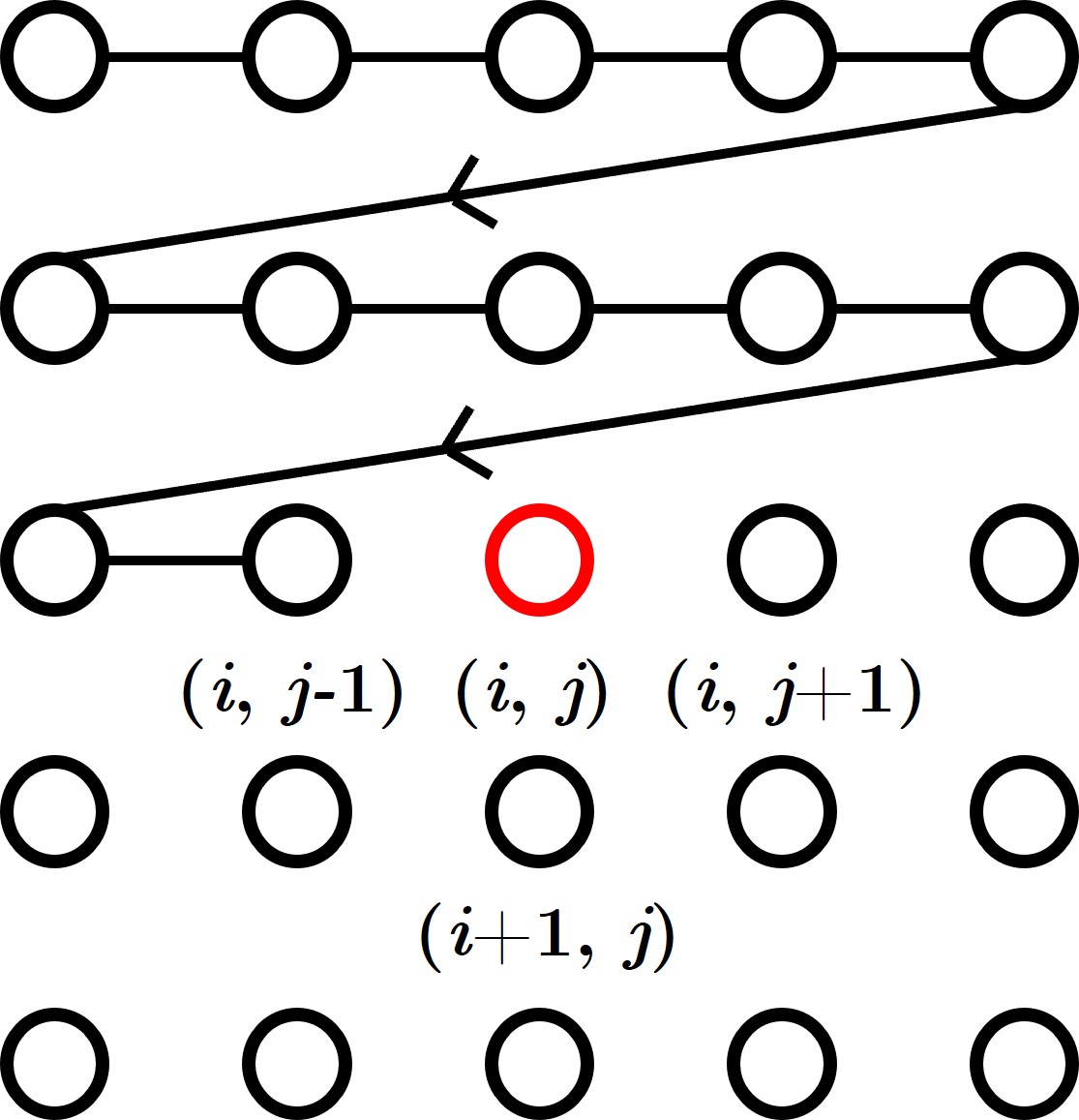}
    \caption{Illustration of the site labeling for the 2D Hubbard model. Site $(i,j)$ is highlighted in red. The solid line illustrates the choice of Jordan-Wigner string.}
    \label{fig:Grid}
\end{figure}

The 2D Hubbard model is~\cite{Fazekas1999},
\begin{equation}
    \begin{split}
    \label{eq:2DHubHam}
    H^H = - t \sum_{<ij,kl>} \sum_{s\in\{\uparrow,\downarrow\}} &\left( c^{\dagger}_{ij,s}c_{kl,s} + c_{ij,s}c^{\dagger}_{kl,s}\right) \\ &+ U\sum_{ij} n_{ij,\uparrow} n_{ij,\downarrow}
    \end{split}
\end{equation}
where $t$ is the hopping energy, $U$ is the on-site interaction energy, and $c_{j,s}^{\dagger}$ and $c_{j,s}$ are the fermionic creation and annihilation operators obeying the anticommutation relations,
\begin{equation}
    \begin{split}
    & \{c_{ij,s},c_{kl,s'}^{\dagger}\} = \delta_{s,s'}\delta_{ij,kl}, \\
    & \{c_{ij,s}^{\dagger},c_{kl,s'}^{\dagger}\} = \{c_{ij,s},c_{kl,s'}\} = 0.
    \end{split}
\end{equation}
Note that $n_{ij,s} = c_{ij,s}^{\dagger}c_{ij,s}$ and $n_{ij} = n_{ij,\uparrow} + n_{ij,\downarrow}$. We consider a rectangular lattice with $N = N_x N_y$ sites. Each pair of indices ($i,k = 1,\dots,N_y$ and $j,l = 1,\dots,N_x$) denote the $x,y$ position of the site in the lattice, respectively, as illustrated in Fig.~\ref{fig:Grid}. We also assume a closed system with a fixed number of particles and thus do not include a chemical potential term in the Hamiltonian. 

The t-J model can be derived from the Hubbard model by working in the regime where $U/t \gg 1$ and considering only the lower energy subspace of unoccupied and singly-occupied sites~\cite{Spalek1988, Fazekas1999}. The Hamiltonian for the t-J model is,
\begin{equation}
    \label{eq:tJHam}
    \begin{split}
    H^{tJ} = &- t \sum_{<ij,kl>} \sum_{s\in\{\uparrow,\downarrow\}} P \left( c^{\dagger}_{ij,s}c_{kl,s} + c_{ij,s}c^{\dagger}_{kl,s}\right) P \\ 
    &+ J \sum_{<ij,kl>} P \left[S_{ij} \cdot S_{kl} - \frac{n_{ij} n_{kl}}{4}\right] P,
    \end{split}
\end{equation}
where $J \equiv 4t^2/U$. Here the $P$ are the Gutzwiller projection operators that ensure the system does not admit any doubly-occupied sites. In Fig.~\ref{fig:Regimes} we provide a schematic representation demonstrating the conditions under which the t-J model emerges from a perturbative treatment of the Hubbard model.   

\begin{widetext}
For later convenience when applying the Jordan-Wigner transformation it will be useful to express the t-J Hamiltonian entirely in terms of the creation, annihilation, and number operators,
\, \\
\begin{equation}
    \label{eq:2DtJHam}
    \begin{split}
    H^{tJ} = &- t \sum_{<ij,kl>} \sum_{s\in\{\uparrow,\downarrow\}} \left[ \left(1-n_{ij,\bar{s}}\right)\left(c^{\dagger}_{ij,s}c_{kl,s} + c^{\dagger}_{kl,s}c_{ij,s}\right)\left(1-n_{kl,\bar{s}}\right)\right] \\ &+ \frac{J}{2}\sum_{<ij,kl>} \sum_{s\in\{\uparrow,\downarrow\}} \left[c^{\dagger}_{ij,s}c_{ij,\bar{s}}c^{\dagger}_{kl,\bar{s}}c_{kl,s} - \left(1-n_{ij,\bar{s}}\right)n_{ij,s}n_{kl,\bar{s}}\left(1-n_{kl,s}\right)\right].
    \end{split}
\end{equation}
\, \\
\, \\
The full derivation for this expression is provided in Appendix~\ref{Appendix B}. 
\end{widetext}
\, \\
\, \\
Examining Eq.~\eqref{eq:2DtJHam} we see that the local projection operators leave the Heisenberg terms unchanged. This is expected, as the spin-spin interaction only arises under the condition of single occupancy. A natural question is then whether the projection operators need to be applied to the $J$ terms at all. Assuming an initial state with no double occupancy, the existence of the projection operators on the hopping terms is sufficient to keep the state restricted to the subspace consisting of only unoccupied and singly-occupied sites. However, simulating the dynamics of the t-J model on a real quantum device will inevitably introduce errors, some of which may lead to the system straying out of the restricted Hilbert space. With this in mind, we leave the projection operators in place on the $J$ terms as a limited form of protection against this class of errors. 

A deeper degree of protection that would also guard against initial states with doubly occupied sites can be implemented using global projection operators that also project the implicit identity operators that act on every other site in the lattice for each term in Eq.~\eqref{eq:2DtJHam}. However, these global projection operators are highly nonlocal and come at a prohibitive cost in terms of Pauli depth. Thus, for the purpose of this analysis, we assume the initial state used for any simulation of the t-J model does not contain any double occupancy.   

\section{The Jordan-Wigner Transformation}
\label{sec:4}

It is well established that fermionic creation and annihilation operators can be mapped to Pauli (qubit) operators by means of the Jordan-Wigner transformation. In 1D this transformation takes the form of~\cite{Jordan1928, Whitfield2011},
\begin{equation}
    \label{eq:JW}
    \begin{split}
    & c_{j}^{\dagger} = \exp\left(i \pi \sum_{k=1}^{j-1}\sigma_k^+ \sigma_k^-\right)\sigma_j^+, \\
    & c_{j} = \exp\left(-i \pi \sum_{k=1}^{j-1}\sigma_k^+ \sigma_k^-\right)\sigma_j^-, \\
    & n_j = \frac{1}{2}\left(\sigma^0_j + \sigma_j^z\right).
    \end{split}
\end{equation}
Here $\sigma^+ \equiv (\sigma^x + i \sigma^y)/2$, $\sigma^- \equiv (\sigma^x - i \sigma^y)/2$ with $\sigma^x, \sigma^y$, and $\sigma^z$ being the usual Pauli matrices and $\sigma^0$ being the identity matrix.

In order to capture the full Jordan-Wigner string in two dimensions, as illustrated in Fig.~\ref{fig:Grid}, each fermionic operator must transform as follows, 
\begin{equation}
    \label{eq:JW2D}
    \begin{split}
    & c_{i,j}^{\dagger} = \exp\left(i \pi \sum_{k=1}^{i-1} \sum_{l=1}^{N_x} \sigma_{k,l}^+ \sigma_{k,l}^-\right) \exp\left(i \pi \sum_{l=1}^{j-1} \sigma_{i,l}^+ \sigma_{i,l}^-\right) \sigma_{i,j}^+ \\
    & c_{i,j} = \exp\left(-i \pi \sum_{k=1}^{i-1} \sum_{l=1}^{N_x} \sigma_{k,l}^+ \sigma_{k,l}^-\right) \exp\left(-i \pi \sum_{l=1}^{j-1} \sigma_{i,l}^+ \sigma_{i,l}^-\right) \sigma_{i,j}^- \\
    & n_{i,j} = \frac{1}{2}\left(\sigma^0_{i,j} + \sigma_{i,j}^z\right)
    \end{split}
\end{equation}
Note that this implementation of the 2D Jordan-Wigner transformation matches with the result derived in Ref.~\cite{Azzouz1993}.   

Using Eq.~\eqref{eq:JW2D} we have,
\begin{equation}
    \label{eq:OpProd}
    \begin{split}
    & c_{i,j}^{\dagger}c_{k,l} = \\ & \exp\left(i \pi \sum_{\alpha=1}^{i-1} \sum_{\beta=1}^{N_x} \sigma_{\alpha,\beta}^+ \sigma_{\alpha,\beta}^-\right)  \exp\left(i \pi \sum_{\beta=1}^{j-1} \sigma_{i,\beta}^+ \sigma_{i,\beta}^-\right) \sigma_{i,j}^+ \\ &
    \exp\left(-i \pi \sum_{\alpha=1}^{k-1} \sum_{\beta=1}^{N_x} \sigma_{\alpha,\beta}^+ \sigma_{\alpha,\beta}^-\right) \exp\left(-i \pi \sum_{\beta=1}^{l-1} \sigma_{k,\beta}^+ \sigma_{k,\beta}^-\right) \sigma_{k,l}^-
    \end{split}
\end{equation}
Without loss of generality we assume $i<k$ and $j<l$. Equation~\eqref{eq:OpProd} then becomes,
\begin{equation}
    \label{eq:OpProdSimp}
    \begin{split}
    c_{i,j}^{\dagger}c_{k,l} = \, &\sigma_{i,j}^+ \exp\left(-i \pi \sum_{\beta=j}^{N_x} \sigma_{i,\beta}^+ \sigma_{i,\beta}^-\right) \\&  \times \exp\left(-i \pi \sum_{\alpha=i+1}^{k-1} \sum_{\beta=1}^{N_x} \sigma_{\alpha,\beta}^+ \sigma_{\alpha,\beta}^-\right) \\&  \times \exp\left(-i \pi \sum_{\beta=1}^{l-1} \sigma_{k,\beta}^+ \sigma_{k,\beta}^-\right) \sigma_{k,l}^-. 
    \end{split}
\end{equation}
We note that Eq.~\eqref{eq:OpProdSimp} has three distinct ``segments" of Jordan-Wigner strings. We will refer to these segments as string A,
\begin{equation}
    \exp\left(-i \pi \sum_{\beta=j}^{N_x} \sigma_{i,\beta}^+ \sigma_{i,\beta}^-\right),
\end{equation}
string B,
\begin{equation}
    \exp\left(-i \pi \sum_{\alpha=i+1}^{k-1} \sum_{\beta=1}^{N_x} \sigma_{\alpha,\beta}^+ \sigma_{\alpha,\beta}^-\right),
\end{equation}
and string C,
\begin{equation}
    \exp\left(-i \pi \sum_{\beta=1}^{l-1} \sigma_{k,\beta}^+ \sigma_{k,\beta}^-\right),
\end{equation}
respectively. From a quick inspection of the bounds on the summations for each string we see that together the three segments account for all the sites between $(i,j)$ and $(k,l)$. This is illustrated in Fig.~\ref{fig:StringSeg}.

We are now ready to apply the Jordan-Wigner transformation to the Hubbard and t-J Hamiltonians. The first step in doing so is to remove the spin dependence from the Hamiltonians. We can do this by mapping from a spinful lattice to a bipartite spinless lattice. This technique is illustrated graphically in Fig.~\ref{fig:SpinlessGrid}. 

\begin{widetext}

\begin{figure}
    \begin{minipage}[c]{0.45\linewidth}
    \centering
    \includegraphics[width=0.8\linewidth]{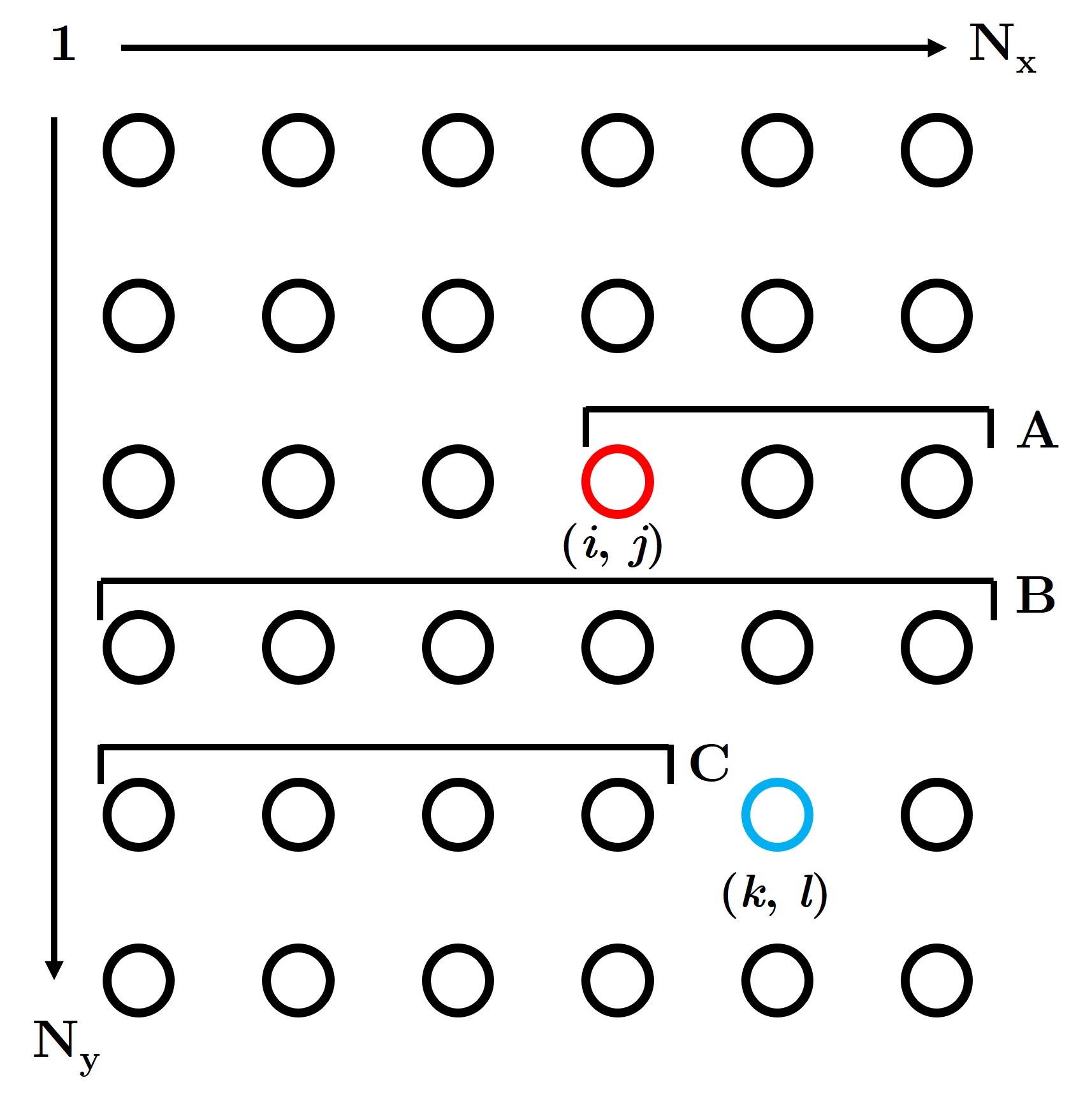}
    \caption{Illustration of the three ``segments" of Jordan-Wigner string found in Eq.~\eqref{eq:OpProdSimp} between sites $(i,j)$ (highlighted in red) and $(k,l)$ (highlighted in blue).}
    \label{fig:StringSeg}
    \end{minipage} \hfill
    \begin{minipage}[c]{0.45\linewidth}
    \centering
    \includegraphics[width=0.8\linewidth]{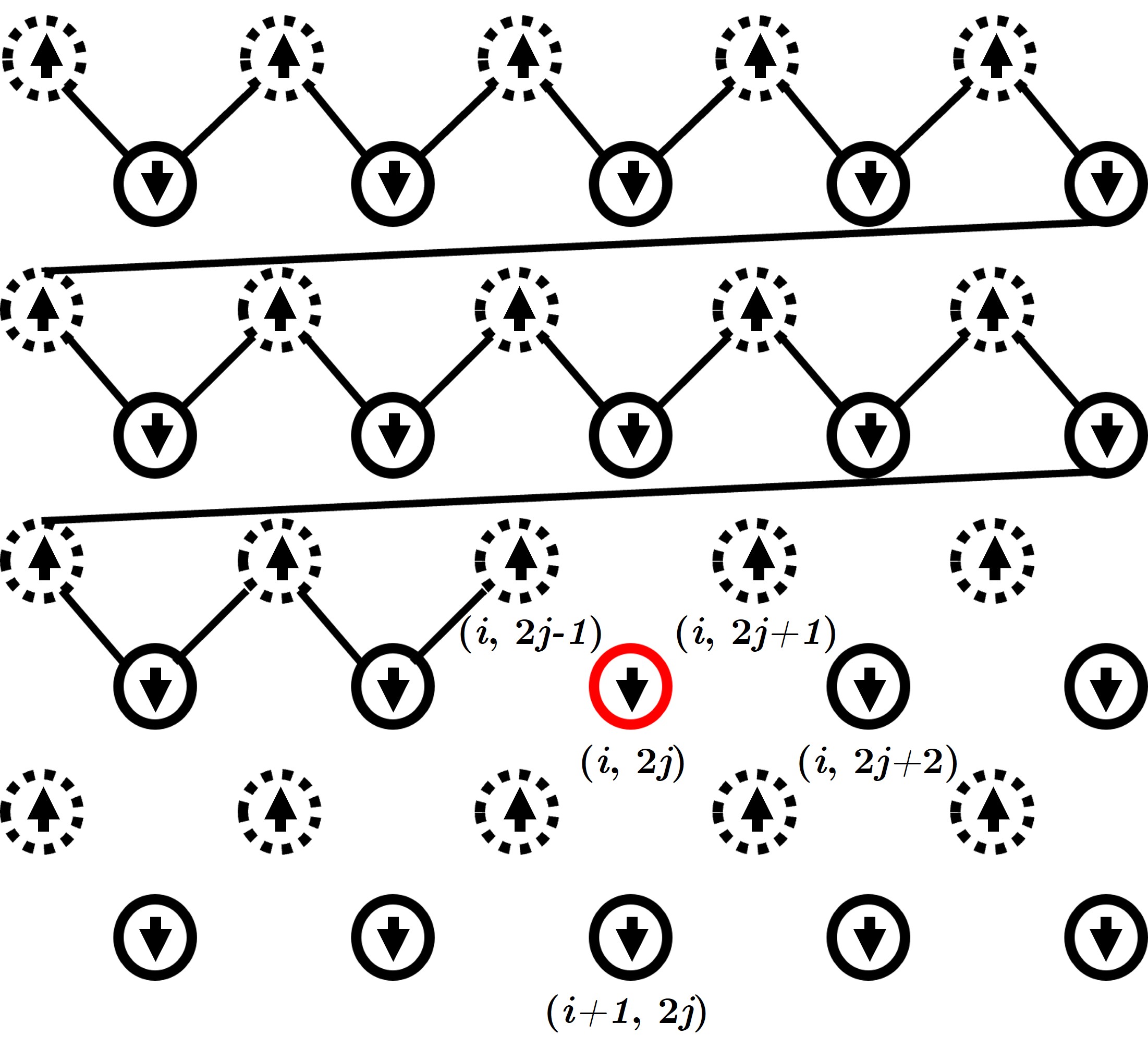}
    \caption{Jordan-Wigner string (solid line) for mapping a 2D grid of spinful fermions to two grids of spinless fermions. Note that each pair of dotted and solid rows represents one row of the spinful grid. Site $(i,2j)$ is highlighted in red.}
    \label{fig:SpinlessGrid}
    \end{minipage}
\end{figure}
Making the substitutions,
\begin{equation}
\begin{split}
    c_{i,j,\uparrow} \rightarrow c_{i,2j-1}, \qquad
    c_{i,j,\downarrow} \rightarrow c_{i,2j}, \qquad
    c_{i,j+1,\uparrow} \rightarrow c_{i,2j+1}, \qquad
    c_{i,j+1,\downarrow} \rightarrow c_{i,2j+2}.
\end{split}
\end{equation}
we arrive at
\begin{equation}
    H^H = -t \sum_j \Big(c_{2j-1}^{\dagger}c_{2j+1} + c_{2j+1}^{\dagger}c_{2j-1}
    + c_{2j}^{\dagger}c_{2j+2} + c_{2j+2}^{\dagger}c_{2j} \Big)+ U \sum_j n_{2j-1}n_{2j},
\end{equation}
for the Hubbard model and, 
\begin{equation}
    \label{eq:tJspinless}
    \begin{split}
    H^{tJ} = & - t \sum_{j} \Big[ \left( 1 - n_{2j}\right)\left(c^{\dagger}_{2j-1}c_{2j+1} + c_{2j-1}c^{\dagger}_{2j+1}\right)\left( 1 - n_{2j+2}\right) + \left( 1 - n_{2j-1}\right)\left(c^{\dagger}_{2j}c_{2j+2} + c_{2j}c^{\dagger}_{2j+2}\right)\left( 1 - n_{2j+1}\right)\Big] \\
     & + \frac{J}{2}\sum_{j} \Big[c^{\dagger}_{2j-1}c_{2j}c^{\dagger}_{2j+2}c_{2j+1} + c^{\dagger}_{2j} c_{2j-1}c^{\dagger}_{2j+1}c_{2j+2}- (1-n_{2j})n_{2j-1}n_{2j+2}(1-n_{2j+1}) \\
     & \qquad \qquad \quad - (1-n_{2j-1})n_{2j}n_{2j+1}(1-n_{2j+2}) \Big].
    \end{split}
\end{equation}
for the t-J model. 

Applying the Jordan-Wigner transformation as given in Eq.~\eqref{eq:JW} and simplifying using 
\begin{equation}
    \sigma_j^+ \sigma_j^z = -\sigma_j^+ \quad \text{and} \quad \sigma_j^z \sigma_j^- = - \sigma_j^-
\end{equation}
as well as,
\begin{equation}
    \exp\left(\pm i \pi \sum_{k = 1}^{j-1}\sigma_k^+ \sigma_k^{-}\right) = (-1)^{j-1} \prod_{k=1}^{j-1} \sigma_k^z
\end{equation}
we arrive at,
\begin{equation}
    \label{eq:HubbardHJW}
    \begin{split}
    H^H = & \frac{t}{2} \sum_{i,j} \Bigg[ (-1)^{N_x}\Big\{\left(\sigma_{i,2j-1}^x \sigma_{i+1,2j-1}^x + \sigma_{i+1,2j-1}^y \sigma_{i,2j-1}^y\right) \sigma_{i,2j}^z \\
    &+ \left(\sigma_{i,2j}^x \sigma_{i+1,2j}^x + \sigma_{i+1,2j}^y \sigma_{i,2j}^y \right) \sigma_{i+1,2j-1}^z \Big\} \left( \prod_{\beta = 2j+1}^{N_x} \sigma_{i,\beta}^z \right) \left(\prod_{\beta = 1}^{2j-2} \sigma_{i+1,\beta}^z \right)\\
    & + \left( \sigma_{i,2j-1}^x \sigma_{i,2j+1}^x + \sigma_{i,2j+1}^y \sigma_{i,2j-1}^y \right) \sigma_{i,2j}^z + \left(\sigma_{i,2j}^x \sigma_{i,2j+2}^x + \sigma_{i,2j+2}^y \sigma_{i,2j}^y \right)  \sigma_{i,2j+1}^z \Bigg] \\ 
    & + \frac{U}{4} \sum_{i,j} \left(\sigma^0_{i,2j-1} + \sigma_{i,2j-1}^z\right) \left(\sigma^0_{i,2j} + \sigma_{i,2j}^z\right)
    \end{split}
\end{equation}
For the Hubbard model. Similarly, for the t-J model we find,
\begin{equation}
    \label{eq:tJHJW}
    \begin{split}
    &H^{tJ} = - \frac{t}{8} \sum_{i,j} \Bigg[(-1)^{N_x}\Big\{\left(\sigma^0_{i,2j} - \sigma^z_{i,2j}\right) \left(\sigma^x_{i,2j-1} \sigma^x_{i+1,2j-1} + \sigma^y_{i,2j-1} \sigma^y_{i+1,2j-1} \right) \left(\sigma^0_{i+1,2j} - \sigma^z_{i+1,2j}\right) \\& + \left(\sigma ^0_{i,2j-1} - \sigma ^z_{i,2j-1}\right) \left( \sigma^x_{i,2j} \sigma^x_{i+1,2j} + \sigma^y_{i,2j} \sigma^y_{i+1,2j} \right) \left(\sigma^0_{i+1,2j-1} - \sigma^z_{i+1,2j-1}\right) \Big\} \left( \prod_{\beta=2j+1}^{N_x} \sigma^z_{i,\beta} \right) \left(\prod_{\beta=1}^{2j-2} \sigma^z_{i+1,\beta} \right)\\ & 
    +\left(\sigma^0_{i,2j} - \sigma^z_{i,2j}\right) \left(\sigma^x_{i,2j-1} \sigma^x_{i,2j+1} + \sigma^y_{i,2j+1} \sigma^y_{i,2j-1}
    \right) \left(\sigma^0_{i,2j+2} - \sigma^z_{i,2j+2}\right) \\ &+
    \left(\sigma^0_{i,2j-1} - \sigma^z_{i,2j-1}\right) \left( \sigma^x_{i,2j} \sigma^x_{i,2j+2} + \sigma^y_{i,2j+2}\sigma^y_{i,2j}
    \right) \left(\sigma^0_{i,2j+1} - \sigma^z_{i,2j+1}\right) \Bigg] \\ 
     &+ \frac{J}{2} \sum_{i,j} \Bigg[\sigma^+_{i,2j-1} \sigma^-_{i,2j} \left(\sigma^+_{i+1,2j} \sigma^-_{i+1,2j-1} + \sigma^+_{i,2j+2} \sigma^-_{i,2j+1}\right) + \sigma^+_{i,2j} \sigma^-_{i,2j-1} \left(\sigma^+_{i+1,2j-1} \sigma^-_{i+1,2j} + \sigma^+_{i,2j+1} \sigma^-_{i,2j+2}\right) \\ 
     & \qquad - \frac{1}{16} \Big\{ \left(\sigma^0_{i,2j} - \sigma^z_{i,2j}\right)\left(\sigma^0_{i,2j-1} + \sigma^z_{i,2j-1}\right)\left(\sigma^0_{i+1,2j} + \sigma^z_{i+1,2j}\right)\left(\sigma^0_{i+1,2j-1} - \sigma^z_{i+1,2j-1}\right) \\
     & \qquad \qquad + \left(\sigma^0_{i,2j} - \sigma^z_{i,2j}\right)\left(\sigma^0_{i,2j-1} + \sigma^z_{i,2j-1}\right)\left(\sigma^0_{i,2j+2} + \sigma^z_{i,2j+2}\right)\left(\sigma^0_{i,2j+1} - \sigma^z_{i,2j+1}\right) \\
     & \qquad \qquad + \left(\sigma^0_{i,2j-1} - \sigma^z_{i,2j-1}\right)\left(\sigma^0_{i,2j} + \sigma^z_{i,2j}\right)\left(\sigma^0_{i+1,2j-1} + \sigma^z_{i+1,2j-1}\right)\left(\sigma^0_{i+1,2j} - \sigma^z_{i+1,2j}\right) \\ 
     & \qquad \qquad +\left(\sigma^0_{i,2j-1} - \sigma^z_{i,2j-1}\right)\left(\sigma^0_{i,2j} + \sigma^z_{i,2j}\right)\left(\sigma^0_{i,2j+1} + \sigma^z_{i,2j+1}\right)\left(\sigma^0_{i,2j+2} - \sigma^z_{i,2j+2}\right) \Big\}\Bigg]
    \end{split}
\end{equation}
\end{widetext}

\begin{figure}
    \centering
    \includegraphics[width=0.45\textwidth]{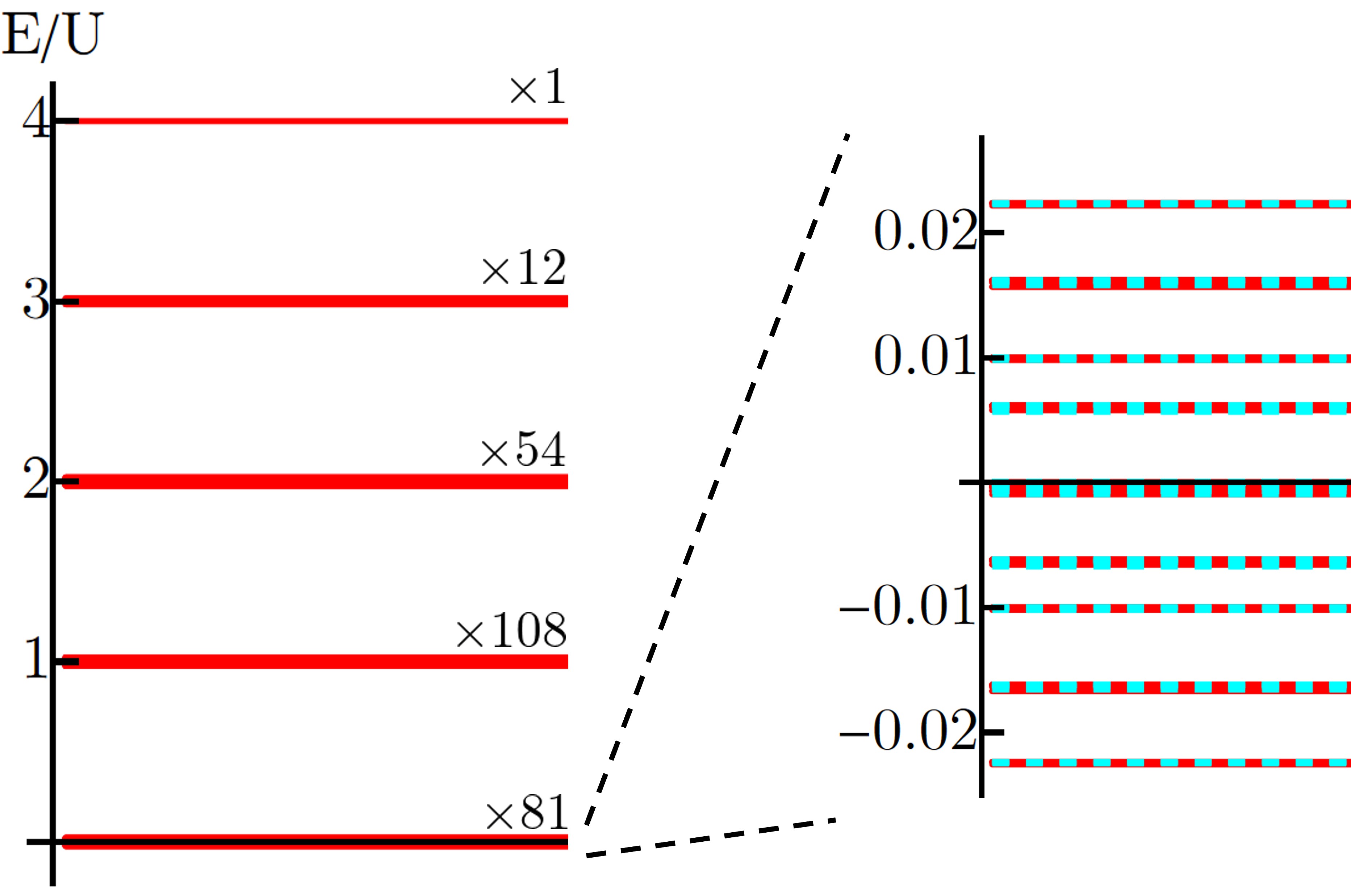}
    \caption{Energy spectrum for a 1D four-site Hubbard model model with open boundary conditions (red, solid) superimposed with the energy spectrum for a 1D four-site t-J model (cyan, dashed). We have set $U/t = 100$ such that we see full overlap between the low energy spectrum of the Hubbard model with the t-J model spectrum. 
}
    \label{fig:Spectrum}
\end{figure}

The t-J model arises from the low energy Hubbard model in the limit $U/t \gg 1$. As a check for the consistency of the Jordan-Wigner transformations, we numerically calculate the energy spectra of a four-site, four-particle 1D Hubbard and t-J model using the Pauli representation we derived in section~\ref{sec:4}. We plot the resulting spectra for $U = 10$, $t=0.1$ and $J = 4t^2/U = 0.004$ in Fig.~\ref{fig:Spectrum}. We see that, as expected, the t-J energy spectrum overlaps with that of the low energy states of the Hubbard spectrum.            

As an additional verification, we also calculate the energy eigenstates for the Hubbard and t-J models under the same parameters using the ALPS  (Algorithms and Libraries for Physics Simulations) software package~\cite{Bauer2011}. To within numerical precision, we obtain the same spectrum as from the exact diagonlization of our Jordan-Wigner transformed Hamiltonians. 

Comparing Eqs.~\eqref{eq:HubbardHJW} and~\eqref{eq:tJHJW} we see that the maximum Pauli depth of the $t$-dependent terms is identical between the two models. Both contain at least one term that consists of a product of $N_x$ Pauli operators. In contrast, the maximum Pauli depth of the $U$-dependent term in the Hubbard model is only two, half that of the $J$-dependent term in the t-J model.     

\section{\lowercase{t}-J and Hubbard Trotter Depth}
\label{sec:5}

With the full Jordan-Wigner transformed Hamiltonians, we can now apply Eq.~\eqref{eq:ComExpanded} to bound the number of Trotter steps necessary to simulate the time evolution of the Hubbard and t-J models. Note that, for completeness, we examine the 1-norm scaling in Appendix~\ref{Appendix C}. This comparison emphasizes the improvement offered by the commutator scaling, particularly for the Hubbard model, which can be over an order of magnitude lower, even for small system sizes. In the following analysis we also consider specifically the case of open boundary conditions. In Appendix~\ref{Appendix D} we extend these results to the case of periodic boundary conditions.

We begin by considering some common features of both models. As both the Hubbard and t-J models are restricted to only nearest neighbor hopping, we have $A^{i_1,j_1}_{i_2,j_2} = 0$ for all $\left(i_2,j_2\right) \neq \left(i_1+1, 1 \leq j_2 \leq j_1 \right)$ and for all $\left(i_2,j_2\right) \neq \left(i_1,j_1 \leq j_2 \leq N_x\right)$. These conditions account for the segments of Jordan-Wigner string that are not shared between site $(i,j)$ and its vertical neighbor $(i+1,j)$, as discussed in Section~\ref{sec:4}. The first set of nonzero $A^{i_1,j_1}_{i_2,j_2}$ accounts for the segment of Jordan-Wigner string stretching from site $(i,j)$ rightward to site $(i,N_x)$, and the second accounts for the segment stretching leftward from site $(i+1,1)$ to site $(i+1,j)$. 

Furthermore, by inspecting Eqs.~\eqref{eq:HubbardHJW} and~\eqref{eq:tJHJW} we find that all $A^{i,j}_{i,j+p>1}$ and $A^{i+1,j}_{i,j+p>1}$ vanish. This is due to the fact that the only Pauli operators with the same indices $H_{i,j}$ will share with $H_{i,j+p>1}$ arise from the Jordan-Wigner strings, which consist of only $\sigma^z$ operators. Thus, each commutator will take the form of either,
\begin{equation}
    \label{eq:PauliComm}
    \begin{split}
    &\left[\sigma^z \sigma^x, \sigma^x  \sigma^z\right], \quad \left[\sigma^z  \sigma^y, \sigma^y  \sigma^z\right], \quad \left[\sigma^z  \sigma^x, \sigma^y  \sigma^z\right], \\
    &\left[\sigma^z  \sigma^z, \sigma^x  \sigma^x\right], \quad \left[\sigma^z  \sigma^z, \sigma^y  \sigma^y\right].
    \end{split}
\end{equation}
all of which are zero by the properties of the Pauli operators.

Thus, for the case of the Hubbard and t-J models, Eq.~\eqref{eq:ComExpanded} simplifies to, 
\begin{equation}
    \label{eq:ComAH}
    \begin{split}
    r^{H,\,tJ}_{\mathrm{com}} = &\frac{\tau^2}{\epsilon} \Big\{N_x N_y A^{1,1}_{1,1} + 2 N_y \left(N_x-1\right)A^{1,1}_{1,2} \\
    &\qquad + 2 N_x \left(N_y-1\right)A^{1,1}_{2,1} \\ & \qquad \quad
    + 2 \left(N_y-1\right) \left(N_x-1\right)A^{2,1}_{1,2} \Big\}
    \end{split}
\end{equation}
We reiterate here that, like Eq.~\eqref{eq:ComExpanded}, this equation is valid for any choice of $i$ and $j$, and that for any choice other than $i=j=1$ all $j$ indices should be considered modulo $N_x$ and all $i$ indices modulo $N_y$.

\subsection{Hubbard}

In order to get an expression for Eq.~\eqref{eq:ComAH} in terms of our model parameters ($t$ and $U$) we need to compute each $A^{i_2,j_2}_{i_1,j_1}$. In Table~\ref{tbl:Hubbardterms} we break up each single site term in the Hubbard Hamiltonian, Eq.~\eqref{eq:HubbardHJW}, into terms that consist only of products of Pauli operators, along with some common prefactors. Then using, 
\begin{equation}
    \label{eq:AHub}
    A^{i_1,j_1}_{i_2,j_2} \equiv \sum_{\delta_1, \delta_2 = 1}^{12} || \left[H_{i_1, j_1}^{\delta_1},H_{i_2, j_2}^{\delta_2}\right]||,
\end{equation}
where each $H_{i,j}^{\delta}$ is listed in Table~\ref{tbl:Hubbardterms}, we calculate each $A^{i_2,j_2}_{i_1,j_1}$. This process is considerably simplified since each $H_{i,j}^{\delta}$ consists of a tensor product of Pauli operators. Since the norm of a tensor product of Pauli operators is always one, we immediately know each $A^{i_2,j_2}_{i_1,j_1}$ will be of the form,
\begin{equation}
    \label{eq:HApoly}
   A^{i_2,j_2}_{i_1,j_1} =  2\left(a_H \frac{t^2}{4} + b_H \frac{|t U|}{8} + c_H \frac{U^2}{16} \right),
\end{equation}
where $a_H$, $b_H$ and $c_H$ are positive integers determined by the number of nonzero commutators in $A^{i_2,j_2}_{i_1,j_1}$ with the corresponding prefactor, and the factor of 2 arises from the Pauli commutation relations.

Inspecting the terms in Table~\ref{tbl:Hubbardterms} we first note that for all $A^{i_2,j_2}_{i_1,j_1}$ we have $c_H=0$, as the potential terms all consist only of $\sigma^z$ and identity operators and thus will always commute with each other. To determine $a_H$ and $b_H$ let us consider each $A^{i_2,j_2}_{i_1,j_1}$ term in Eq.~\eqref{eq:ComAH}. 

Of the 144 commutators in Eq.~\eqref{eq:AHub}, we see that only 40 are non-zero for $A^{i,j}_{i,j}$. Of these 40 commutators, 8 of them have the prefactor $t^2/4$ and the other 32 have the prefactor $|tU|/8$. Thus, for $A^{i,j}_{i,j}$ we have $a_H = 8$, $b_H = 32$, and $c_H =0$. Combining these coefficients with  Eq.~\eqref{eq:HApoly} arrive at,
\begin{equation}
    A^{i,j}_{i,j} = A^{1,1}_{1,1} =  4t^2 + 8 |t U| .
\end{equation}
We can repeat the same process for $A^{i,j}_{i,j+1}$ and $A^{i,j}_{i+1,j}$ where in both cases we find 16 non-zero commutators, 8 with the prefactor $t^2/4$ and 8 with $|tU|/8$. Thus,
\begin{equation}
    A^{i,j}_{i,j+1} =  A^{i,j}_{i+1,j} = A^{1,1}_{1,2} = A^{1,1}_{2,1} = 4t^2 + 2 |t U|
\end{equation}
Finally, we see that for $A^{i+1,j}_{i,j+1}$ there are only 4 nonzero commutators, all corresponding to $t^2/4$. Thus,  
\begin{equation}
    A^{i+1,j}_{i,j+1} = A^{2,1}_{1,2} = 2t^2.
\end{equation}

We can now combine all these results with Eq.~\eqref{eq:ComAH} to yield,
\begin{equation}
    \label{eq:2DcommH}
    \begin{split}
    r^H_{\mathrm{com}} = &N_x N_y \left(4 t^2+8 t U\right) +2 N_y (N_x-1)  \left(4 t^2+2 t U\right) \\
    &+ 2 N_x (N_y-1) \left(4 t^2+2 t U\right)+4 (N_y-1) (N_x-1) t^2
   \end{split}
\end{equation}
Here we note several important features of this result. First, we see that the bound scales quadratically with $t$ and linearly with $U$. However, even more notably, it scales linearly with the total number of lattice sites, $N \equiv N_x N_y$, despite the presence of the lattice spanning Jordan-Wigner strings in Eq.~\eqref{eq:HubbardHJW}. This linear scaling arises due to the fact that all commutators between non-nearest-neighbor sites will be of the form given in Eq.~\eqref{eq:PauliComm} and will thus vanish.    

\begin{figure*}
	\subfigure[]{
		\includegraphics[width=.28\textwidth]{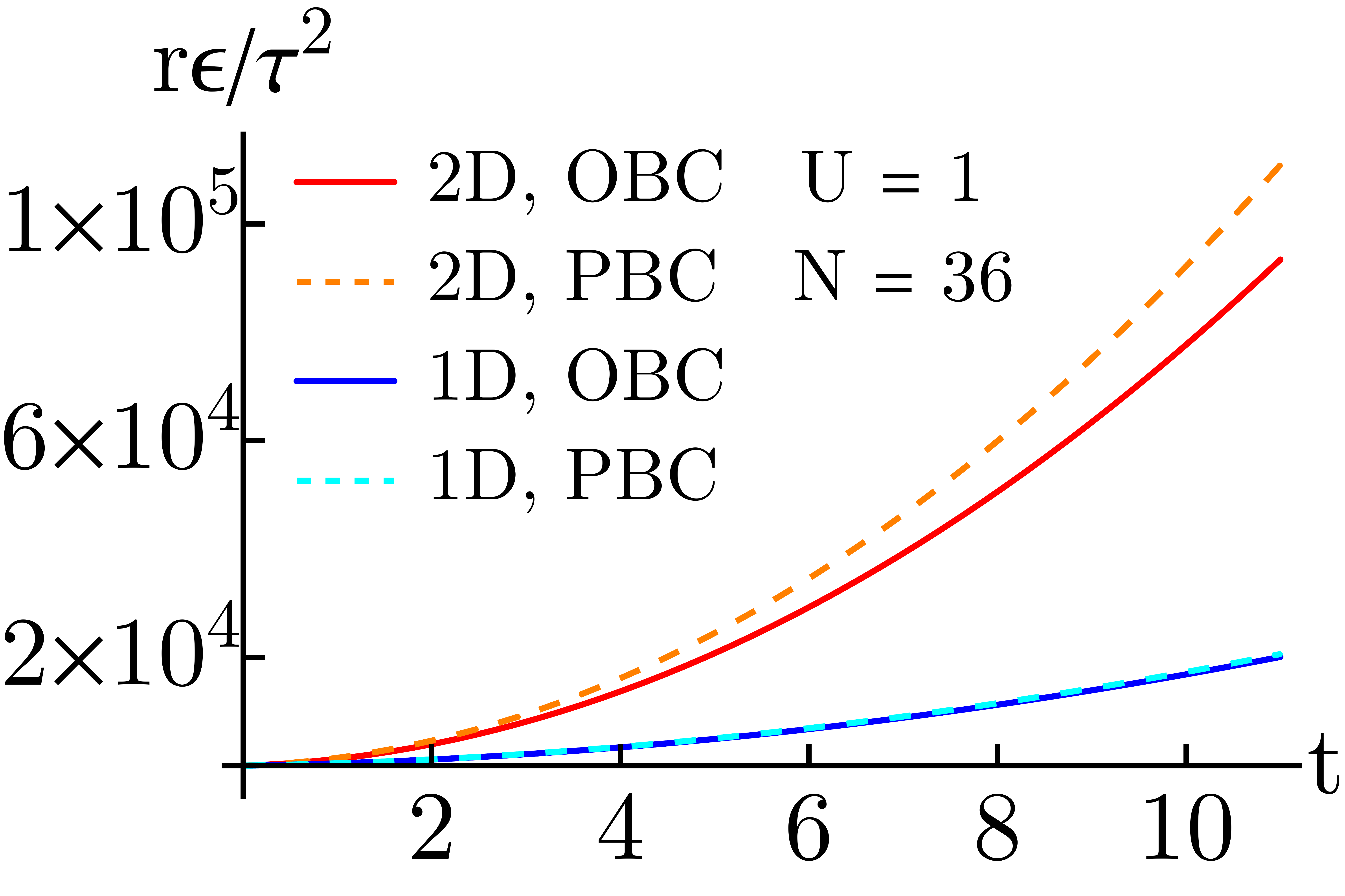}
	}
	\hspace{5mm}
	\subfigure[]{
		\includegraphics[width=.28\textwidth]{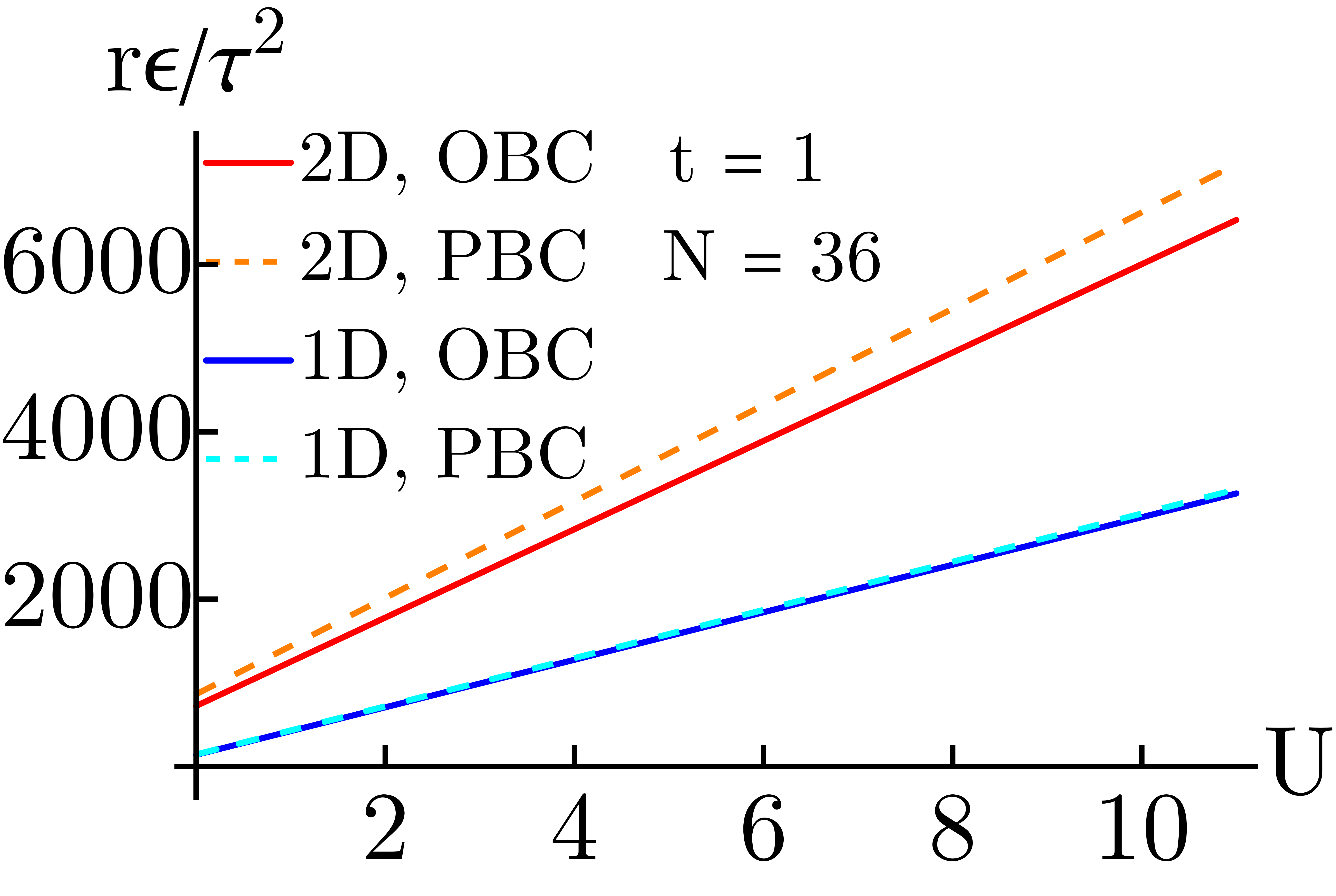}
	}
	\hspace{5mm}
	\subfigure[]{
		\includegraphics[width=.28\textwidth]{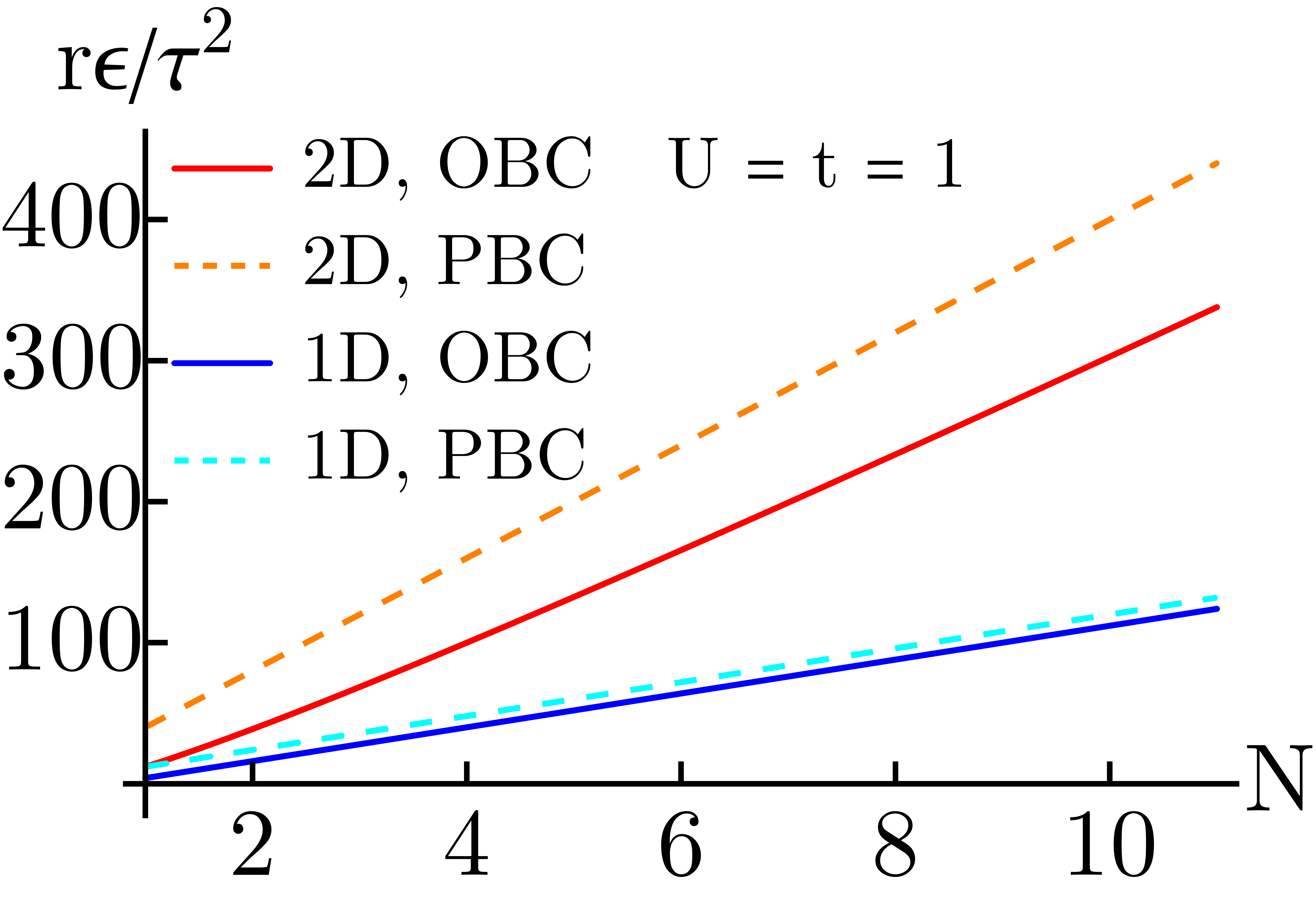}
	}
    \caption{Commutator bound on the Trotter depth for the Hubbard model as a function of (a) $t$ with $U=1$, $N=36$ (b) $U$ with $t=1$, $N=36$ and (c) $N$ with $U=t=1$. The upper pair of lines show the Trotter depth for a 2D model with periodic (orange, dotted) and open (red, solid) boundary conditions. The lower pair of lines correspond to a 1D model with periodic boundary conditions (dark blue, dashed) and open boundary conditions (light blue, solid).}
    \label{fig:HubTrotterDepth}
\end{figure*}

In Fig.~\ref{fig:HubTrotterDepth} we plot the functional form of the Trotter depth from Eq.~\eqref{eq:2DcommH} as a function of $t$, $U$, and $N$ for the case of a 6 by 6 lattice. As the choices for values of $\epsilon$ and $\tau$ will generally depend on the observables of interest for the simulation, we quantify the overall Trotter cost in terms of the problem-independent quantity $r\epsilon/\tau^2$. For comparison, we also include the Trotter depth for the case of periodic boundary conditions (Appendix~\ref{Appendix D}) and for a 1D Hubbard model with an equivalent number of lattice sites (Appendix~\ref{Appendix F}). Here we see the quadratic scaling with $t$ and linear scaling with $U$ and $N$ illustrated for both 1D and 2D models. Significantly, the difference in cost between 1D and 2D simulations is less than a factor of ten for small system sizes, but still larger than what can be efficiently simulated classically.       

\subsection{t-J}

We can follow an exactly analogous process for the t-J model. In Table~\ref{tbl:tJterms} we break up each single site term in the t-J Hamiltonian, Eq.~\eqref{eq:tJHJW}, into terms that consist only of products of Pauli operators, along with some common prefactors. Then using, 
\begin{equation}
    A^{i_1,j_1}_{i_2,j_2} \equiv \sum_{\delta_1, \delta_2 = 1}^{64} || \left[H_{i_1, j_1}^{\delta_1},H_{i_2, j_2}^{\delta_2}\right]||,
\end{equation}
where each $H_{i,j}^{\delta}$ is listed in Table~\ref{tbl:tJterms}, we calculate each $A^{i_2,j_2}_{i_1,j_1}$ in Eq.~\eqref{eq:ComAH}.

For the t-J model, each $A^{i_2,j_2}_{i_1,j_1}$ will be of the form,
\begin{equation}
    \label{eq:AtJ}
   A^{i_2,j_2}_{i_1,j_1} =  2\left(a_{tJ} \frac{t^2}{64} + b_{tJ} \frac{|t J|}{128} + c_{tJ} \frac{J^2}{256} \right).
\end{equation}
Unlike the Hubbard model, neither $a_{tJ}$, $b_{tJ}$, or $c_{tJ}$ will always be zero, since the $J$ terms arise from a second-order perturbation that mixes both hopping and interaction. Let us again consider each $A^{i_1,j_1}_{i_2,j_2}$ individually.

For $A^{i,j}_{i,j}$, of the 4096 commutators in~\eqref{eq:AtJ}, 1600 are non-zero. Of these 1600 commutators, 384 of them correspond to $t^2/32$, 1024 to $|tJ|/128$, and 192 to $J^2/256$. Thus we have $a_{tJ} = 384$, $b_{tJ}= 1024$, and $c_{tJ} = 192$ for$A^{i,j}_{i,j}$. Combining these coefficients with Eq.~\eqref{eq:AtJ} we arrive at,
\begin{equation}
    A^{i,j}_{i,j} = A^{1,1}_{1,1} =  12 t^2 + 16 |tJ|+ \frac{3}{2} J^2.
\end{equation}
For $A^{i,j}_{i,j+1}$ and $A^{i,j}_{i+1,j}$ we find 960 nonzero commutators, with 256 corresponding to $t^2/32$, 512 to $|tJ|/128$, and 192 to $J^2/256$. Thus,
\begin{equation}
    A^{i,j}_{i,j+1} =  A^{i,j}_{i+1,j} = A^{1,1}_{1,2} = A^{1,1}_{2,1} = 8 t^2 + 8 |tJ|+ \frac{3}{2} J^2.
\end{equation}
For $A^{i+1,j}_{i,j+1}$ there are only 480 nonzero commutators, with 128 corresponding to $t^2/32$, 256 to $|tJ|/128$, and 96 to $J^2/256$. Thus,
\begin{equation}
   A^{i+1,j}_{i,j+1} = A^{2,1}_{1,2} =  4 t^2 + 4 |tJ|+ \frac{3}{4} J^2.
\end{equation}

Combining these results with Eq.~\eqref{eq:ComAH}, we find that the functional form of the commutator bound on the Trotter depth for the 2D t-J model is,
\begin{equation}
    \label{eq:2DcommtJ}
    \begin{split}
    r^{tJ}_{\mathrm{com}} = &N_x N_y \left(12 t^2 + 16 |tJ|+ \frac{3}{2} J^2\right) \\&+2 N_y (N_x-1)  \left(8 t^2 + 8 |tJ|+ \frac{3}{2} J^2\right) \\ &+ 2 N_x (N_y-1) \left(8 t^2 + 8 |tJ|+ \frac{3}{2} J^2\right) \\& +2 (N_y-1) (N_x-1) \left(4 t^2 + 4 |tJ|+ \frac{3}{4} J^2\right) 
   \end{split}
\end{equation}
As in the case of the Hubbard model, we see that the bound on Trotter depth for the 2D t-J model scales linearly with $N \equiv N_x N_y$ and quadratically with $t$. However, while the Hubbard model scaled linearly with $U$, the t-J model scales quadratically with $J$. This is expected, as the $J$ term in the t-J model arises from a second-order perturbation that mixes both the hopping and interaction.  

\begin{figure*}
	\subfigure[]{
		\includegraphics[width=.28\textwidth]{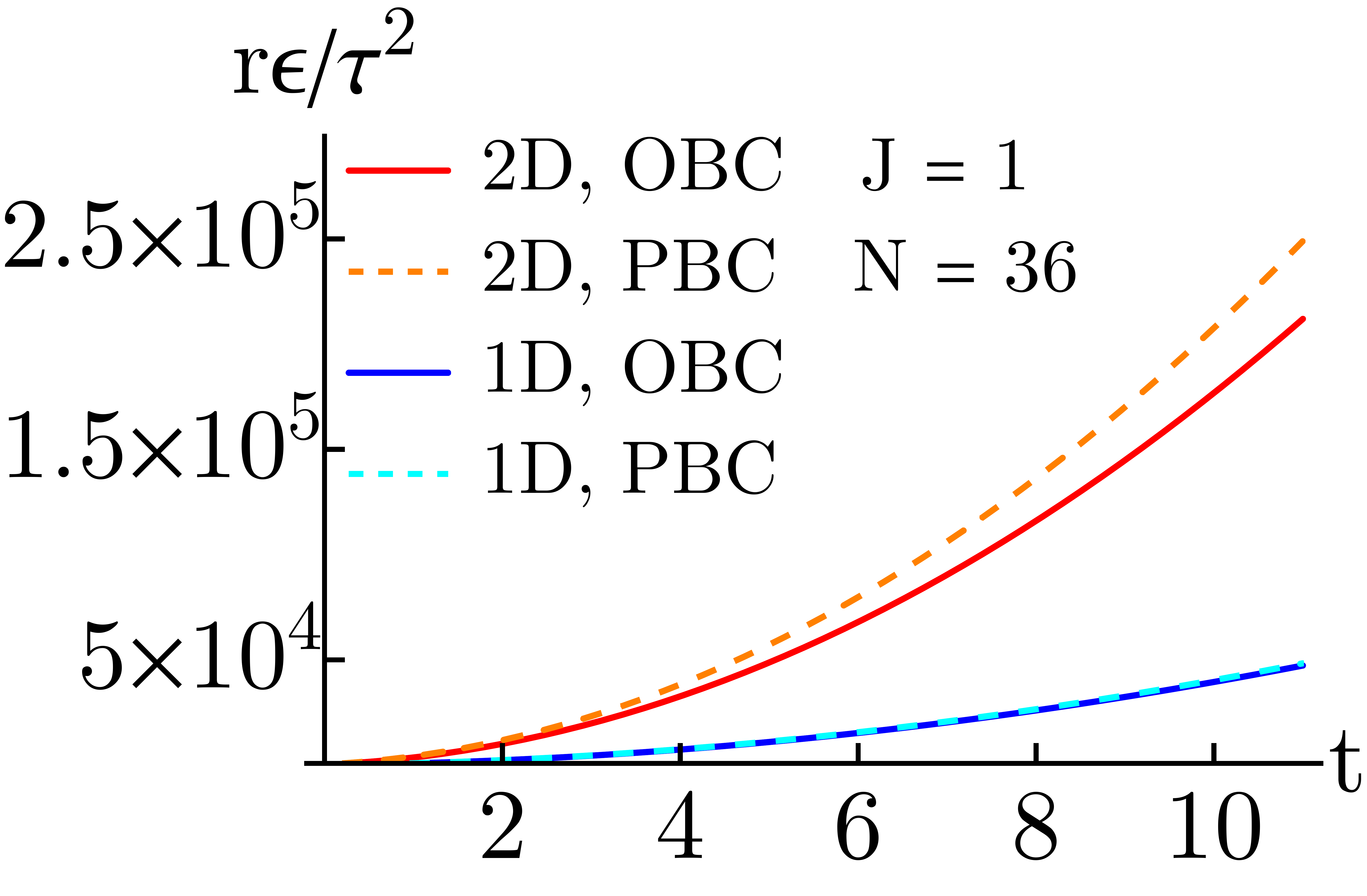}
	}
	\hspace{5mm}
	\subfigure[]{
		\includegraphics[width=.28\textwidth]{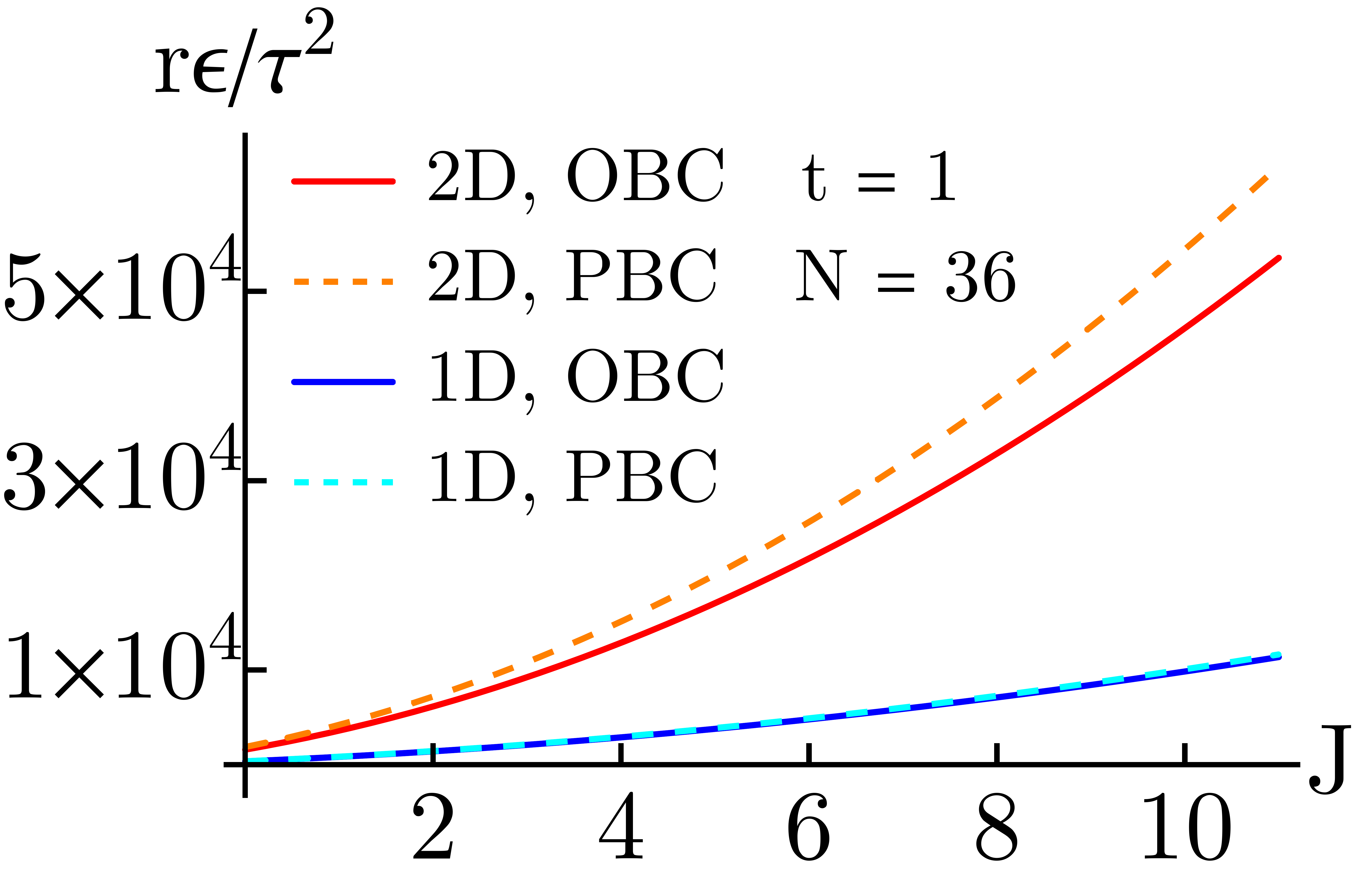}
	}
	\hspace{5mm}
	\subfigure[]{
		\includegraphics[width=.28\textwidth]{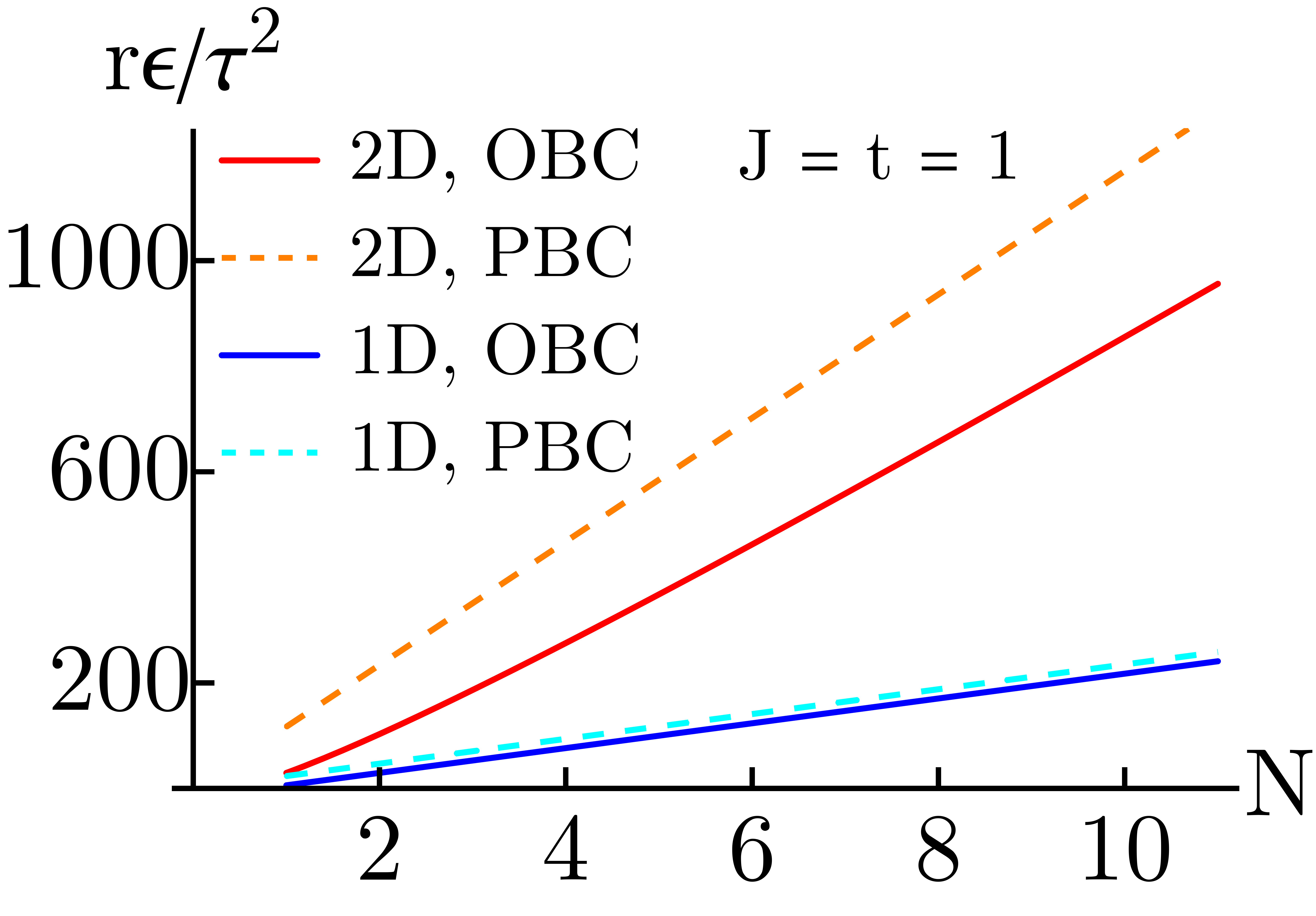}
	}
    \caption{Commutator bound on the Trotter depth for the t-J model as a function of (a) $t$ with $J=1$, $N=36$ (b) $J$ with $t=1$, $N=36$ and (c) $N$ with $J=t=1$. The upper pair of lines show the Trotter depth for a 2D model with periodic (orange, dotted) and open (red, solid) boundary conditions. The lower pair of lines correspond to a 1D model with periodic boundary conditions (dark blue, dashed) and open boundary conditions (light blue, solid).}
    \label{fig:tJTrotterDepth}
\end{figure*}

In Fig.~\ref{fig:tJTrotterDepth} we plot the functional form of the Trotter depth from Eq.~\eqref{eq:2DcommtJ} as a function of $t$, $J$, and $N$, again for a 6 by 6 lattice. As before, we include the Trotter depth for the case of periodic boundary conditions (and for a 1D t-J model with an equivalent number of lattice sites. As we observed for the Hubbard model, we see that the Trotter cost for both the 1D and 2D models scales linearly with $N$, with less than a factor of ten difference between them at $N=36$. 

\subsection{Comparison: Hubbard vs t-J}

\begin{figure*}
	\subfigure[]{
		\includegraphics[width=.28\textwidth]{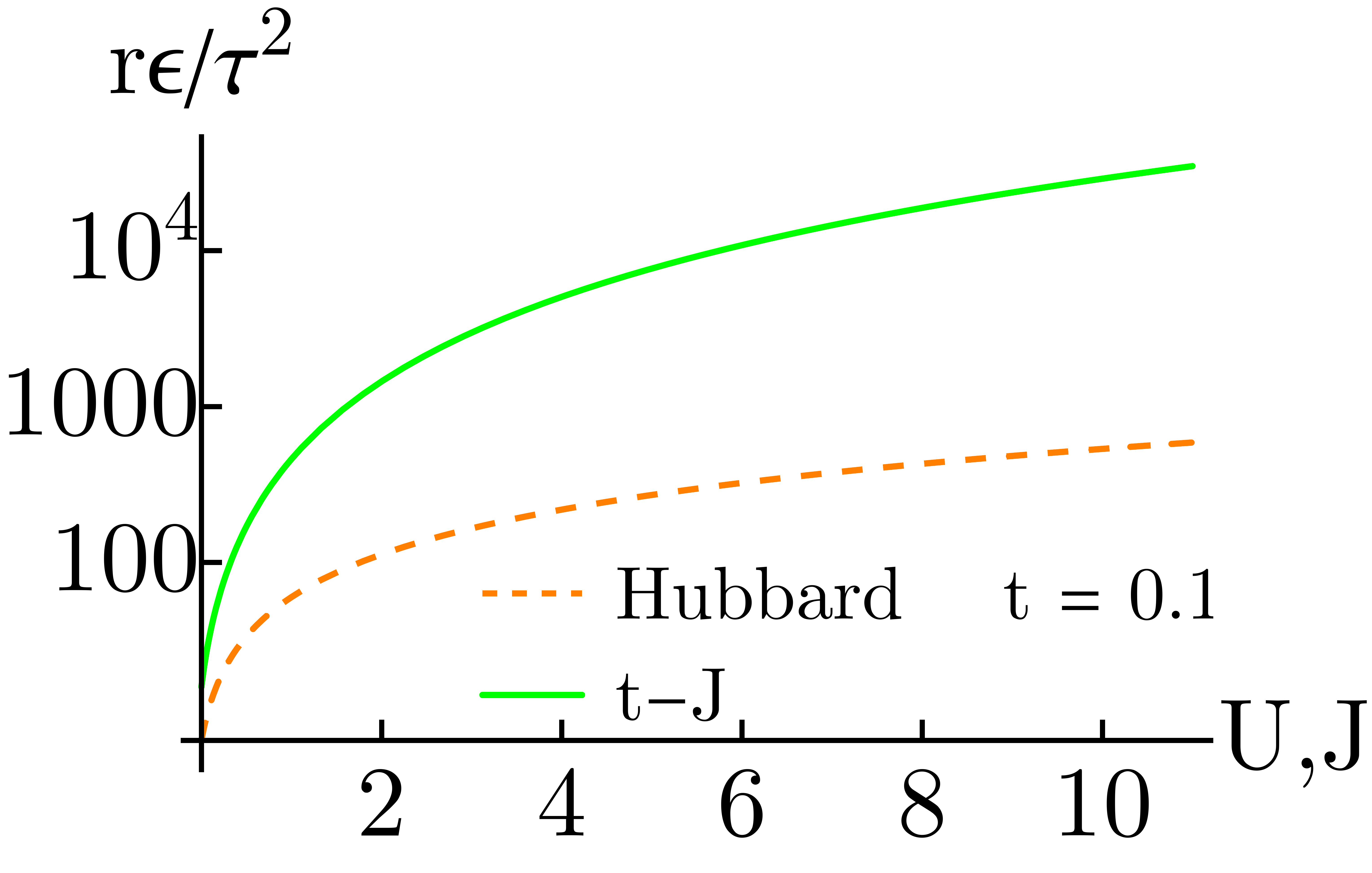}
	}
	\hspace{5mm}
	\subfigure[]{
		\includegraphics[width=.28\textwidth]{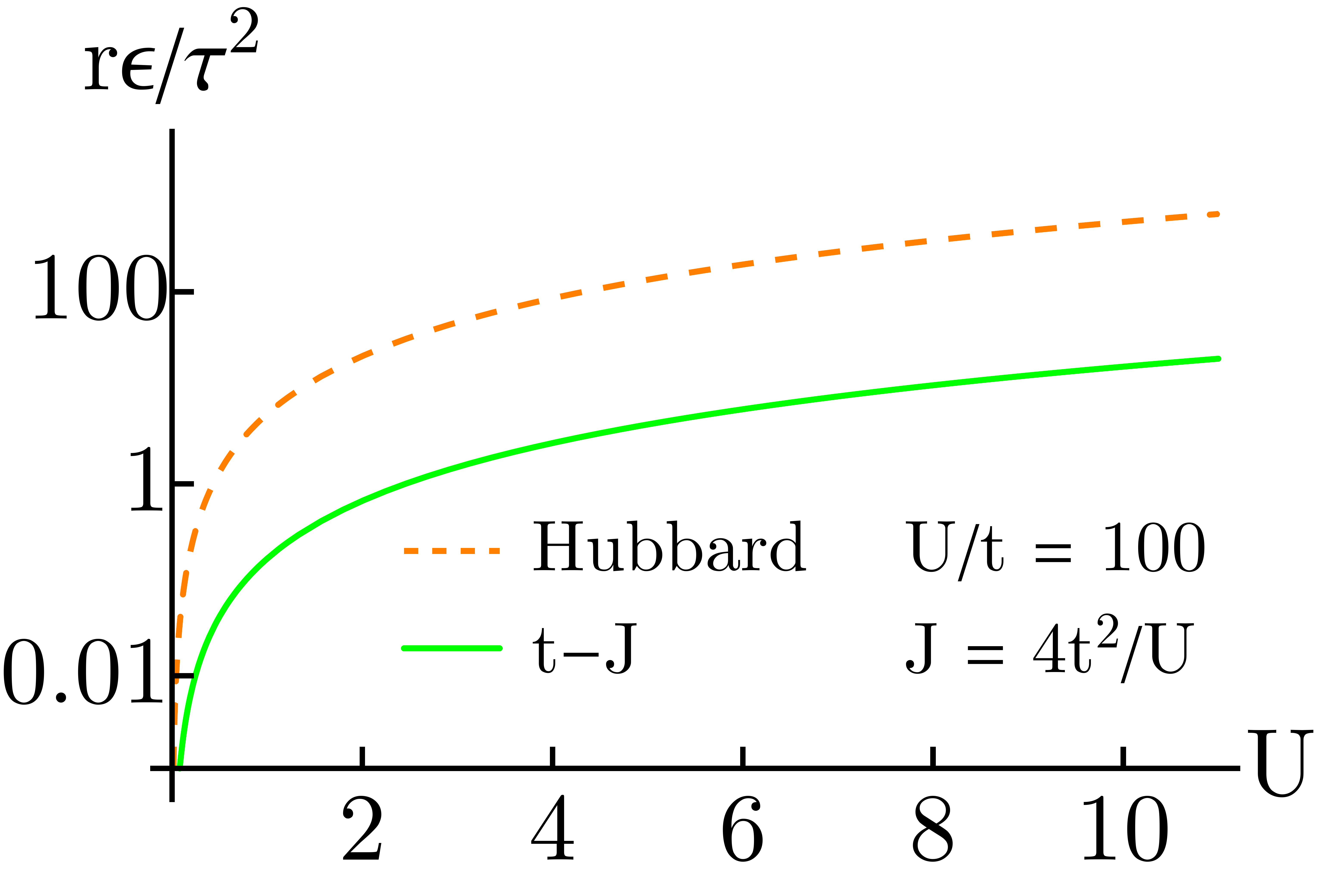}
	}
	\hspace{5mm}
	\subfigure[]{
		\includegraphics[width=.28\textwidth]{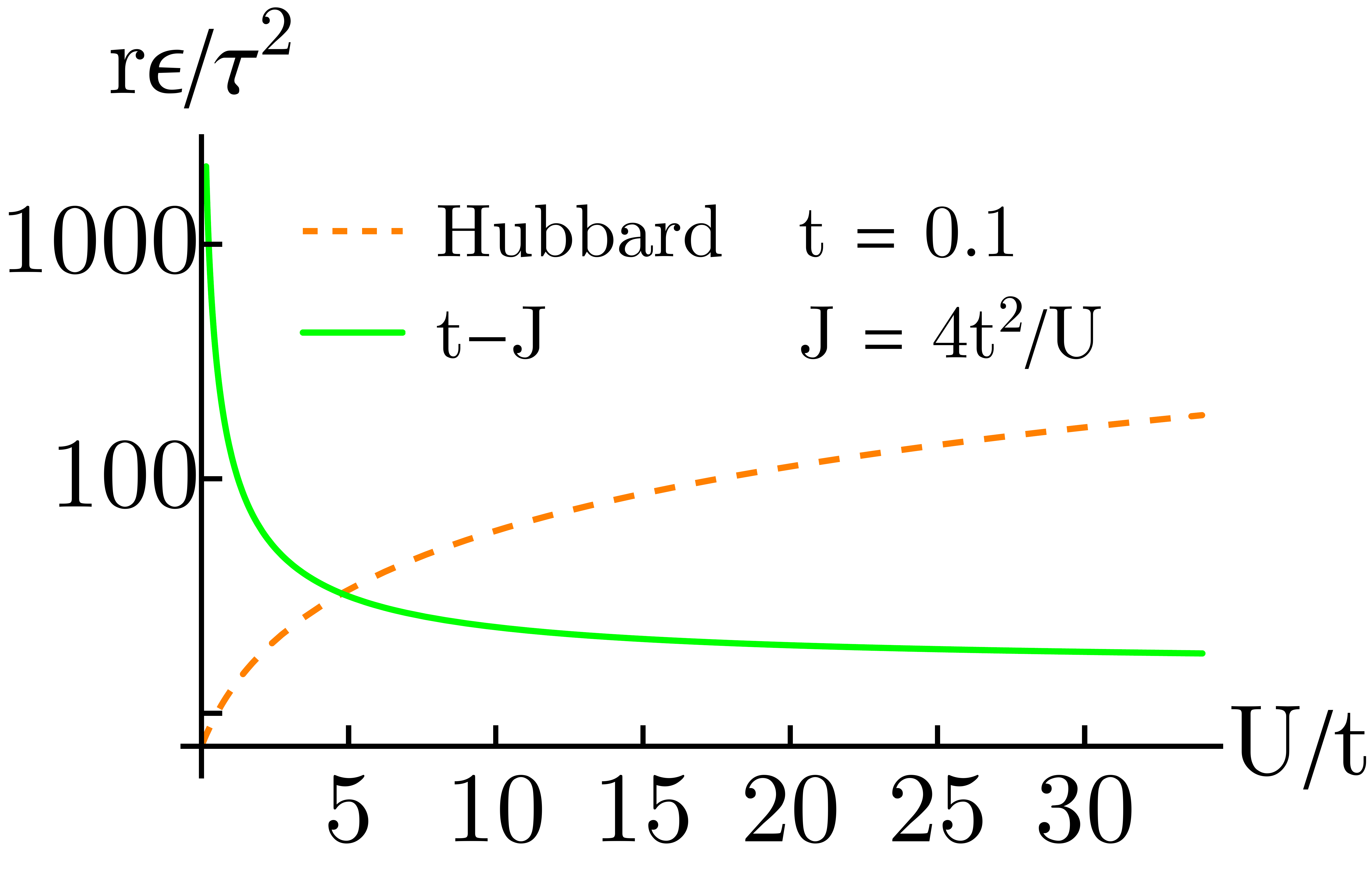}
	}
    \caption{Log scale comparison for the commutator bound on Trotter depth for the 2D Hubbard (orange, dashed) and t-J (green, solid) models. In (a) we fix $t=0.1$ and plot the bound as a function of $U$ for the Hubbard model and as a function of $J$ for the t-J model. In (b) we fix $U/t = 100$ and take $J = 4t^2/U$ and plot the bound on both models as a function of $U$. In (c) we fix $t=0.1$, take $J = 4t^2/U$ and plot both models as a function of $U/t$ For all plots we have taken $N_x=N_y=6$.}
    \label{fig:tJHubComp}
\end{figure*}

With the functional form of the Trotter bound for both Hubbard and t-J, we can now make a direct comparison between the models. As we do so, we must be careful in making sure our comparison is fair, as the relevant parameters in the Hubbard model are $t$ and $U$ while in the t-J model the relevant parameters are $t$ and $J$. In Fig.~\ref{fig:tJHubComp}a we compare the two bounds treating $t$ as fixed and $U$ and $J$ as independent parameters that are varied over the same range of values. In this case, we see that the bound on the Hubbard model is significantly lower than the t-J model. 

However, this comparison is somewhat disingenuous, as we know that $U$ and $J$ are not independent, but are related by $J \equiv 4t^2/U$. Furthermore, the t-J model is a valid approximation to the Hubbard model only under the condition that $U/t \gg 1$. To account for both of these factors, in Fig.~\ref{fig:tJHubComp}b we replace $J$ in the t-J model bound with $4t^2/U$ and fix $U/t = 100$. In this case, we see that the bound for the t-J model is over an order of magnitude lower than for the Hubbard model. The reason for this behavior can be intuitively seen when considering Fig.~\ref{fig:Spectrum}. For the Hubbard model, $U$ sets the energy scale, with the gaps between relevant energy levels being on this order. For the t-J model, $J$ sets the energy scale. Thus, a Hubbard simulation that wants to capture the behavior of the model at the energy scale of t-J must be able to resolve energy gaps on the order of $J$, which for the $t$ and $U$ parameters used in Figs.~\ref{fig:Spectrum} and~\ref{fig:tJHubComp}, are two orders of magnitude smaller than $U$. For the same accuracy, the Hubbard simulation must have much greater fidelity, and therefore requires a larger Trotter depth\footnote{While the numerical calculation used to produce Fig.~\ref{fig:Spectrum} was done for 1D Hubbard and t-J models, the same logic applies in the 2D case.}. 

It is important to note that in Fig.~\ref{fig:tJHubComp}, as in our other comparison plots, we have plotted the bound in terms of the dimensionless quantity $r \epsilon/\tau^2$, which is the maximum number of Trotter steps multiplied by the allowed error divided by the square of the dimensionless time parameter. When comparing the Hubbard and t-J model, the implicit assumption of this approach is that the allowed error, $\epsilon$, is the same for both models. If we wish to treat $U$ and $J$ as independent parameters, as in Fig.~\ref{fig:tJHubComp}a, an alternative approach that maintains the fairness of the comparison is to realize that $\epsilon$ must be different for each model. For the Hubbard model it is logical to set $\epsilon$ as a fraction of $U$ while for the t-J model it is logical to set it as a fraction of $J$. Since the Hubbard model must resolve energy spacing on the order of $J$ to have accuracy comparable to the t-J model, and since $U \gg J$, this means $\epsilon_{H} \ll \epsilon_{tJ}$.   

As an illustration of this behavior, let us consider a specific example for parameters of $t=0.1$ and $U=10$. In order to resolve the energy spectra of both models with equal accuracy, we need to pick an $\epsilon$ significantly lower than the smallest characteristic energy scale, which in this case is given by $J = 4t^2/U$. Let us choose $\epsilon = 0.1 J$. Thus we see that for the Hubbard model we have an $\epsilon$ that is five orders of magnitude smaller than the characteristic model energy scale, while for the t-J model we have an $\epsilon$ that is only one order of magnitude smaller than the characteristic energy scale. Assuming an evolution time of $\tau = 10t$ and plugging all our parameters into Eqs. ~\eqref{eq:2DcommH} and~\eqref{eq:2DcommtJ} we arrive at $r^H_{\mathrm{com}} \approx 1.34 \times 10^6$ and $r^{tJ}_{\mathrm{com}} \approx 4.15 \times 10^4$, indicating that the Hubbard model is over 30 times more costly to simulate than the t-J model at these parameters.   

From these results, it is clear that the best choice of model to simulate depends on the desired parameter range, namely the ratio of $U/t$. With this in mind, in Fig.~\ref{fig:tJHubComp}c we plot the bound on the Trotter depth as a function of this ratio. We see that as $U/t \rightarrow 0$ the bound for the Hubbard model vanishes. This is expected, as in the limit of vanishing $U$ with finite $t$ all terms in the Hubbard Hamiltonian commute. Conversely, as $U/t \rightarrow 0$ the bound for the t-J model blows up. This occurs since the t-J bound grows with $J$ and $J$ is inversely related to $U/t$. Physically, this can be seen as corollary to the behavior discussed in the previous two paragraphs. As we move beyond the parameter range satisfying $U \gg t$, the energy scale grows much larger than the  typical spacing of the t-J model, and eventually the assumptions going into the derivation of the model break down.      

As the number of Trotter steps is directly proportional to the circuit depth, this advantage translates to a significant reduction in overall gate count for simulating the t-J model in comparison to Hubbard. These results indicate that, on NISQ devices where gate errors and decoherence are limiting factors on circuit design, the t-J model is a more amenable target for near-term simulation.   

\section{Concluding Remarks} 
\label{sec:6}

In this work, we constructed a lattice-fermion formalism for studing the Trotter depth of important condensed models. We applied our approach to the Hubbard and t-J models, motivated by the supposition that the reduced Hilbert space of the t-J model due to the elimination of any doubly-occupied states would naturally lead to lower computational costs. Our results show that the situation is significantly more nuanced than this straightforward hypothesis. 

In terms of the Pauli depth, we find that we gain no advantage from the reduced Hilbert space. While it is true that the t-J model has only three basis states for each orbital, $\ket{0}$, $\ket{\uparrow}$, and $\ket{\downarrow}$, in comparison to the four basis states of the Hubbard model, which also allows $\ket{\uparrow \downarrow}$, both cases still require two qubits to represent their respective bases. In this case, the t-J model is actually disadvantaged, as extra qubit overhead is required to implement the projection operators that ensure the system never strays out of the reduced Hilbert space. We note, however, that this is naturally a consequence of using a fermion-\textit{qubit} mapping. If we instead applied a \textit{qutrit} based mapping for the t-J model, the additional overhead would be mitigated (an example of a fermion-qutrit mapping for the 2D t-J model can be found in Ref.~\cite{Po2021}). While the difference in Pauli depth is not significant in the 2D case, as the Jordan-Wigner transformed Hamiltonians for both the Hubbard and t-J models contain terms that span the full width of the lattice, it plays a larger role in the 1D case where these non-local Jordan-Wigner strings cancel out (see Appendix~\ref{Appendix F}).

In terms of the bound on Trotter depth, we find the situation to be more in-line with our initial hypothesis. If we naively treat $U$ and $J$ on the same footing, the bound first appears to favor the Hubbard model by around two orders of magnitude. However, if we properly account for the fact that the t-J model is a valid approximation to the Hubbard model only 
under the condition $U >> t$, and that $J$ is in reality a function of $U$ and $t$, then we see that the bound significantly favors the t-J model. Thus, the most efficient choice of model depends on the energy scale that the simulation needs to resolve.    

We provide the full form of the Jordan-Wigner transformed Hamiltonians for the  Hubbard and t-J models. In particular, we include an explicit qubit representation of the 2D t-J model without relying on alternative approaches, such as the auxiliary particle scheme~\cite{Barnes2002}. While this work provides a first benchmark of comparison between the Hubbard and t-J models, there remain numerous avenues for future exploration. Chief among these is the extension to higher order product formulas, using the generalized bounds provided in Ref.~\cite{Childs2021}. More compact fermion-qubit mappings that reduce the Pauli depth through the use of ancilla qubits~\cite{Bravyi2002, Ball2005, Verstraete2005, Whitfield2016, Havlicek2017, Steudtner2019, Setia2019, Derby2021} or via circuit design that leads to the cancellation of Jordan-Wigner strings~\cite{Hastings2015} could also be considered. However, we note that, as the Pauli depth for both the Hubbard and t-J models is equivalent outside of 1D, any compact mapping based on reducing the Jordan-Wigner string will benefit both models equally and will not impact that advantage in Trotter depth that the t-J model possesses.

A more detailed accounting of the resource overhead could also be carried out in order to provide specific gate count estimates for each model, which will serve as a necessary prerequisite to implementing simulations on real-world devices. Furthermore, it would be of interest to determine if the t-J model shares a similar advantage under other simulation schemes, such as hybrid quantum-classical variational algorithms \cite{Lyu2020, Keever2022}. Finally, we note that the general nature of the formalism developed in Section~\ref{sec:2} makes it straightforward to extend this approach to other lattice models with different geometries or to orbital models for application to quantum chemistry simulations.

\begin{acknowledgments}

N. M. M., R.S., and V.W.S. acknowledge support from AFOSR (FA2386-21-1-4081, FA9550-19-1-0272,FA9550-23-1-0034) and ARO (W911NF2210247, W911NF2010013). This work was in part supported by the KIAS Individual Grant, PG032303 (K. P.), and also by the National Research Foundation of Korea (NRF) funded by the Ministry of Science and ICT of the Korean Government, NRF-2021M3H3A1085208 (K. P.).

\end{acknowledgments}
\hfill \break
\appendix

\onecolumngrid

\section{Spectral Norm of Pauli Products}
\label{Appendix A}

In this appendix we prove the spectral norm of a Kronecker product of any number of Pauli matrices is always unity. Let us consider the Kronecker product of $N$ Pauli matrices,
\begin{equation}
    \label{eq:PauliProd}
    \prod_{i = 1}^{N} \sigma_i^{\alpha}
\end{equation}
where $\alpha = 0,x,y,z$ with $\sigma^0$ being the identity matrix. As transposition is distributive over the Kronecker product, and as the Pauli Matrices are Hermitian, the matrix in Eq.~\eqref{eq:PauliProd} must be Hermitian. Thus, the spectral norm will be given by the absolute value of its largest magnitude eigenvalue. 

The eigenvalues of $\sigma^x$, $\sigma^y$, and $\sigma^z$ are each $\{-1,1\}$ while the eigenvalues of $\sigma^0$ are $\{1,1\}$. As the eigenvalues of the Kronecker product of two square matrices are given by the products of the eigenvalues of the individual matrices, we can see that the eigenvalues of Eq.~\eqref{eq:PauliProd} will be $-1$ and $1$, each with multiplicity $2^{N/2}$. Thus,
\begin{equation}
    \left|\left|\prod_{i = 1}^{N} \sigma_i^{\alpha} \right|\right| = 1
\end{equation}

\section{Fermionic representation of the t-J model}
\label{Appendix B}

In this appendix we derive an expression for the 2D t-J model entirely in terms of the fermionic creation, annihilation, and number operators. Starting from the typical t-J Hamiltonian,
\begin{equation}
    \label{eq:tJHamA}
    H^{tJ} = - t \sum_{<ij,kl>} \sum_{s\in\{\uparrow,\downarrow\}} P \left( c^{\dagger}_{ij,s}c_{kl,s} + c_{ij,s}c^{\dagger}_{kl,s}\right) P \\ 
    + J \sum_{<ij,kl>} P \Big[S_{ij} \cdot S_{kl} - \frac{n_{ij} n_{kl}}{4}\Big] P,
\end{equation}
we note,
\begin{equation}
    S_{ij} \cdot S_{kl} = S_{ij}^x S_{kl}^x + S_{ij}^y S_{kl}^y + S_{ij}^z S_{kl}^z,
\end{equation}
and
\begin{equation}
    S_{kl}^+ = S_{kl}^x + i S_{kl}^y \quad \text{and} \quad S_{kl}^- = S_{kl}^x - i S_{kl}^y.
\end{equation}
Thus we have,
\begin{equation}
    \label{eq:spinTerm}
    S_{ij} \cdot S_{kl} = \frac{1}{2}\left(S_{ij}^+ S_{kl}^- + S_{ij}^- S_{kl}^+ \right) + S_{ij}^z S_{kl}^z.
\end{equation}
Now using~\cite{Fazekas1999},
\begin{equation}
    S_{kl}^+ = c_{kl,s}^{\dagger} c_{kl, \bar{s}}, \quad S_{kl}^- = c_{kl, \bar{s}}^{\dagger} c_{kl,s}, \quad S_{kl}^z = \frac{1}{2}(n_{kl,s} - n_{kl, \bar{s}}),
\end{equation}
we can express Eq.~\eqref{eq:spinTerm} as,
\begin{equation}
    S_{ij} \cdot S_{kl} = \frac{1}{2}\left( c_{ij,\uparrow}^{\dagger} c_{ij,\downarrow} c_{kl,\downarrow}^{\dagger} c_{kl,\uparrow} +  c_{ij,\downarrow}^{\dagger} c_{ij,\uparrow} c_{kl,\uparrow}^{\dagger} c_{kl,\downarrow}\right)
    + \frac{1}{4} \left(n_{ij, \uparrow}n_{kl, \uparrow} - n_{ij, \uparrow} n_{kl,\downarrow} - n_{ij, \downarrow} n_{kl,\uparrow} + n_{ij, \downarrow} n_{kl,\downarrow} \right)
\end{equation}
Thus,
\begin{equation}
    S_{ij} \cdot S_{kl} - \frac{n_{ij} n_{kl}}{4} = S_{ij} \cdot S_{kl} - \frac{(n_{ij, \uparrow}+n_{ij, \downarrow})(n_{kl, \uparrow}+ n_{kl, \downarrow})}{4} = \frac{1}{2} \sum_{s\in\{\uparrow,\downarrow\}} \left( c^{\dagger}_{ij,s} c_{ij,\bar{s}} c^{\dagger}_{kl,\bar{s}}c_{kl,s} - n_{ij,s} n_{kl, \bar{s}}\right)
\end{equation}  

The projection operators can be explicitly accounted for by replacing the fermionic operators in Eq.~\eqref{eq:tJHamA} with the projected operators~\cite{Spalek1988},
\begin{equation}
    \label{eq:projOps}
    \begin{split}
    c^{\dagger}_{ij,s} \rightarrow c^{\dagger}_{ij, s} (1 - n_{ij, \bar{s}}), \\ 
    \quad c_{ij, s} \rightarrow c_{ij, s} (1 - n_{ij, \bar{s}}), \\ \quad n_{ij, s} \rightarrow n_{ij, s} (1 - n_{ij, \bar{s}}).
    \end{split}
\end{equation}
Thus we can rewrite Eq.~\eqref{eq:tJHam} as,
\begin{equation}
    \label{eq:projtj}
    \begin{split}
    H^{tJ} = &- t \sum_{<ij,kl>} \sum_{s\in\{\uparrow,\downarrow\}} \left[ \left(1-n_{ij,\bar{s}}\right)\left(c^{\dagger}_{ij,s}c_{kl,s} + c^{\dagger}_{kl,s}c_{ij,s}\right)\left(1-n_{kl,\bar{s}}\right)\right] \\ &+ \frac{J}{2}\sum_{<ij,kl>} \sum_{s\in\{\uparrow,\downarrow\}} \left[c^{\dagger}_{ij,s}c_{ij,\bar{s}}c^{\dagger}_{kl,\bar{s}}c_{kl,s} - \left(1-n_{ij,\bar{s}}\right)n_{ij,s}n_{kl,\bar{s}}\left(1-n_{kl,s}\right)\right]
    \end{split}
\end{equation}
Note that the projected operators in Eq.~\eqref{eq:projOps} provide a \textit{local} implementation of the the Gutzwiller projection through the elimination of any terms that would ever produce a doubly-occupied site. However, the projectors in Eq.~\eqref{eq:projOps} will not protect against an initial state of the system that contains doubly-occupied sites. In other words, the implicit identity operators that act on every site $pq \neq ij,kl$ contained in each term in Eq.~\eqref{eq:projtj} are not projected.

\section{1-norm bound on Trotter depth}
\label{Appendix C}

In this appendix we determine a looser bound on the Trotter depth for the Hubbard and t-J models by applying the operator norm scaling derived in Ref.~\cite{Childs2021}. For the case of the 1-norm, the bound for a first order product formula is given by,
\begin{equation}
    \label{eq:1norm}
    r_{1-\mathrm{norm}}  = \mathcal{O} \left(\frac{\left(\sum_{\gamma}^{\Gamma} ||H_{\gamma}||\tau \right)^{2}}{\epsilon}\right)
\end{equation}
As in the calculation of the commutator bound, we begin by decomposing the full Hamiltonian into terms that consist only of products of Pauli matrices,
\begin{equation}
    H = \sum_{ij} \sum_{\delta} H_{ij}^{\delta}.
\end{equation}
For the purposes of Eq.~\eqref{eq:1norm}, $H_{\gamma} = H_{ij}^{\delta}$. Thus we have,
\begin{equation}
     r_{1-\mathrm{norm}}  = \frac{\tau^2}{\epsilon} \left(\sum_{ij} \sum_{\delta} ||H_{ij}^{\delta}|| \right)^{2} = \frac{\tau^2}{\epsilon} N^2 \left( \sum_{\delta} ||H_{ij}^{\delta}|| \right)^{2},
\end{equation}
where $N \equiv N_x N_y$ is the total number of sites in the lattice. Since each  $H_{ij}^{\delta}$ consists of a product of Pauli matrices, along with some prefactor, its norm will simply be the absolute value of the prefactor.  

For the 2D Hubbard model there are 12 distinct $H_{ij}^{\delta}$ terms, listed in Appendix~\ref{Appendix E}. Eight of these terms have a prefactor of $t/2$ and four have a prefactor of $U/4$. However, one of the $U/4$ terms is simply an identity term and can be disregarded. Thus for the Hubbard model we have,
\begin{equation}
    \label{eq:2D1normH}
     r^{H}_{1-\mathrm{norm}}  = \frac{\tau^2}{\epsilon} N^2 \left(4t + \frac{3}{4}U\right)^{2}.
\end{equation}

For the 2D t-J model there are 64 distinct $H_{ij}^{\delta}$ terms, also tabulated in Appendix~\ref{Appendix E}. Half of these terms have a prefactor of $t/16$ and the other half have a prefactor of $J/16$. However, two of the $J/16$ terms are identity terms and can be disregarded. Thus for the t-J model we have,
\begin{equation}
    \label{eq:2D1normtJ}
     r^{tJ}_{1-\mathrm{norm}}  = \frac{\tau^2}{\epsilon} N^2 \left(2t + \frac{15}{8}J\right)^{2}.
\end{equation}

\begin{figure*}
	\subfigure[]{
		\includegraphics[width=.28\textwidth]{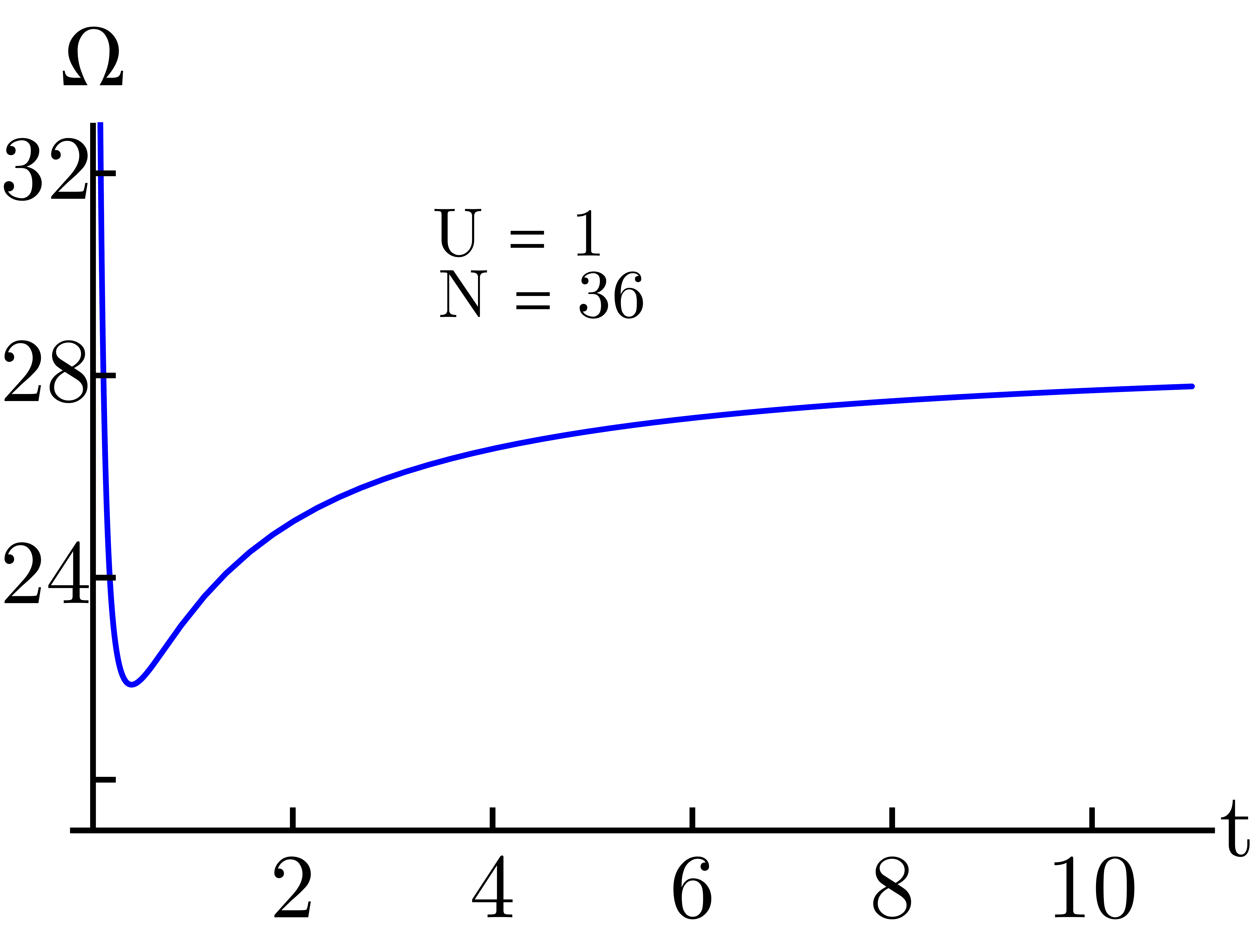}
	}
	\hspace{5mm}
	\subfigure[]{
		\includegraphics[width=.28\textwidth]{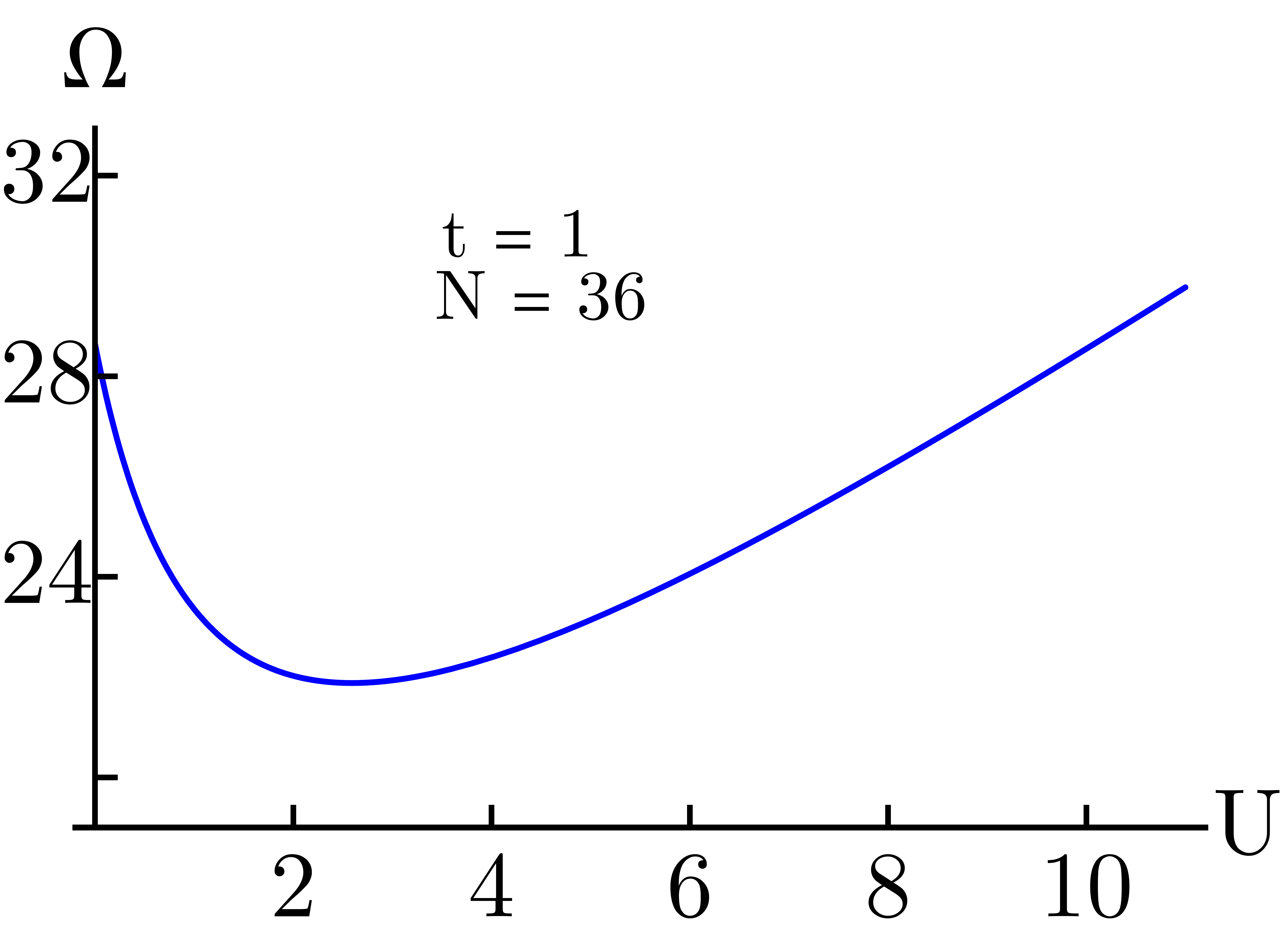}
	}
	\hspace{5mm}
	\subfigure[]{
		\includegraphics[width=.28\textwidth]{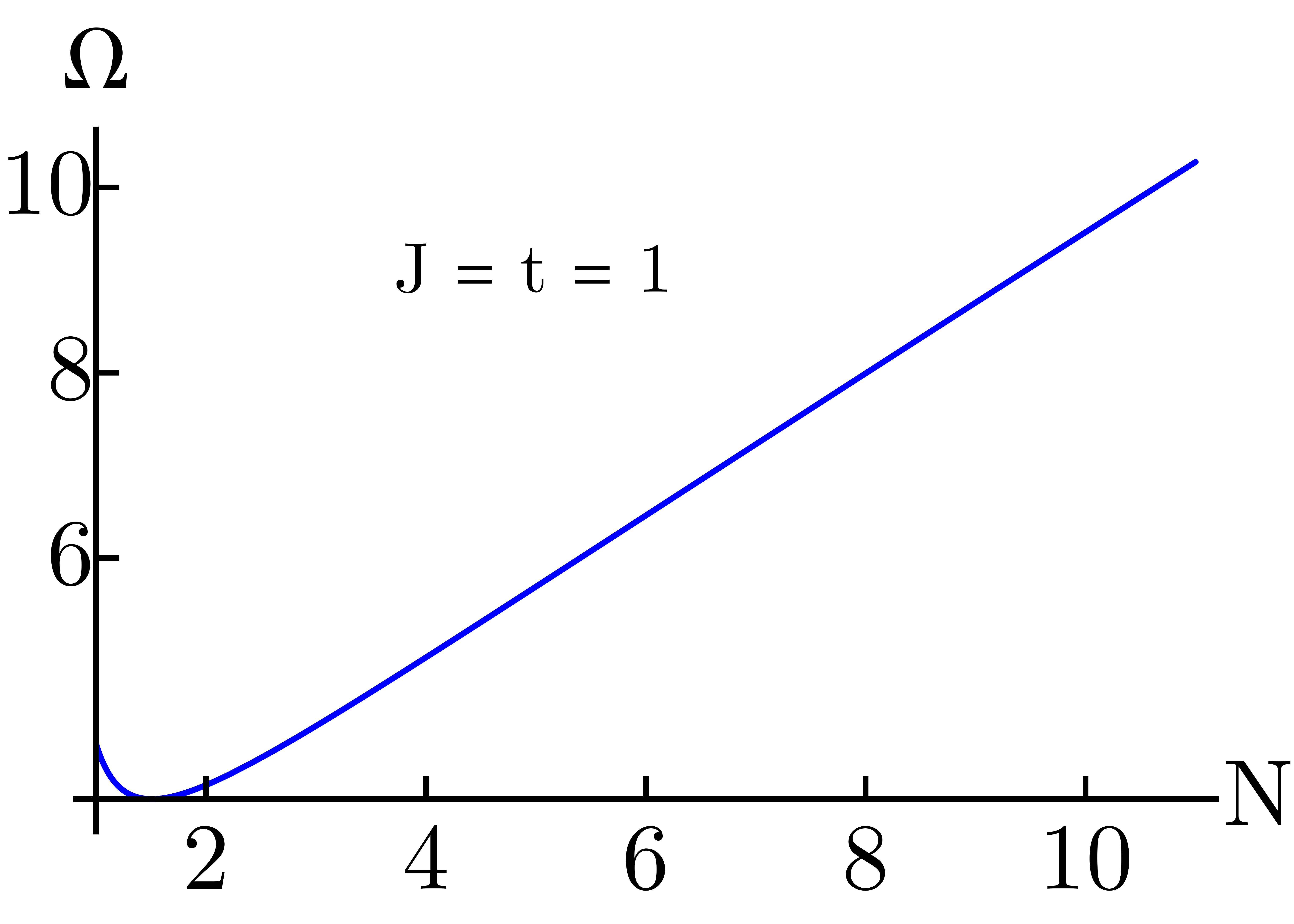}
	}
    \caption{Ratio of the bound on Trotter depth for the 2D Hubbard model arising from the 1-norm to that of the commutator scaling as a function of (a) $t$ with $U = 1$, $N \equiv N_x N_y = 36$, (b) $U$ with $t = 1$, $N = 36$, and (c) $N$ with $U=t=1$.}
    \label{fig:1normH}
\end{figure*}

\begin{figure*}
	\subfigure[]{
		\includegraphics[width=.28\textwidth]{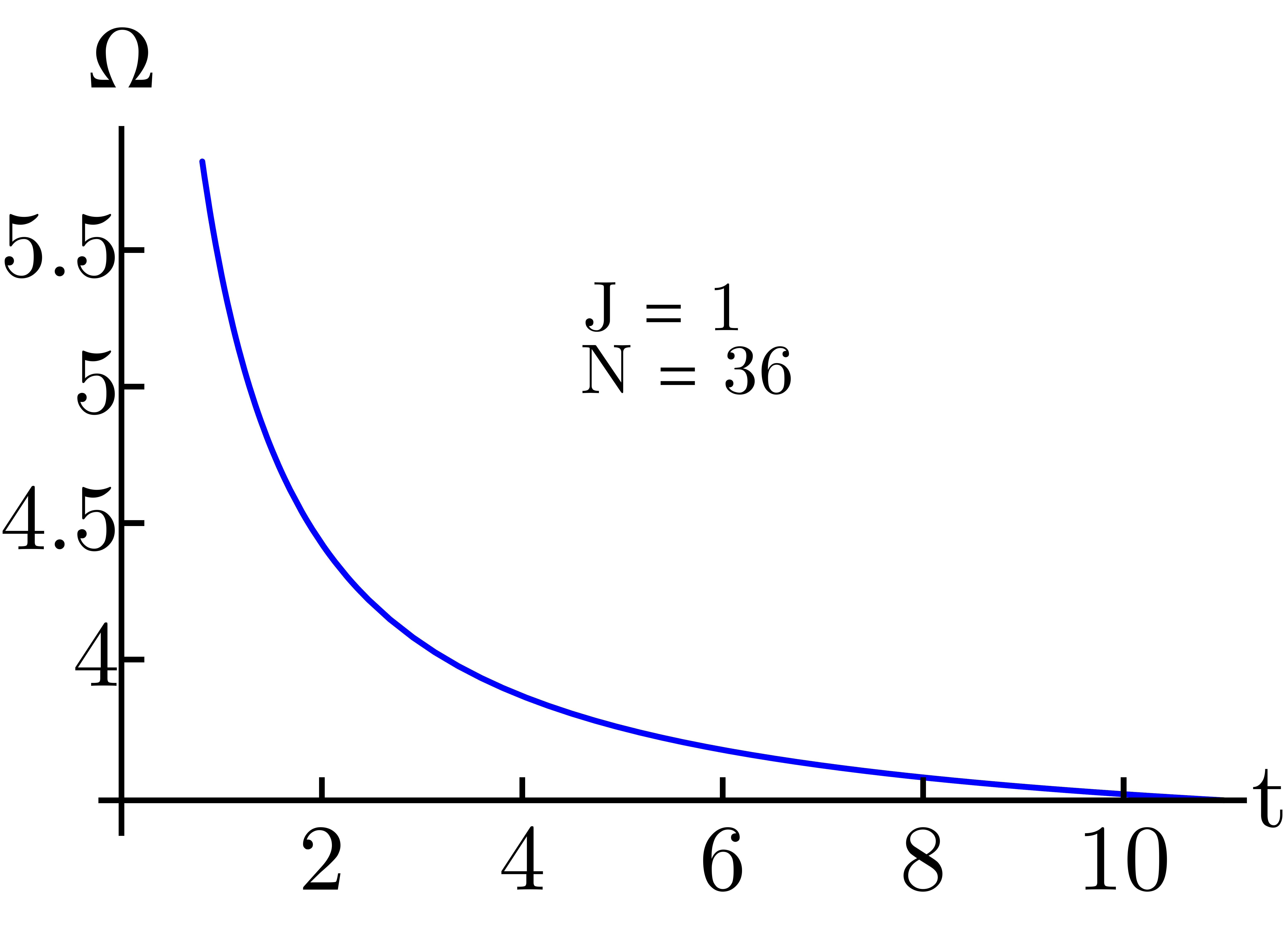}
	}
	\hspace{5mm}
	\subfigure[]{
		\includegraphics[width=.28\textwidth]{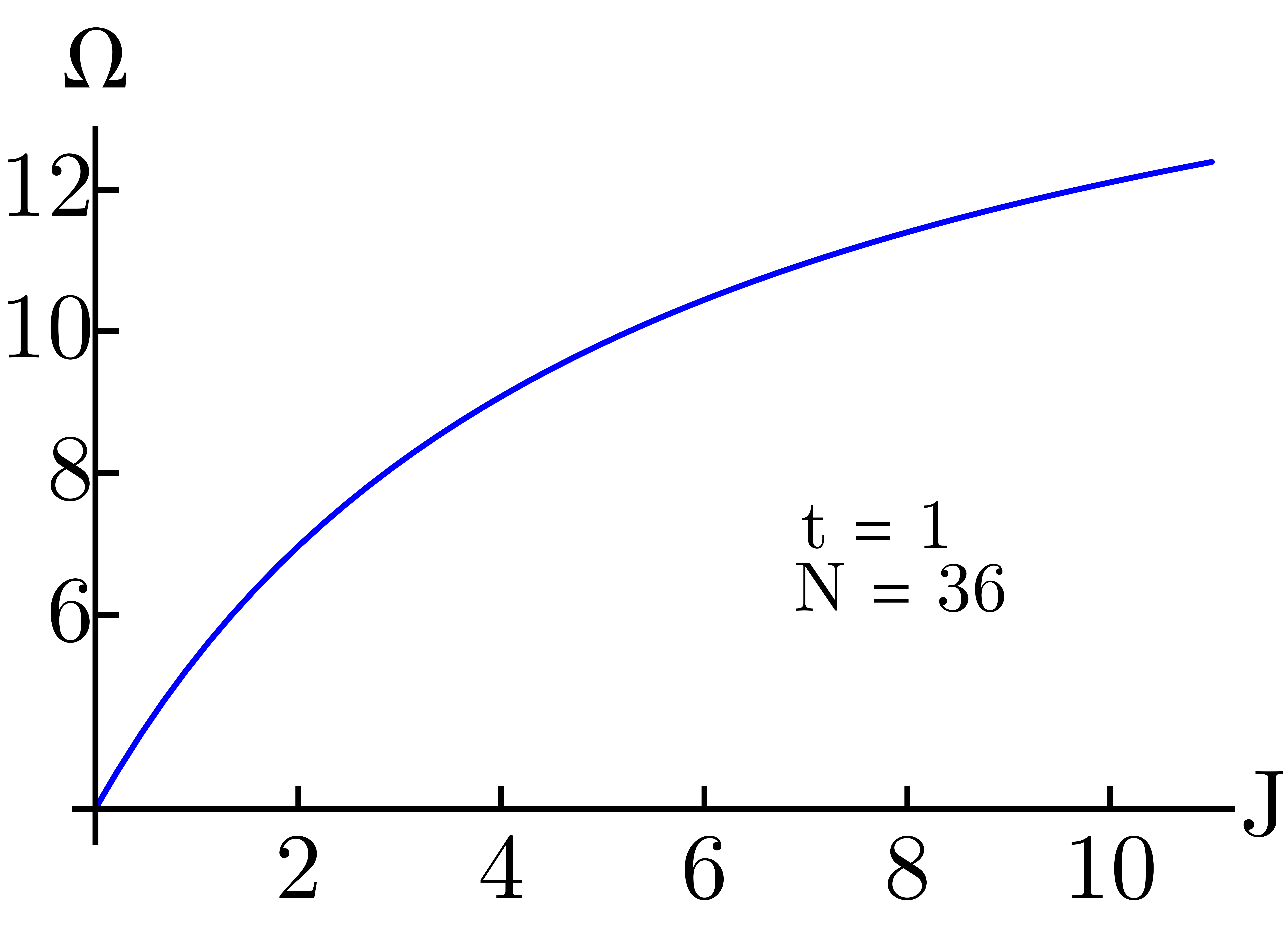}
	}
	\hspace{5mm}
	\subfigure[]{
		\includegraphics[width=.28\textwidth]{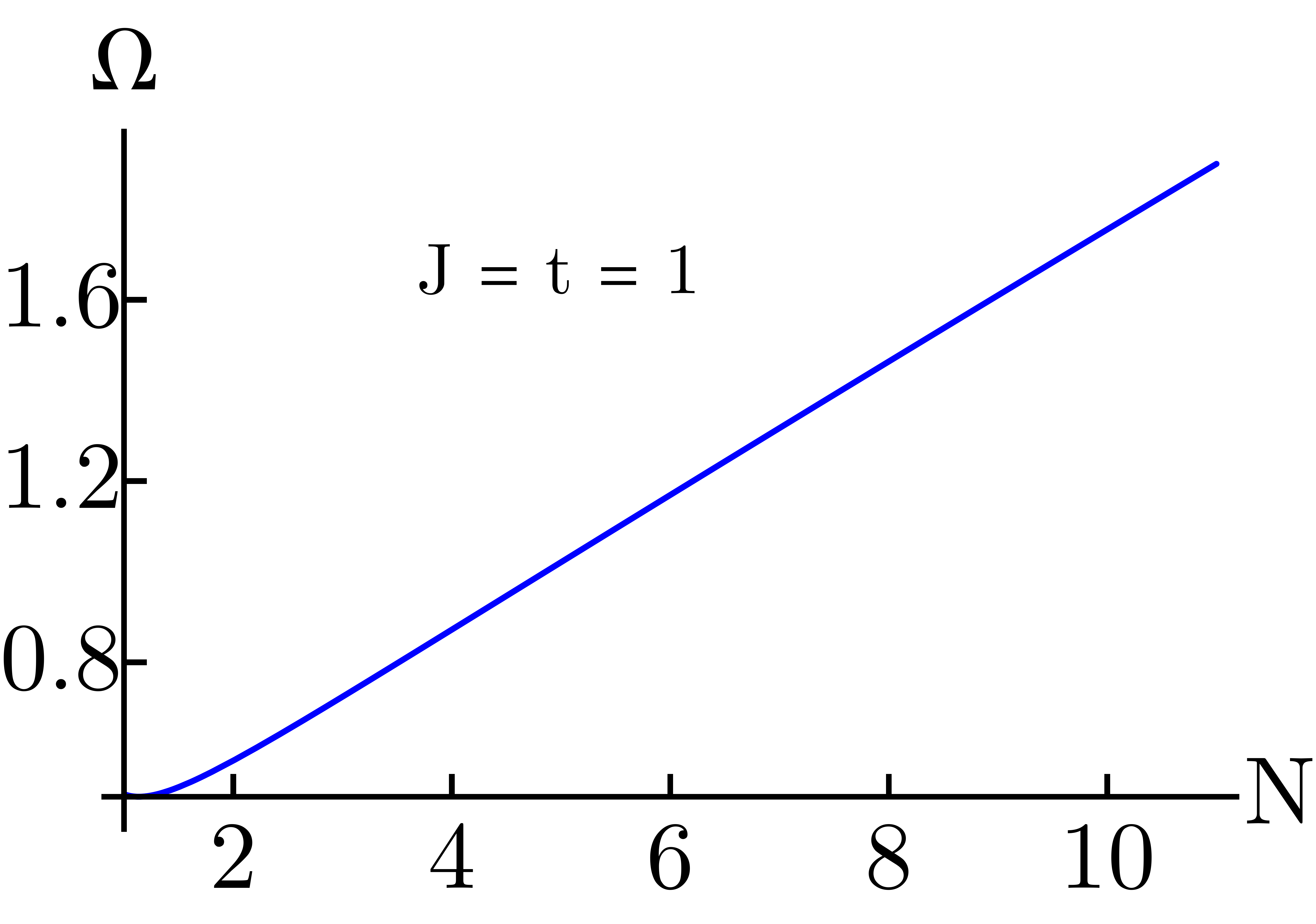}
	}
    \caption{Ratio of the bound on Trotter depth for the 2D t-J model arising from the 1-norm to that of the commutator scaling as a function of (a) $t$ with $J = 1$, $N \equiv N_x N_y = 36$, (b) $J$ with $t = 1$, $N = 36$, and (c) $N$ with $J=t=1$.}
    \label{fig:1normtJ}
\end{figure*}

In order to compare the bounds given by the 1-norm scaling to those given by the commutator scaling, we consider the ratio,
\begin{equation}
    \Omega \equiv r_{1-\mathrm{norm}}/r_{\mathrm{comm}}
\end{equation}
In Fig.~\ref{fig:1normH} and~\ref{fig:1normtJ} we plot this ratio for the 2D Hubbard and t-J models, respectively. We see that, at fixed $N$, the commutator bound is significantly tighter for both models. Notably, the commutator bound grows linearly with $N$ while the 1-norm bound grows quadratically with $N$. Comparing Eqs.~\eqref{eq:2DcommH} and~\eqref{eq:2D1normH} we see that for any fixed $U$ and $t$ there is no lattice size, $N$, where the 1-norm bound will be tighter. However, this is not true for the case of the t-J model. Comparing Eqs.~\eqref{eq:2DcommtJ} and~\eqref{eq:2D1normtJ} we see that at small lattice sizes the 1-norm bound will be tighter due to the large polynomial of $t$ and $J$ in the commutator bound. This is illustrated in Fig.~\ref{fig:1normtJ}c where at small $N$ with fixed $t$ and $J$ the ratio of the bounds is less than one, indicating that the commutator bound is larger. Furthermore, we also see that the ratio of the bounds is monotonic as a function of $t$ and $J$ for the t-J model, but that the Hubbard model displays a clear minimum as a function of both $t$ and $U$. While a minimum in the ratio as a function of $N$ also occurs between $N=1$ and $N=2$ for both models, this does have physical significance, as in practice $N$ is a discrete quantity.             

\section{Periodic boundary conditions}
\label{Appendix D}

\begin{figure*}
	\subfigure[]{
		\includegraphics[width=.34\textwidth]{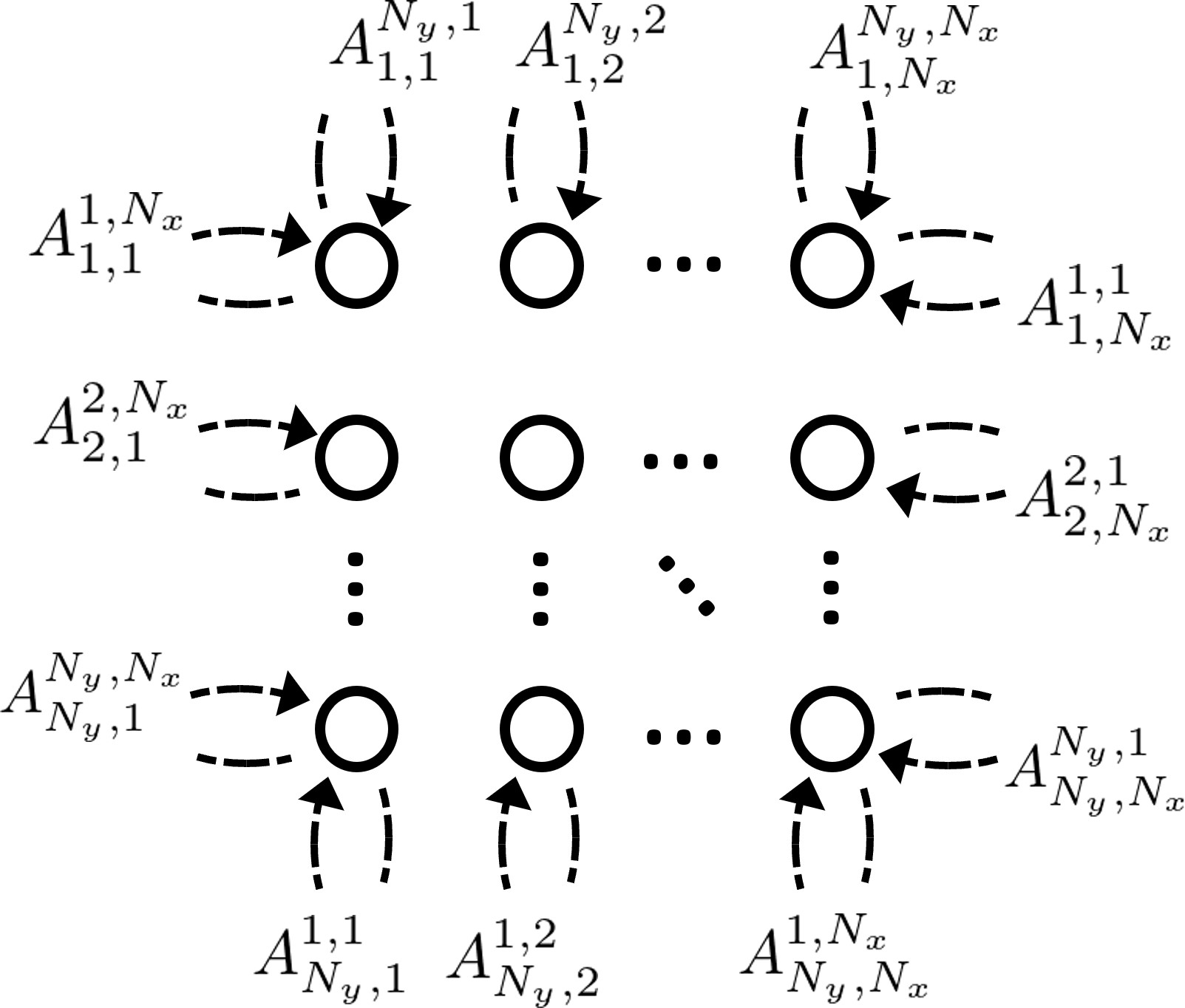}
	}
	\hspace{5mm}
	\subfigure[]{
		\includegraphics[width=.3\textwidth]{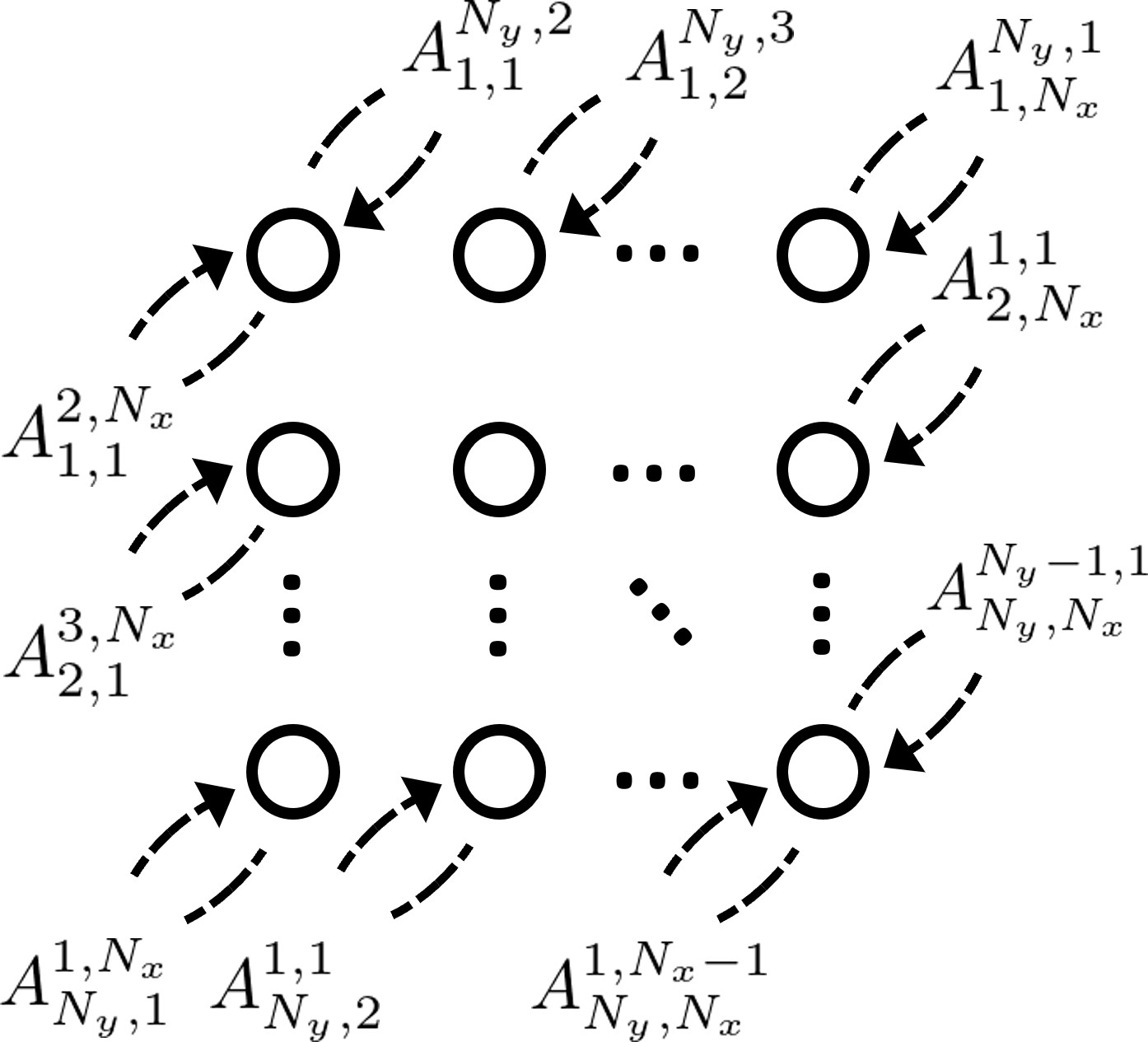}
	}
    \caption{Illustration of the nonzero boundary terms that must be accounted for the case of nearest neighbor hopping under periodic boundary conditions on a $N_x$ by $N_y$ rectangular lattice. In (a) we illustrate terms of the form $A^{i,j}_{i,j+1}$ and $A^{i,j}_{i+1,j}$ and in (b) we illustrate terms of the form $A^{i+1,j}_{i,j+1}$.}
    \label{fig:PeriodicBT}
\end{figure*}

In this appendix we determine the Trotter depth for the 2D Hubbard and t-J models for the case of periodic boundary conditions. We begin by simplifying Eq.~\eqref{eq:ComExpanded} to include only nearest neighbor hopping, as is the case in both the Hubbard and t-J models,
\begin{equation}
    \label{eq:rperiod}
    \begin{split}
    r_{\mathrm{com}} &= \frac{\tau^2}{\epsilon} \Bigg[N_x N_y A^{1,1}_{1,1} + 2 N_y\left(N_x-1\right)A^{1,1}_{1,2} + 2 N_x \left(N_y-1\right)A^{1,1}_{2,1}
    \\&+ 2 \left(N_x-1\right)\left(N_y-1\right)A^{1,1}_{2,2} + 2 \left(N_x-1\right)\left(N_y-1\right)A^{2,1}_{1,2} + \mathrm{b.t.} \Bigg],
    \end{split}
\end{equation}
where b.t. denotes the contribution from the boundary terms. In Fig.~\ref{fig:PeriodicBT} we illustrate the boundary terms that arise for a rectangular lattice of dimension $N_x$ by $N_y$. Let us first consider the terms shown in Fig.~\ref{fig:PeriodicBT}a. We can see that we have $N_x$ terms that take the form $A^{N_y,j}_{1,j}$ and $A^{1,j}_{N_y,j}$ as well as $N_y$ terms that take the form $A^{i,N_x}_{i,1}$ and $A^{i,1}_{i,N_x}$. Now we consider the terms shown in Fig.~\ref{fig:PeriodicBT}b. We can see that we have $N_x$ terms that take the form $A^{N_y,j+1}_{1,j}$ and $A^{1,j}_{N_y,j+1}$ as well as $N_y-1$ terms that take the form $A^{i+1,N_x}_{i,1}$ and $A^{i,1}_{i+1,N_x}$. Thus in total we have,
\begin{equation}
\begin{split}
    \mathrm{b.t.} &= N_x \left(A^{N_y,1}_{1,1} + A^{1,1}_{N_y,1} \right) + N_y \left(A^{1,N_x}_{1,1} + A^{1,1}_{1,N_x} \right) + N_x \left(A^{N_y,2}_{1,1} + A^{1,1}_{N_y,2} \right) + \left(N_y-1\right)\left(A^{2,N_x}_{1,1} + A^{1,1}_{2,N_x}\right) \\
    & = 2 N_x A^{N_y,1}_{1,1} + 2 N_y A^{1,N_x}_{1,1} + 2 N_x A^{N_y,2}_{1,1} + 2 \left(N_y-1\right)A^{2,N_x}_{1,1}   
\end{split}
\end{equation}
where we have simplified using the property $A^{i_1,j_1}_{i_2,j_2} = A^{i_2,j_2}_{i_1,j_1}$. Furthermore, we note that $A^{N_y,j}_{1,j} = A^{i,j}_{i+1,j}$, $A^{i,N_x}_{i,1} = A^{i,j}_{i,j+1}$, $A^{N_y,j+1}_{1,j} = A^{i,j+1}_{i+1,j}$, and $A^{i+1,N_x}_{i,1} = A^{i+1,j}_{i,j+1}$.

Combining all of this, we arrive at,
\begin{equation}
    \mathrm{b.t.} = 2 N_x A^{1,1}_{2,1} + 2 N_y A^{1,N_x}_{1,1} + \left(2 N_x + 2 N_y -2 \right) A^{2,1}_{1,2}.  
\end{equation}
Plugging this expression for the boundary terms into~\eqref{eq:rperiod} we find,
\begin{equation}
    \label{eq:rperiodfinal}
    \begin{split}
    r^{PBC}_{\mathrm{com}} = \frac{\tau^2}{\epsilon} N_x N_y \Bigg(A^{1,1}_{1,1} + 2 A^{1,1}_{1,2} + 2 A^{1,1}_{2,1} + 2 A^{1,1}_{2,2} + 2 A^{2,1}_{1,2} \Bigg).
    \end{split}
\end{equation}
Thus, comparing Eqs.~\eqref{eq:ComAH} and~\eqref{eq:rperiodfinal} we see that the only difference in the Trotter depth between open and periodic boundary conditions is to change the factors of $N_x-1$ and $N_y-1$ to $N_x$ and $N_y$ in the multi-site terms.   

We can now replace each $A^{i_1,j_1}_{i_2,j_2}$ with the functional forms found in section~\ref{sec:5}. The full expressions for the commutator Trotter bounds with periodic boundary conditions are thus,
\begin{equation}
    r^{H,PBC}_{\mathrm{com}} = \frac{\tau^2}{\epsilon} N_x N_y \left[ \left(4 t^2+8 t U\right) +2 \left(4 t^2+2 t U\right) + 2 \left(4 t^2+2 t U\right)+4 t^2 \right]
\end{equation}
for the Hubbard model and,
\begin{equation}
    \begin{split}
    r^{tJ, PBC}_{\mathrm{com}} = \frac{\tau^2}{\epsilon} N_x N_y &\Big[ \left(12 t^2 + 16 |tJ|+ \frac{3}{2} J^2\right) +2 \left(8 t^2 + 8 |tJ|+ \frac{3}{2} J^2\right) \\ &+2 \left(8 t^2 + 8 |tJ|+ \frac{3}{2} J^2\right) +2 \left(4 t^2 + 4 |tJ|+ \frac{3}{4} J^2\right) \Big]
    \end{split}
\end{equation}
for the t-J model.

\section{Decomposition of single-site Hubbard and t-J Hamiltonians}
\label{Appendix E}

In this appendix we provide explicit decompositions of the Hubbard and t-J Hamiltonians into terms that consist of solely products of Pauli operators. For the Hubbard model each single-site term, $H^H_{i,j}$, in Eq.~\eqref{eq:HubbardHJW} can be decomposed as listed in Table~\ref{tbl:Hubbardterms}. Similarly, for the t-J model each single-site term, $H_{i,j}^{tJ}$, in Eq.~\eqref{eq:tJHJW} can be decomposed as listed in Table~\ref{tbl:tJterms}.

\begin{table}
\centering
\begin{tabular}{ |c|c|c| }
\hline
\multicolumn{3}{|c|}{$H_{i,j}^{\delta}$ for 2D Hubbard} \\
\hline
$\times \frac{t}{2}(-1)^{N_x} \prod_{\beta = 2j+1}^{N_x} \sigma_{i,\beta}^z \prod_{\beta = 1}^{2j-2} \sigma_{i+1,\beta}^z$ & $\times t/2$ & $ \times U/4$ \\ 
\hline
$H_{i,j}^1 = \sigma_{i,2j-1}^x  \sigma_{i+1,2j-1}^x \sigma_{i,2j}^z$ &  $H_{i,j}^5 = \sigma_{i,2j}^x \sigma_{i,2j+2}^x \sigma_{i,2j+1}^z$ & $H_{i,j}^9 = \sigma^0$\\
$H_{i,j}^2 = \sigma_{i,2j-1}^y  \sigma_{i+1,2j-1}^y \sigma_{i,2j}^z$ & $H_{i,j}^6 = \sigma_{i,2j+2}^y \sigma_{i,2j}^y \sigma_{i,2j+1}^z$ & $H_{i,j}^{10} = \sigma_{i, 2j-1}^z$\\
$H_{i,j}^3 = \sigma_{i,2j}^x \sigma_{i+1,2j}^x \sigma_{i+1,2j-1}^z$ & $H_{i,j}^7 = \sigma_{i,2j-1}^x \sigma_{i,2j+1}^x \sigma_{i,2j}^z$  & $ H_{i,j}^{11} = \sigma_{i, 2j}^z$\\
$H_{i,j}^4 = \sigma_{i,2j}^y \sigma_{i+1,2j}^y \sigma_{i+1, 2j-1}^z$ & $H_{i,j}^8 = \sigma_{i,2j+1}^y \sigma_{i,2j-1}^y \sigma_{i,2j}^z$ & $H_{i,j}^{12} = \sigma_{i, 2j-1}^z \sigma_{i, 2j}^z$ \\
\hline
\end{tabular}
\caption{Decomposition of each single-site term in the 2D Jordan-Wigner transformed Hubbard Hamiltonian,  Eq.~\eqref{eq:HubbardHJW}, into terms that consist only of products of Pauli operators. The terms in each column share the common factor listed in the top row of the column.}
\label{tbl:Hubbardterms}
\end{table}

\begin{turnpage}
\begin{table}
\centering
\begin{tabular}{ |c|c|c|c| }
\hline
\multicolumn{4}{|c|}{$H_{i,j}^{\delta}$ for 2D t-J} \\
\hline
$\times \frac{t}{8}(-1)^{N_x} \prod_{\beta=2j+1}^{N_x} \sigma^z_{i,\beta}  \prod_{\beta=1}^{2j-2} \sigma^z_{i+1,\beta}$  & $\times t/8$ & $ \times J/16$ & $ \times J/16$\\ 
\hline
$H_{i,j}^1 = -\sigma_{i, 2j-1}^x \sigma_{i+1, 2j-1}^x$ & $H_{i,j}^{17} = -\sigma_{i, 2j-1}^x \sigma_{i, 2j+1}^x$ & $H_{i,j}^{33} = \sigma_{i, 2j-1}^x \sigma_{i, 2j}^x \sigma_{i+1, 2j-1}^x \sigma_{i+1, 2j}^x$ & $H_{i,j}^{49} = -\sigma^0$\\
$H_{i,j}^2 = -\sigma_{i, 2j-1}^y \sigma_{i+1, 2j-1}^y$ & $H_{i,j}^{18} = -\sigma_{i, 2j-1}^y \sigma_{i, 2j+1}^y$ & $H_{i,j}^{34} = \sigma_{i, 2j-1}^x \sigma_{i, 2j}^x \sigma_{i+1, 2j-1}^y \sigma_{i+1, 2j}^y$ & $H_{i,j}^{50} = \sigma_{i, 2j-1}^z \sigma_{i, 2j}^z$\\
$H_{i,j}^3 = \sigma_{i, 2j-1}^x \sigma_{i+1, 2j-1}^x \sigma^z_{i+1,2j}$ & $H_{i,j}^{19} = \sigma_{i, 2j-1}^x \sigma_{i, 2j+1}^x \sigma_{i,2j}^z$ & $H_{i,j}^{35} = \sigma_{i, 2j-1}^x \sigma_{i, 2j}^x \sigma_{i, 2j+1}^x \sigma_{i, 2j+2}^x$ & $H_{i,j}^{51} = \sigma_{i+1, 2j-1}^z \sigma_{i+1, 2j}^z$\\
$H_{i,j}^4 = \sigma_{i, 2j-1}^y \sigma_{i+1, 2j-1}^y \sigma^z_{i+1,2j}$ & $H_{i,j}^{20} = \sigma_{i, 2j-1}^y \sigma_{i, 2j+1}^y \sigma_{i,2j}^z$ & $H_{i,j}^{36} = \sigma_{i, 2j-1}^x \sigma_{i, 2j}^x \sigma_{i, 2j+1}^y \sigma_{i, 2j+2}^y$ & $H_{i,j}^{52} = \sigma_{i, 2j+1}^z \sigma_{i, 2j+2}^z$\\
$H_{i,j}^5 = \sigma_{i, 2j-1}^x \sigma_{i+1, 2j-1}^x \sigma^z_{i,2j}$ & $H_{i,j}^{21} = \sigma_{i, 2j-1}^x \sigma_{i, 2j+1}^x \sigma_{i,2j+2}^z$ & $H_{i,j}^{37} = \sigma_{i, 2j-1}^x \sigma_{i, 2j}^y \sigma_{i+1, 2j-1}^x \sigma_{i+1, 2j}^y$ & $H_{i,j}^{53} = \sigma_{i, 2j}^z \sigma_{i+1, 2j}^z$\\
$H_{i,j}^6 = \sigma_{i, 2j-1}^y \sigma_{i+1, 2j-1}^y \sigma^z_{i,2j}$ & $H_{i,j}^{22} = \sigma_{i, 2j-1}^y \sigma_{i, 2j+1}^y \sigma_{i,2j+2}^z$ & $H_{i,j}^{38} = -\sigma_{i, 2j-1}^x \sigma_{i, 2j}^y \sigma_{i+1, 2j-1}^y \sigma_{i+1, 2j}^x$ & $H_{i,j}^{54} = -\sigma_{i, 2j}^z \sigma_{i+1, 2j-1}^z$\\
$H_{i,j}^7 = -\sigma_{i, 2j-1}^x \sigma_{i+1, 2j-1}^x \sigma^z_{i,2j} \sigma^z_{i+1,2j}$ &  $H_{i,j}^{23} = -\sigma_{i, 2j-1}^x \sigma_{i, 2j+1}^x \sigma_{i,2j}^z\sigma_{i,2j+2}^z$ & $H_{i,j}^{39} = \sigma_{i, 2j-1}^x \sigma_{i, 2j}^y \sigma_{i, 2j+1}^x \sigma_{i, 2j+2}^y$ & $H_{i,j}^{55} = \sigma_{i, 2j}^z \sigma_{i, 2j+2}^z$\\
$H_{i,j}^8 = -\sigma_{i, 2j-1}^y \sigma_{i+1, 2j-1}^y \sigma^z_{i,2j} \sigma^z_{i+1,2j}$ &  $H_{i,j}^{24} = -\sigma_{i, 2j-1}^y \sigma_{i, 2j+1}^y \sigma_{i,2j}^z\sigma_{i,2j+2}^z$ & $H_{i,j}^{40} = -\sigma_{i, 2j-1}^x \sigma_{i, 2j}^y \sigma_{i, 2j+1}^y \sigma_{i, 2j+2}^x$ & $H_{i,j}^{56} = -\sigma_{i, 2j}^z \sigma_{i, 2j+1}^z$\\
$H_{i,j}^9 = -\sigma_{i, 2j}^x \sigma_{i+1, 2j}^x$ &  $H_{i,j}^{25} = -\sigma_{i, 2j}^x \sigma_{i, 2j+2}^x$ & $H_{i,j}^{41} = -\sigma_{i, 2j-1}^y \sigma_{i, 2j}^x \sigma_{i+1, 2j-1}^x \sigma_{i+1, 2j}^y$ & $H_{i,j}^{57} = -\sigma_{i, 2j-1}^z \sigma_{i+1, 2j}^z$\\
$H_{i,j}^{10} = -\sigma_{i, 2j}^y \sigma_{i+1, 2j}^y$ &  $H_{i,j}^{26} = -\sigma_{i, 2j}^y \sigma_{i, 2j+2}^y$ & $H_{i,j}^{42} = \sigma_{i, 2j-1}^y \sigma_{i, 2j}^x \sigma_{i+1, 2j-1}^y \sigma_{i+1, 2j}^x$ & $H_{i,j}^{58} = \sigma_{i, 2j-1}^z \sigma_{i+1, 2j-1}^z$\\
$H_{i,j}^{11} = \sigma_{i, 2j}^x \sigma_{i+1, 2j}^x \sigma^z_{i+1,2j-1}$ &  $H_{i,j}^{27} = \sigma_{i, 2j}^x \sigma_{i, 2j+2}^x \sigma_{i,2j-1}^z$ & $H_{i,j}^{43} = -\sigma_{i, 2j-1}^y \sigma_{i, 2j}^x \sigma_{i, 2j+1}^x \sigma_{i, 2j+2}^y$ & $H_{i,j}^{59} = -\sigma_{i, 2j-1}^z \sigma_{i, 2j+2}^z$\\
$H_{i,j}^{12} = \sigma_{i, 2j}^y \sigma_{i+1, 2j}^y \sigma^z_{i+1,2j-1}$ &  $H_{i,j}^{28} = \sigma_{i, 2j}^y \sigma_{i, 2j+2}^y \sigma_{i,2j-1}^z$ & $H_{i,j}^{44} = \sigma_{i, 2j-1}^y \sigma_{i, 2j}^x \sigma_{i, 2j+1}^y \sigma_{i, 2j+2}^x$ & $H_{i,j}^{60} = \sigma_{i, 2j-1}^z \sigma_{i, 2j+1}^z$\\
$H_{i,j}^{13} = \sigma_{i, 2j}^x \sigma_{i+1, 2j}^x \sigma^z_{i,2j-1}$ &  $H_{i,j}^{29} = \sigma_{i, 2j}^x \sigma_{i, 2j+2}^x \sigma_{i,2j+1}^z$ & $H_{i,j}^{45} = \sigma_{i, 2j-1}^y \sigma_{i, 2j}^y \sigma_{i+1, 2j-1}^x \sigma_{i+1, 2j}^x$ & $H_{i,j}^{61} = \sigma_{i, 2j-1}^z \sigma_{i, 2j}^z$\\
$H_{i,j}^{14} = \sigma_{i, 2j}^y \sigma_{i+1, 2j}^y \sigma^z_{i,2j-1}$ &  $H_{i,j}^{30} = \sigma_{i, 2j}^y \sigma_{i, 2j+2}^y \sigma_{i,2j+1}^z$ & $H_{i,j}^{46} = \sigma_{i, 2j-1}^y \sigma_{i, 2j}^y \sigma_{i+1, 2j-1}^y \sigma_{i+1, 2j}^y$ & $H_{i,j}^{62} = -\sigma^0$\\
$H_{i,j}^{15} = -\sigma_{i, 2j}^x \sigma_{i+1, 2j}^x \sigma^z_{i,2j-1} \sigma^z_{i+1,2j-1}$ &  $H_{i,j}^{31} = -\sigma_{i, 2j}^x \sigma_{i, 2j+2}^x \sigma_{i,2j-1}^z\sigma_{i,2j+1}^z$ & $H_{i,j}^{47} = \sigma_{i, 2j-1}^y \sigma_{i, 2j}^y \sigma_{i, 2j+1}^x \sigma_{i, 2j+2}^x$ & $H_{i,j}^{63} = -\sigma_{i, 2j-1}^z \sigma_{i, 2j}^z \sigma_{i+1, 2j-1}^z \sigma_{i+1, 2j}^z$\\
$H_{i,j}^{16} = -\sigma_{i, 2j}^y \sigma_{i+1, 2j}^y \sigma^z_{i,2j-1} \sigma^z_{i+1,2j-1}$ &  $H_{i,j}^{32} = -\sigma_{i, 2j}^y \sigma_{i, 2j+2}^y \sigma_{i,2j-1}^z\sigma_{i,2j+1}^z$ & $H_{i,j}^{48} = \sigma_{i, 2j-1}^y \sigma_{i, 2j}^y \sigma_{i, 2j+1}^y \sigma_{i, 2j+2}^y$ & $H_{i,j}^{64} = -\sigma_{i, 2j-1}^z \sigma_{i, 2j}^z \sigma_{i, 2j+1}^z \sigma_{i, 2j+2}^z$\\
\hline
\end{tabular}
\caption{Decomposition of each single-site term in the Jordan-Wigner transformed 2D t-J model Hamiltonian, Eq.~\eqref{eq:tJHJW}, into terms that consist only of products of Pauli operators. The terms in each column share the common factor listed in the top row of the column.}
\label{tbl:tJterms}
\end{table}
\end{turnpage}

\section{Trotter depth for 1D Hubbard and t-J models}
\label{Appendix F}

In this appendix we determine the functional form of the Trotter depth for the 1D Hubbard and t-J models. 

For reference, in 1D the Hamiltonian for the Hubbard model is, 
\begin{equation}
    H_H = -t \sum_j \Big(c_{2j-1}^{\dagger}c_{2j+1} + c_{2j+1}^{\dagger}c_{2j-1}
    + c_{2j}^{\dagger}c_{2j+2} + c_{2j+2}^{\dagger}c_{2j} \Big)+ U \sum_j n_{2j-1}n_{2j},
\end{equation}
Similarly, the Hamiltonian for the 1D t-J model is,
\begin{equation}
    \label{eq:tJspinless1D}
    \begin{split}
    H_{\mathrm{tJ}} = & - t \sum_{j} \Big[ \left( 1 - n_{2j}\right)\left(c^{\dagger}_{2j-1}c_{2j+1} + c_{2j-1}c^{\dagger}_{2j+1}\right)\left( 1 - n_{2j+2}\right) + \left( 1 - n_{2j-1}\right)\left(c^{\dagger}_{2j}c_{2j+2} + c_{2j}c^{\dagger}_{2j+2}\right)\left( 1 - n_{2j+1}\right)\Big] \\
     & + \frac{J}{2}\sum_{j} \Big[c^{\dagger}_{2j-1}c_{2j}c^{\dagger}_{2j+2}c_{2j+1} + c^{\dagger}_{2j} c_{2j-1}c^{\dagger}_{2j+1}c_{2j+2}- (1-n_{2j})n_{2j-1}n_{2j+2}(1-n_{2j+1}) \\
     & \qquad \qquad \quad - (1-n_{2j-1})n_{2j}n_{2j+1}(1-n_{2j+2}) \Big].
    \end{split}
\end{equation}
Note that we have expressed both these Hamiltonians in spinless form, which can be derived in much the same manner as in the 2D case, by mapping a single spinful chain to two spinless chains.

Applying the standard Jordan-Wigner transformation to the above Hamiltonians we arrive at,
\begin{equation}
\label{eq:HubbardJW}
\begin{split}
    H_H = \frac{t}{2} & \sum_{j = 1}^N \Big[ \big(\sigma^x_{2j-1} \sigma^x_{2j+1} + \sigma^y_{2j-1}\sigma^y_{2j+1}\big)\sigma^z_{2j} + \big(\sigma^x_{2j} \sigma^x_{2j+2} + \sigma^y_{2j}\sigma^y_{2j+2}\big)\sigma^z_{2j+1}\Big] \\
    & + \frac{U}{4} \sum_{j=1}^N \Big[(\sigma^0_{2j-1}+\sigma^z_{2j-1})(\sigma^0_{2j}+\sigma^z_{2j}) \Big] 
\end{split}
\end{equation}
For the Hubbard model. Similarly, for the tJ model we find,
\begin{equation}
    \begin{split}
    H_{tJ} = - \frac{t}{4} & \sum_{j=1}^N \Big[\left(I_{2j}-\sigma_{2j}^z\right) \left(\sigma^+_{2j-1} \sigma_{2j+1}^- + \sigma^+_{2j+1} \sigma_{2j-1}^-\right) \left(I_{2j+2}-\sigma_{2j+2}^z\right) \\ &
    + \left(I_{2j-1}-\sigma_{2j-1}^z\right) \left(\sigma^+_{2j} \sigma_{2j+2}^- + \sigma^+_{2j+2} \sigma_{2j}^-\right) \left(I_{2j+1}-\sigma_{2j+1}^z\right) \Big] \\ &
    + \frac{J}{32} \sum_{j=1}^N \Big[ 16 \sigma^+_{2j-1} \sigma^-_{2j} \sigma_{2j+2}^+ \sigma_{2j+1}^- + 16 \sigma^+_{2j} \sigma^-_{2j-1} \sigma_{2j+1}^+ \sigma_{2j+2}^- \\ &
    - \left(\sigma^0_{2j} - \sigma_{2j}^z\right) \left(\sigma^0_{2j-1} + \sigma_{2j-1}^z\right) \left(\sigma^0_{2j+2} + \sigma_{2j+2}^z\right) \left(\sigma^0_{2j+1} - \sigma_{2j+1}^z\right) \\ &
    - \left(\sigma^0_{2j-1} - \sigma_{2j-1}^z\right) \left(\sigma^0_{2j} + \sigma_{2j}^z\right) \left(\sigma^0_{2j+1} + \sigma_{2j+1}^z\right) \left(\sigma^0_{2j+2} - \sigma_{2j+2}^z\right) \Big] 
    \end{split}
\end{equation}
Note that, unlike the 2D case, for the 1D models we find that the Jordan-Wigner strings cancel out, leading to significantly reduced Pauli depth. In this case, the presence of the projection operators in the t-J model plays a more significant role, leading to a maximum Pauli depth of four for the t-J model in comparison to three for Hubbard. 

Taking the case $N_y = 1$ in Eq.~\eqref{eq:ComAH} we find the general form,
\begin{equation}
    r_{\mathrm{com,1D}} = \frac{\tau^2}{\epsilon} \left\{N A_{1,1}+2\sum_{p=1}^{N-1}\left[(N-p)A_{1,1+p}\right]\right\}.
\end{equation}
Noting that for both the 1D Hubbard and t-J models $A_{j,j+p>1}=0$, this simplifies to, 
\begin{equation}
    r_{\mathrm{com,1D}} = \frac{\tau^2}{\epsilon} \left[N A_{1,1}+2(N-1)A_{1,2}\right].
\end{equation}
In the case of the 1D Hubbard model we have $A_{1,1} = 4 |tU|$ and $A_{1,2} = 2 t^2 + 2 |t U|$. Thus, 
\begin{equation}
    r^H_{\mathrm{com,1D}} = \frac{\tau^2}{\epsilon} \left[4 N |tU|+2(N-1)(2t^2+2 |tU|)\right].
\end{equation}
In the case of the 1D t-J model we have $A_{1,1} = 2t^2 + 4|tJ|$ and $A_{1,2} = 4 t^2 + 4 |t J|+ 3J^2/4$. Thus,
\begin{equation}
    r^{tJ}_{\mathrm{com,1D}} = \frac{\tau^2}{\epsilon} \left[N \left(2t^2 + 4|tJ|\right)+2(N-1)\left(4 t^2 + 4 |t J|+ \frac{3}{4}J^2\right)\right].
\end{equation}

\twocolumngrid

\bibliography{JWtJ}

%apsrev4-2.bst 2019-01-14 (MD) hand-edited version of apsrev4-1.bst
%Control: key (0)
%Control: author (8) initials jnrlst
%Control: editor formatted (1) identically to author
%Control: production of article title (0) allowed
%Control: page (0) single
%Control: year (1) truncated
%Control: production of eprint (0) enabled
\begin{thebibliography}{60}%
\makeatletter
\providecommand \@ifxundefined [1]{%
 \@ifx{#1\undefined}
}%
\providecommand \@ifnum [1]{%
 \ifnum #1\expandafter \@firstoftwo
 \else \expandafter \@secondoftwo
 \fi
}%
\providecommand \@ifx [1]{%
 \ifx #1\expandafter \@firstoftwo
 \else \expandafter \@secondoftwo
 \fi
}%
\providecommand \natexlab [1]{#1}%
\providecommand \enquote  [1]{``#1''}%
\providecommand \bibnamefont  [1]{#1}%
\providecommand \bibfnamefont [1]{#1}%
\providecommand \citenamefont [1]{#1}%
\providecommand \href@noop [0]{\@secondoftwo}%
\providecommand \href [0]{\begingroup \@sanitize@url \@href}%
\providecommand \@href[1]{\@@startlink{#1}\@@href}%
\providecommand \@@href[1]{\endgroup#1\@@endlink}%
\providecommand \@sanitize@url [0]{\catcode `\\12\catcode `\$12\catcode
  `\&12\catcode `\#12\catcode `\^12\catcode `\_12\catcode `\%12\relax}%
\providecommand \@@startlink[1]{}%
\providecommand \@@endlink[0]{}%
\providecommand \url  [0]{\begingroup\@sanitize@url \@url }%
\providecommand \@url [1]{\endgroup\@href {#1}{\urlprefix }}%
\providecommand \urlprefix  [0]{URL }%
\providecommand \Eprint [0]{\href }%
\providecommand \doibase [0]{https://doi.org/}%
\providecommand \selectlanguage [0]{\@gobble}%
\providecommand \bibinfo  [0]{\@secondoftwo}%
\providecommand \bibfield  [0]{\@secondoftwo}%
\providecommand \translation [1]{[#1]}%
\providecommand \BibitemOpen [0]{}%
\providecommand \bibitemStop [0]{}%
\providecommand \bibitemNoStop [0]{.\EOS\space}%
\providecommand \EOS [0]{\spacefactor3000\relax}%
\providecommand \BibitemShut  [1]{\csname bibitem#1\endcsname}%
\let\auto@bib@innerbib\@empty
%</preamble>
\bibitem [{\citenamefont {{Feynman, R. P.}}(1982)}]{Feynman1982}%
  \BibitemOpen
  \bibfield  {author} {\bibinfo {author} {\bibnamefont {{Feynman, R. P.}}},\
  }\bibfield  {title} {\bibinfo {title} {Simulating physics with computers},\
  }\href {https://doi.org/10.1007/BF02650179} {\bibfield  {journal} {\bibinfo
  {journal} {\textit{Int. J. Theor. Phys.}}\ }\textbf {\bibinfo {volume}
  {21}},\ \bibinfo {pages} {467} (\bibinfo {year} {1982})}\BibitemShut
  {NoStop}%
\bibitem [{\citenamefont {Lloyd}(1996)}]{Lloyd1996}%
  \BibitemOpen
  \bibfield  {author} {\bibinfo {author} {\bibfnamefont {S.}~\bibnamefont
  {Lloyd}},\ }\bibfield  {title} {\bibinfo {title} {Universal quantum
  simulators},\ }\href {https://doi.org/10.1126/science.273.5278.1073}
  {\bibfield  {journal} {\bibinfo  {journal} {Science}\ }\textbf {\bibinfo
  {volume} {273}},\ \bibinfo {pages} {1073} (\bibinfo {year}
  {1996})}\BibitemShut {NoStop}%
\bibitem [{\citenamefont {Abrams}\ and\ \citenamefont
  {Lloyd}(1997)}]{Abrams1997}%
  \BibitemOpen
  \bibfield  {author} {\bibinfo {author} {\bibfnamefont {D.~S.}\ \bibnamefont
  {Abrams}}\ and\ \bibinfo {author} {\bibfnamefont {S.}~\bibnamefont {Lloyd}},\
  }\bibfield  {title} {\bibinfo {title} {Simulation of many-body {F}ermi
  systems on a universal quantum computer},\ }\href
  {https://doi.org/10.1103/PhysRevLett.79.2586} {\bibfield  {journal} {\bibinfo
   {journal} {Phys. Rev. Lett.}\ }\textbf {\bibinfo {volume} {79}},\ \bibinfo
  {pages} {2586} (\bibinfo {year} {1997})}\BibitemShut {NoStop}%
\bibitem [{\citenamefont {Georgescu}\ \emph {et~al.}(2014)\citenamefont
  {Georgescu}, \citenamefont {Ashhab},\ and\ \citenamefont
  {Nori}}]{Georgescu2014}%
  \BibitemOpen
  \bibfield  {author} {\bibinfo {author} {\bibfnamefont {I.~M.}\ \bibnamefont
  {Georgescu}}, \bibinfo {author} {\bibfnamefont {S.}~\bibnamefont {Ashhab}},\
  and\ \bibinfo {author} {\bibfnamefont {F.}~\bibnamefont {Nori}},\ }\bibfield
  {title} {\bibinfo {title} {Quantum simulation},\ }\href
  {https://doi.org/10.1103/RevModPhys.86.153} {\bibfield  {journal} {\bibinfo
  {journal} {Rev. Mod. Phys.}\ }\textbf {\bibinfo {volume} {86}},\ \bibinfo
  {pages} {153} (\bibinfo {year} {2014})}\BibitemShut {NoStop}%
\bibitem [{\citenamefont {Kassal}\ \emph {et~al.}(2011)\citenamefont {Kassal},
  \citenamefont {Whitfield}, \citenamefont {Perdomo-Ortiz}, \citenamefont
  {Yung},\ and\ \citenamefont {Aspuru-Guzik}}]{Kassal2011}%
  \BibitemOpen
  \bibfield  {author} {\bibinfo {author} {\bibfnamefont {I.}~\bibnamefont
  {Kassal}}, \bibinfo {author} {\bibfnamefont {J.~D.}\ \bibnamefont
  {Whitfield}}, \bibinfo {author} {\bibfnamefont {A.}~\bibnamefont
  {Perdomo-Ortiz}}, \bibinfo {author} {\bibfnamefont {M.-H.}\ \bibnamefont
  {Yung}},\ and\ \bibinfo {author} {\bibfnamefont {A.}~\bibnamefont
  {Aspuru-Guzik}},\ }\bibfield  {title} {\bibinfo {title} {Simulating chemistry
  using quantum computers},\ }\href
  {https://doi.org/10.1146/annurev-physchem-032210-103512} {\bibfield
  {journal} {\bibinfo  {journal} {Annu. Rev. Phys. Chem.}\ }\textbf {\bibinfo
  {volume} {62}},\ \bibinfo {pages} {185} (\bibinfo {year} {2011})}\BibitemShut
  {NoStop}%
\bibitem [{\citenamefont {Cao}\ \emph {et~al.}(2019)\citenamefont {Cao},
  \citenamefont {Romero}, \citenamefont {Olson}, \citenamefont {Degroote},
  \citenamefont {Johnson}, \citenamefont {Kieferov{\'a}}, \citenamefont
  {Kivlichan}, \citenamefont {Menke}, \citenamefont {Peropadre}, \citenamefont
  {Sawaya}, \citenamefont {Sim}, \citenamefont {Veis},\ and\ \citenamefont
  {Aspuru-Guzik}}]{Cao2019}%
  \BibitemOpen
  \bibfield  {author} {\bibinfo {author} {\bibfnamefont {Y.}~\bibnamefont
  {Cao}}, \bibinfo {author} {\bibfnamefont {J.}~\bibnamefont {Romero}},
  \bibinfo {author} {\bibfnamefont {J.~P.}\ \bibnamefont {Olson}}, \bibinfo
  {author} {\bibfnamefont {M.}~\bibnamefont {Degroote}}, \bibinfo {author}
  {\bibfnamefont {P.~D.}\ \bibnamefont {Johnson}}, \bibinfo {author}
  {\bibfnamefont {M.}~\bibnamefont {Kieferov{\'a}}}, \bibinfo {author}
  {\bibfnamefont {I.~D.}\ \bibnamefont {Kivlichan}}, \bibinfo {author}
  {\bibfnamefont {T.}~\bibnamefont {Menke}}, \bibinfo {author} {\bibfnamefont
  {B.}~\bibnamefont {Peropadre}}, \bibinfo {author} {\bibfnamefont {N.~P.~D.}\
  \bibnamefont {Sawaya}}, \bibinfo {author} {\bibfnamefont {S.}~\bibnamefont
  {Sim}}, \bibinfo {author} {\bibfnamefont {L.}~\bibnamefont {Veis}},\ and\
  \bibinfo {author} {\bibfnamefont {A.}~\bibnamefont {Aspuru-Guzik}},\
  }\bibfield  {title} {\bibinfo {title} {Quantum chemistry in the age of
  quantum computing},\ }\href {https://doi.org/10.1021/acs.chemrev.8b00803}
  {\bibfield  {journal} {\bibinfo  {journal} {Chem. Rev.}\ }\textbf {\bibinfo
  {volume} {119}},\ \bibinfo {pages} {10856} (\bibinfo {year}
  {2019})}\BibitemShut {NoStop}%
\bibitem [{\citenamefont {McArdle}\ \emph {et~al.}(2020)\citenamefont
  {McArdle}, \citenamefont {Endo}, \citenamefont {Aspuru-Guzik}, \citenamefont
  {Benjamin},\ and\ \citenamefont {Yuan}}]{McArdle2020}%
  \BibitemOpen
  \bibfield  {author} {\bibinfo {author} {\bibfnamefont {S.}~\bibnamefont
  {McArdle}}, \bibinfo {author} {\bibfnamefont {S.}~\bibnamefont {Endo}},
  \bibinfo {author} {\bibfnamefont {A.}~\bibnamefont {Aspuru-Guzik}}, \bibinfo
  {author} {\bibfnamefont {S.~C.}\ \bibnamefont {Benjamin}},\ and\ \bibinfo
  {author} {\bibfnamefont {X.}~\bibnamefont {Yuan}},\ }\bibfield  {title}
  {\bibinfo {title} {Quantum computational chemistry},\ }\href
  {https://doi.org/10.1103/RevModPhys.92.015003} {\bibfield  {journal}
  {\bibinfo  {journal} {Rev. Mod. Phys.}\ }\textbf {\bibinfo {volume} {92}},\
  \bibinfo {pages} {015003} (\bibinfo {year} {2020})}\BibitemShut {NoStop}%
\bibitem [{\citenamefont {Babbush}\ \emph
  {et~al.}(2018{\natexlab{a}})\citenamefont {Babbush}, \citenamefont {Wiebe},
  \citenamefont {McClean}, \citenamefont {McClain}, \citenamefont {Neven},\
  and\ \citenamefont {Chan}}]{Babbush2018}%
  \BibitemOpen
  \bibfield  {author} {\bibinfo {author} {\bibfnamefont {R.}~\bibnamefont
  {Babbush}}, \bibinfo {author} {\bibfnamefont {N.}~\bibnamefont {Wiebe}},
  \bibinfo {author} {\bibfnamefont {J.}~\bibnamefont {McClean}}, \bibinfo
  {author} {\bibfnamefont {J.}~\bibnamefont {McClain}}, \bibinfo {author}
  {\bibfnamefont {H.}~\bibnamefont {Neven}},\ and\ \bibinfo {author}
  {\bibfnamefont {G.~K.-L.}\ \bibnamefont {Chan}},\ }\bibfield  {title}
  {\bibinfo {title} {Low-depth quantum simulation of materials},\ }\href
  {https://doi.org/10.1103/PhysRevX.8.011044} {\bibfield  {journal} {\bibinfo
  {journal} {Phys. Rev. X}\ }\textbf {\bibinfo {volume} {8}},\ \bibinfo {pages}
  {011044} (\bibinfo {year} {2018}{\natexlab{a}})}\BibitemShut {NoStop}%
\bibitem [{\citenamefont {Jordan}\ and\ \citenamefont
  {Wigner}(1928)}]{Jordan1928}%
  \BibitemOpen
  \bibfield  {author} {\bibinfo {author} {\bibfnamefont {P.}~\bibnamefont
  {Jordan}}\ and\ \bibinfo {author} {\bibfnamefont {E.}~\bibnamefont
  {Wigner}},\ }\bibfield  {title} {\bibinfo {title} {{\"U}ber das paulische
  {\"a}quivalenzverbot},\ }\href {https://doi.org/10.1007/BF01331938}
  {\bibfield  {journal} {\bibinfo  {journal} {Z. Phys.}\ }\textbf {\bibinfo
  {volume} {47}},\ \bibinfo {pages} {631} (\bibinfo {year} {1928})}\BibitemShut
  {NoStop}%
\bibitem [{\citenamefont {Whitfield}\ \emph {et~al.}(2011)\citenamefont
  {Whitfield}, \citenamefont {Biamonte},\ and\ \citenamefont
  {Aspuru-Guzik}}]{Whitfield2011}%
  \BibitemOpen
  \bibfield  {author} {\bibinfo {author} {\bibfnamefont {J.~D.}\ \bibnamefont
  {Whitfield}}, \bibinfo {author} {\bibfnamefont {J.}~\bibnamefont
  {Biamonte}},\ and\ \bibinfo {author} {\bibfnamefont {A.}~\bibnamefont
  {Aspuru-Guzik}},\ }\bibfield  {title} {\bibinfo {title} {Simulation of
  electronic structure {H}amiltonians using quantum computers},\ }\href
  {https://doi.org/10.1080/00268976.2011.552441} {\bibfield  {journal}
  {\bibinfo  {journal} {Mol. Phys.}\ }\textbf {\bibinfo {volume} {109}},\
  \bibinfo {pages} {735} (\bibinfo {year} {2011})}\BibitemShut {NoStop}%
\bibitem [{\citenamefont {Hastings}\ \emph {et~al.}(2015)\citenamefont
  {Hastings}, \citenamefont {Wecker}, \citenamefont {Bauer},\ and\
  \citenamefont {Troyer}}]{Hastings2015}%
  \BibitemOpen
  \bibfield  {author} {\bibinfo {author} {\bibfnamefont {M.~B.}\ \bibnamefont
  {Hastings}}, \bibinfo {author} {\bibfnamefont {D.}~\bibnamefont {Wecker}},
  \bibinfo {author} {\bibfnamefont {B.}~\bibnamefont {Bauer}},\ and\ \bibinfo
  {author} {\bibfnamefont {M.}~\bibnamefont {Troyer}},\ }\bibfield  {title}
  {\bibinfo {title} {Improving quantum algorithms for quantum chemistry},\
  }\href {https://dl.acm.org/doi/10.5555/2685188.2685189} {\bibfield  {journal}
  {\bibinfo  {journal} {Quantum Info. Comput.}\ }\textbf {\bibinfo {volume}
  {15}},\ \bibinfo {pages} {1} (\bibinfo {year} {2015})}\BibitemShut {NoStop}%
\bibitem [{\citenamefont {Troyer}\ and\ \citenamefont
  {Wiese}(2005)}]{Troyer2005}%
  \BibitemOpen
  \bibfield  {author} {\bibinfo {author} {\bibfnamefont {M.}~\bibnamefont
  {Troyer}}\ and\ \bibinfo {author} {\bibfnamefont {U.-J.}\ \bibnamefont
  {Wiese}},\ }\bibfield  {title} {\bibinfo {title} {Computational complexity
  and fundamental limitations to fermionic quantum monte carlo simulations},\
  }\href {https://doi.org/10.1103/PhysRevLett.94.170201} {\bibfield  {journal}
  {\bibinfo  {journal} {Phys. Rev. Lett.}\ }\textbf {\bibinfo {volume} {94}},\
  \bibinfo {pages} {170201} (\bibinfo {year} {2005})}\BibitemShut {NoStop}%
\bibitem [{\citenamefont {Trotter}(1959)}]{Trotter1959}%
  \BibitemOpen
  \bibfield  {author} {\bibinfo {author} {\bibfnamefont {H.~F.}\ \bibnamefont
  {Trotter}},\ }\bibfield  {title} {\bibinfo {title} {On the product of
  semi-groups of operators},\ }\href
  {https://doi.org/10.1090/S0002-9939-1959-0108732-6} {\bibfield  {journal}
  {\bibinfo  {journal} {Proc. Am. Math. Soc.}\ }\textbf {\bibinfo {volume}
  {10}},\ \bibinfo {pages} {545} (\bibinfo {year} {1959})}\BibitemShut
  {NoStop}%
\bibitem [{\citenamefont {Suzuki}(1976)}]{Suzuki1976}%
  \BibitemOpen
  \bibfield  {author} {\bibinfo {author} {\bibfnamefont {M.}~\bibnamefont
  {Suzuki}},\ }\bibfield  {title} {\bibinfo {title} {Generalized {T}rotter's
  formula and systematic approximants of exponential operators and inner
  derivations with applications to many-body problems},\ }\href
  {https://doi.org/10.1007/BF01609348} {\bibfield  {journal} {\bibinfo
  {journal} {Commun. Math. Phys.}\ }\textbf {\bibinfo {volume} {51}},\ \bibinfo
  {pages} {183} (\bibinfo {year} {1976})}\BibitemShut {NoStop}%
\bibitem [{\citenamefont {Nielsen}\ and\ \citenamefont
  {Chuang}(2010)}]{Nielsen2010}%
  \BibitemOpen
  \bibfield  {author} {\bibinfo {author} {\bibfnamefont {M.~A.}\ \bibnamefont
  {Nielsen}}\ and\ \bibinfo {author} {\bibfnamefont {I.~L.}\ \bibnamefont
  {Chuang}},\ }\href {https://doi.org/10.1017/CBO9780511976667} {\emph
  {\bibinfo {title} {Quantum Computation and Quantum Information}}}\ (\bibinfo
  {publisher} {Cambridge University Press},\ \bibinfo {year}
  {2010})\BibitemShut {NoStop}%
\bibitem [{\citenamefont {Lee}\ \emph {et~al.}(2022)\citenamefont {Lee},
  \citenamefont {Scott},\ and\ \citenamefont {Scarola}}]{Lee2022}%
  \BibitemOpen
  \bibfield  {author} {\bibinfo {author} {\bibfnamefont {W.-R.}\ \bibnamefont
  {Lee}}, \bibinfo {author} {\bibfnamefont {R.}~\bibnamefont {Scott}},\ and\
  \bibinfo {author} {\bibfnamefont {V.~W.}\ \bibnamefont {Scarola}},\
  }\bibfield  {title} {\bibinfo {title} {A {Compact} {Noise}-{Tolerant}
  {Algorithm} for {Unbiased} {Quantum} {Simulation} {Using} {Feynman}'s $i\eta$
  {Prescription}},\ }\href {https://arxiv.org/abs/2212.14039} {\bibfield
  {journal} {\bibinfo  {journal} {arXiv:2212.14039}\ } (\bibinfo {year}
  {2022})}\BibitemShut {NoStop}%
\bibitem [{\citenamefont {Wecker}\ \emph {et~al.}(2014)\citenamefont {Wecker},
  \citenamefont {Bauer}, \citenamefont {Clark}, \citenamefont {Hastings},\ and\
  \citenamefont {Troyer}}]{Wecker2014}%
  \BibitemOpen
  \bibfield  {author} {\bibinfo {author} {\bibfnamefont {D.}~\bibnamefont
  {Wecker}}, \bibinfo {author} {\bibfnamefont {B.}~\bibnamefont {Bauer}},
  \bibinfo {author} {\bibfnamefont {B.~K.}\ \bibnamefont {Clark}}, \bibinfo
  {author} {\bibfnamefont {M.~B.}\ \bibnamefont {Hastings}},\ and\ \bibinfo
  {author} {\bibfnamefont {M.}~\bibnamefont {Troyer}},\ }\bibfield  {title}
  {\bibinfo {title} {Gate-count estimates for performing quantum chemistry on
  small quantum computers},\ }\href
  {https://doi.org/10.1103/PhysRevA.90.022305} {\bibfield  {journal} {\bibinfo
  {journal} {Phys. Rev. A}\ }\textbf {\bibinfo {volume} {90}},\ \bibinfo
  {pages} {022305} (\bibinfo {year} {2014})}\BibitemShut {NoStop}%
\bibitem [{\citenamefont {Babbush}\ \emph
  {et~al.}(2018{\natexlab{b}})\citenamefont {Babbush}, \citenamefont {Gidney},
  \citenamefont {Berry}, \citenamefont {Wiebe}, \citenamefont {McClean},
  \citenamefont {Paler}, \citenamefont {Fowler},\ and\ \citenamefont
  {Neven}}]{Babbush2018PRX2}%
  \BibitemOpen
  \bibfield  {author} {\bibinfo {author} {\bibfnamefont {R.}~\bibnamefont
  {Babbush}}, \bibinfo {author} {\bibfnamefont {C.}~\bibnamefont {Gidney}},
  \bibinfo {author} {\bibfnamefont {D.~W.}\ \bibnamefont {Berry}}, \bibinfo
  {author} {\bibfnamefont {N.}~\bibnamefont {Wiebe}}, \bibinfo {author}
  {\bibfnamefont {J.}~\bibnamefont {McClean}}, \bibinfo {author} {\bibfnamefont
  {A.}~\bibnamefont {Paler}}, \bibinfo {author} {\bibfnamefont
  {A.}~\bibnamefont {Fowler}},\ and\ \bibinfo {author} {\bibfnamefont
  {H.}~\bibnamefont {Neven}},\ }\bibfield  {title} {\bibinfo {title} {Encoding
  electronic spectra in quantum circuits with linear {T} complexity},\ }\href
  {https://doi.org/10.1103/PhysRevX.8.041015} {\bibfield  {journal} {\bibinfo
  {journal} {Phys. Rev. X}\ }\textbf {\bibinfo {volume} {8}},\ \bibinfo {pages}
  {041015} (\bibinfo {year} {2018}{\natexlab{b}})}\BibitemShut {NoStop}%
\bibitem [{\citenamefont {Nam}\ and\ \citenamefont {Maslov}(2019)}]{Nam2019}%
  \BibitemOpen
  \bibfield  {author} {\bibinfo {author} {\bibfnamefont {Y.}~\bibnamefont
  {Nam}}\ and\ \bibinfo {author} {\bibfnamefont {D.}~\bibnamefont {Maslov}},\
  }\bibfield  {title} {\bibinfo {title} {Low-cost quantum circuits for
  classically intractable instances of the {H}amiltonian dynamics simulation
  problem},\ }\href {https://doi.org/10.1038/s41534-019-0152-0} {\bibfield
  {journal} {\bibinfo  {journal} {npj Quantum Inf.}\ }\textbf {\bibinfo
  {volume} {5}},\ \bibinfo {pages} {44} (\bibinfo {year} {2019})}\BibitemShut
  {NoStop}%
\bibitem [{\citenamefont {Gross}\ and\ \citenamefont
  {Bloch}(2017)}]{Gross2017}%
  \BibitemOpen
  \bibfield  {author} {\bibinfo {author} {\bibfnamefont {C.}~\bibnamefont
  {Gross}}\ and\ \bibinfo {author} {\bibfnamefont {I.}~\bibnamefont {Bloch}},\
  }\bibfield  {title} {\bibinfo {title} {Quantum simulations with ultracold
  atoms in optical lattices},\ }\href {https://doi.org/10.1126/science.aal3837}
  {\bibfield  {journal} {\bibinfo  {journal} {Science}\ }\textbf {\bibinfo
  {volume} {357}},\ \bibinfo {pages} {995} (\bibinfo {year}
  {2017})}\BibitemShut {NoStop}%
\bibitem [{\citenamefont {Schneider}\ \emph {et~al.}(2008)\citenamefont
  {Schneider}, \citenamefont {Hackermuller}, \citenamefont {Will},
  \citenamefont {Best}, \citenamefont {Bloch}, \citenamefont {Costi},
  \citenamefont {Helmes}, \citenamefont {Rasch},\ and\ \citenamefont
  {Rosch}}]{Schneider2008}%
  \BibitemOpen
  \bibfield  {author} {\bibinfo {author} {\bibfnamefont {U.}~\bibnamefont
  {Schneider}}, \bibinfo {author} {\bibfnamefont {L.}~\bibnamefont
  {Hackermuller}}, \bibinfo {author} {\bibfnamefont {S.}~\bibnamefont {Will}},
  \bibinfo {author} {\bibfnamefont {T.}~\bibnamefont {Best}}, \bibinfo {author}
  {\bibfnamefont {I.}~\bibnamefont {Bloch}}, \bibinfo {author} {\bibfnamefont
  {T.~A.}\ \bibnamefont {Costi}}, \bibinfo {author} {\bibfnamefont {R.~W.}\
  \bibnamefont {Helmes}}, \bibinfo {author} {\bibfnamefont {D.}~\bibnamefont
  {Rasch}},\ and\ \bibinfo {author} {\bibfnamefont {A.}~\bibnamefont {Rosch}},\
  }\bibfield  {title} {\bibinfo {title} {Metallic and {Insulating} {Phases} of
  {Repulsively} {Interacting} {Fermions} in a {3D} {Optical} {Lattice}},\
  }\href {https://doi.org/10.1126/science.1165449} {\bibfield  {journal}
  {\bibinfo  {journal} {Science}\ }\textbf {\bibinfo {volume} {322}},\ \bibinfo
  {pages} {1520} (\bibinfo {year} {2008})}\BibitemShut {NoStop}%
\bibitem [{\citenamefont {J{\"o}rdens}\ \emph {et~al.}(2008)\citenamefont
  {J{\"o}rdens}, \citenamefont {Strohmaier}, \citenamefont {G{\"u}nter},
  \citenamefont {Moritz},\ and\ \citenamefont {Esslinger}}]{Jordens2008}%
  \BibitemOpen
  \bibfield  {author} {\bibinfo {author} {\bibfnamefont {R.}~\bibnamefont
  {J{\"o}rdens}}, \bibinfo {author} {\bibfnamefont {N.}~\bibnamefont
  {Strohmaier}}, \bibinfo {author} {\bibfnamefont {K.}~\bibnamefont
  {G{\"u}nter}}, \bibinfo {author} {\bibfnamefont {H.}~\bibnamefont {Moritz}},\
  and\ \bibinfo {author} {\bibfnamefont {T.}~\bibnamefont {Esslinger}},\
  }\bibfield  {title} {\bibinfo {title} {A {M}ott insulator of fermionic atoms
  in an optical lattice},\ }\href {https://doi.org/10.1038/nature07244}
  {\bibfield  {journal} {\bibinfo  {journal} {Nature}\ }\textbf {\bibinfo
  {volume} {455}},\ \bibinfo {pages} {204} (\bibinfo {year}
  {2008})}\BibitemShut {NoStop}%
\bibitem [{\citenamefont {J{\"o}rdens}\ \emph {et~al.}(2010)\citenamefont
  {J{\"o}rdens}, \citenamefont {Tarruell}, \citenamefont {Greif}, \citenamefont
  {Uehlinger}, \citenamefont {Strohmaier}, \citenamefont {Moritz},
  \citenamefont {Esslinger}, \citenamefont {De~Leo}, \citenamefont {Kollath},
  \citenamefont {Georges}, \citenamefont {Scarola}, \citenamefont {Pollet},
  \citenamefont {Burovski}, \citenamefont {Kozik},\ and\ \citenamefont
  {Troyer}}]{Jordens2010}%
  \BibitemOpen
  \bibfield  {author} {\bibinfo {author} {\bibfnamefont {R.}~\bibnamefont
  {J{\"o}rdens}}, \bibinfo {author} {\bibfnamefont {L.}~\bibnamefont
  {Tarruell}}, \bibinfo {author} {\bibfnamefont {D.}~\bibnamefont {Greif}},
  \bibinfo {author} {\bibfnamefont {T.}~\bibnamefont {Uehlinger}}, \bibinfo
  {author} {\bibfnamefont {N.}~\bibnamefont {Strohmaier}}, \bibinfo {author}
  {\bibfnamefont {H.}~\bibnamefont {Moritz}}, \bibinfo {author} {\bibfnamefont
  {T.}~\bibnamefont {Esslinger}}, \bibinfo {author} {\bibfnamefont
  {L.}~\bibnamefont {De~Leo}}, \bibinfo {author} {\bibfnamefont
  {C.}~\bibnamefont {Kollath}}, \bibinfo {author} {\bibfnamefont
  {A.}~\bibnamefont {Georges}}, \bibinfo {author} {\bibfnamefont
  {V.}~\bibnamefont {Scarola}}, \bibinfo {author} {\bibfnamefont
  {L.}~\bibnamefont {Pollet}}, \bibinfo {author} {\bibfnamefont
  {E.}~\bibnamefont {Burovski}}, \bibinfo {author} {\bibfnamefont
  {E.}~\bibnamefont {Kozik}},\ and\ \bibinfo {author} {\bibfnamefont
  {M.}~\bibnamefont {Troyer}},\ }\bibfield  {title} {\bibinfo {title}
  {Quantitative determination of temperature in the approach to magnetic order
  of ultracold fermions in an optical lattice},\ }\href
  {https://doi.org/10.1103/PhysRevLett.104.180401} {\bibfield  {journal}
  {\bibinfo  {journal} {Phys. Rev. Lett.}\ }\textbf {\bibinfo {volume} {104}},\
  \bibinfo {pages} {180401} (\bibinfo {year} {2010})}\BibitemShut {NoStop}%
\bibitem [{\citenamefont {Mazurenko}\ \emph {et~al.}(2017)\citenamefont
  {Mazurenko}, \citenamefont {Chiu}, \citenamefont {Ji}, \citenamefont
  {Parsons}, \citenamefont {Kan{\'a}sz-Nagy}, \citenamefont {Schmidt},
  \citenamefont {Grusdt}, \citenamefont {Demler}, \citenamefont {Greif},\ and\
  \citenamefont {Greiner}}]{Mazurenko2017}%
  \BibitemOpen
  \bibfield  {author} {\bibinfo {author} {\bibfnamefont {A.}~\bibnamefont
  {Mazurenko}}, \bibinfo {author} {\bibfnamefont {C.~S.}\ \bibnamefont {Chiu}},
  \bibinfo {author} {\bibfnamefont {G.}~\bibnamefont {Ji}}, \bibinfo {author}
  {\bibfnamefont {M.~F.}\ \bibnamefont {Parsons}}, \bibinfo {author}
  {\bibfnamefont {M.}~\bibnamefont {Kan{\'a}sz-Nagy}}, \bibinfo {author}
  {\bibfnamefont {R.}~\bibnamefont {Schmidt}}, \bibinfo {author} {\bibfnamefont
  {F.}~\bibnamefont {Grusdt}}, \bibinfo {author} {\bibfnamefont
  {E.}~\bibnamefont {Demler}}, \bibinfo {author} {\bibfnamefont
  {D.}~\bibnamefont {Greif}},\ and\ \bibinfo {author} {\bibfnamefont
  {M.}~\bibnamefont {Greiner}},\ }\bibfield  {title} {\bibinfo {title} {A
  cold-atom {Fermi-Hubbard} antiferromagnet},\ }\href
  {https://doi.org/10.1038/nature22362} {\bibfield  {journal} {\bibinfo
  {journal} {Nature}\ }\textbf {\bibinfo {volume} {545}},\ \bibinfo {pages}
  {462} (\bibinfo {year} {2017})}\BibitemShut {NoStop}%
\bibitem [{\citenamefont {Drewes}\ \emph {et~al.}(2017)\citenamefont {Drewes},
  \citenamefont {Miller}, \citenamefont {Cocchi}, \citenamefont {Chan},
  \citenamefont {Wurz}, \citenamefont {Gall}, \citenamefont {Pertot},
  \citenamefont {Brennecke},\ and\ \citenamefont {K\"ohl}}]{Drewes2017}%
  \BibitemOpen
  \bibfield  {author} {\bibinfo {author} {\bibfnamefont {J.~H.}\ \bibnamefont
  {Drewes}}, \bibinfo {author} {\bibfnamefont {L.~A.}\ \bibnamefont {Miller}},
  \bibinfo {author} {\bibfnamefont {E.}~\bibnamefont {Cocchi}}, \bibinfo
  {author} {\bibfnamefont {C.~F.}\ \bibnamefont {Chan}}, \bibinfo {author}
  {\bibfnamefont {N.}~\bibnamefont {Wurz}}, \bibinfo {author} {\bibfnamefont
  {M.}~\bibnamefont {Gall}}, \bibinfo {author} {\bibfnamefont {D.}~\bibnamefont
  {Pertot}}, \bibinfo {author} {\bibfnamefont {F.}~\bibnamefont {Brennecke}},\
  and\ \bibinfo {author} {\bibfnamefont {M.}~\bibnamefont {K\"ohl}},\
  }\bibfield  {title} {\bibinfo {title} {Antiferromagnetic correlations in
  two-dimensional fermionic {M}ott-insulating and metallic phases},\ }\href
  {https://doi.org/10.1103/PhysRevLett.118.170401} {\bibfield  {journal}
  {\bibinfo  {journal} {Phys. Rev. Lett.}\ }\textbf {\bibinfo {volume} {118}},\
  \bibinfo {pages} {170401} (\bibinfo {year} {2017})}\BibitemShut {NoStop}%
\bibitem [{\citenamefont {Hensgens}\ \emph {et~al.}(2017)\citenamefont
  {Hensgens}, \citenamefont {Fujita}, \citenamefont {Janssen}, \citenamefont
  {Li}, \citenamefont {Van~Diepen}, \citenamefont {Reichl}, \citenamefont
  {Wegscheider}, \citenamefont {Das~Sarma},\ and\ \citenamefont
  {Vandersypen}}]{Hensgens2017}%
  \BibitemOpen
  \bibfield  {author} {\bibinfo {author} {\bibfnamefont {T.}~\bibnamefont
  {Hensgens}}, \bibinfo {author} {\bibfnamefont {T.}~\bibnamefont {Fujita}},
  \bibinfo {author} {\bibfnamefont {L.}~\bibnamefont {Janssen}}, \bibinfo
  {author} {\bibfnamefont {X.}~\bibnamefont {Li}}, \bibinfo {author}
  {\bibfnamefont {C.~J.}\ \bibnamefont {Van~Diepen}}, \bibinfo {author}
  {\bibfnamefont {C.}~\bibnamefont {Reichl}}, \bibinfo {author} {\bibfnamefont
  {W.}~\bibnamefont {Wegscheider}}, \bibinfo {author} {\bibfnamefont
  {S.}~\bibnamefont {Das~Sarma}},\ and\ \bibinfo {author} {\bibfnamefont
  {L.~M.~K.}\ \bibnamefont {Vandersypen}},\ }\bibfield  {title} {\bibinfo
  {title} {Quantum simulation of a {Fermi-Hubbard} model using a semiconductor
  quantum dot array},\ }\href {https://doi.org/10.1038/nature23022} {\bibfield
  {journal} {\bibinfo  {journal} {Nature}\ }\textbf {\bibinfo {volume} {548}},\
  \bibinfo {pages} {70} (\bibinfo {year} {2017})}\BibitemShut {NoStop}%
\bibitem [{\citenamefont {Melo}\ \emph {et~al.}(2021)\citenamefont {Melo},
  \citenamefont {Souza}, \citenamefont {Oliveira},\ and\ \citenamefont
  {Sarthour}}]{Melo2021}%
  \BibitemOpen
  \bibfield  {author} {\bibinfo {author} {\bibfnamefont {F.~V.}\ \bibnamefont
  {Melo}}, \bibinfo {author} {\bibfnamefont {A.~M.}\ \bibnamefont {Souza}},
  \bibinfo {author} {\bibfnamefont {I.~S.}\ \bibnamefont {Oliveira}},\ and\
  \bibinfo {author} {\bibfnamefont {R.~S.}\ \bibnamefont {Sarthour}},\
  }\bibfield  {title} {\bibinfo {title} {Quantum simulation of the two-site
  {H}ubbard {Hamiltonian}},\ }\href
  {https://doi.org/https://doi.org/10.1016/j.physo.2020.100053} {\bibfield
  {journal} {\bibinfo  {journal} {Physics Open}\ }\textbf {\bibinfo {volume}
  {6}},\ \bibinfo {pages} {100053} (\bibinfo {year} {2021})}\BibitemShut
  {NoStop}%
\bibitem [{\citenamefont {Cade}\ \emph {et~al.}(2020)\citenamefont {Cade},
  \citenamefont {Mineh}, \citenamefont {Montanaro},\ and\ \citenamefont
  {Stanisic}}]{Cade2020}%
  \BibitemOpen
  \bibfield  {author} {\bibinfo {author} {\bibfnamefont {C.}~\bibnamefont
  {Cade}}, \bibinfo {author} {\bibfnamefont {L.}~\bibnamefont {Mineh}},
  \bibinfo {author} {\bibfnamefont {A.}~\bibnamefont {Montanaro}},\ and\
  \bibinfo {author} {\bibfnamefont {S.}~\bibnamefont {Stanisic}},\ }\bibfield
  {title} {\bibinfo {title} {Strategies for solving the {Fermi-Hubbard} model
  on near-term quantum computers},\ }\href
  {https://doi.org/10.1103/PhysRevB.102.235122} {\bibfield  {journal} {\bibinfo
   {journal} {Phys. Rev. B}\ }\textbf {\bibinfo {volume} {102}},\ \bibinfo
  {pages} {235122} (\bibinfo {year} {2020})}\BibitemShut {NoStop}%
\bibitem [{\citenamefont {Choquette}\ \emph {et~al.}(2021)\citenamefont
  {Choquette}, \citenamefont {Di~Paolo}, \citenamefont {Barkoutsos},
  \citenamefont {S\'en\'echal}, \citenamefont {Tavernelli},\ and\ \citenamefont
  {Blais}}]{Choquette2021}%
  \BibitemOpen
  \bibfield  {author} {\bibinfo {author} {\bibfnamefont {A.}~\bibnamefont
  {Choquette}}, \bibinfo {author} {\bibfnamefont {A.}~\bibnamefont {Di~Paolo}},
  \bibinfo {author} {\bibfnamefont {P.~K.}\ \bibnamefont {Barkoutsos}},
  \bibinfo {author} {\bibfnamefont {D.}~\bibnamefont {S\'en\'echal}}, \bibinfo
  {author} {\bibfnamefont {I.}~\bibnamefont {Tavernelli}},\ and\ \bibinfo
  {author} {\bibfnamefont {A.}~\bibnamefont {Blais}},\ }\bibfield  {title}
  {\bibinfo {title} {{Quantum-optimal-control-inspired} ansatz for variational
  quantum algorithms},\ }\href
  {https://doi.org/10.1103/PhysRevResearch.3.023092} {\bibfield  {journal}
  {\bibinfo  {journal} {Phys. Rev. Res.}\ }\textbf {\bibinfo {volume} {3}},\
  \bibinfo {pages} {023092} (\bibinfo {year} {2021})}\BibitemShut {NoStop}%
\bibitem [{\citenamefont {Stanisic}\ \emph {et~al.}(2022)\citenamefont
  {Stanisic}, \citenamefont {Bosse}, \citenamefont {Gambetta}, \citenamefont
  {Santos}, \citenamefont {Mruczkiewicz}, \citenamefont {O'Brien},
  \citenamefont {Ostby},\ and\ \citenamefont {Montanaro}}]{Stanisic2022}%
  \BibitemOpen
  \bibfield  {author} {\bibinfo {author} {\bibfnamefont {S.}~\bibnamefont
  {Stanisic}}, \bibinfo {author} {\bibfnamefont {J.~L.}\ \bibnamefont {Bosse}},
  \bibinfo {author} {\bibfnamefont {F.~M.}\ \bibnamefont {Gambetta}}, \bibinfo
  {author} {\bibfnamefont {R.~A.}\ \bibnamefont {Santos}}, \bibinfo {author}
  {\bibfnamefont {W.}~\bibnamefont {Mruczkiewicz}}, \bibinfo {author}
  {\bibfnamefont {T.~E.}\ \bibnamefont {O'Brien}}, \bibinfo {author}
  {\bibfnamefont {E.}~\bibnamefont {Ostby}},\ and\ \bibinfo {author}
  {\bibfnamefont {A.}~\bibnamefont {Montanaro}},\ }\bibfield  {title} {\bibinfo
  {title} {Observing ground-state properties of the {Fermi-Hubbard} model using
  a scalable algorithm on a quantum computer},\ }\href
  {https://doi.org/10.1038/s41467-022-33335-4} {\bibfield  {journal} {\bibinfo
  {journal} {Nat. Commun.}\ }\textbf {\bibinfo {volume} {13}},\ \bibinfo
  {pages} {5743} (\bibinfo {year} {2022})}\BibitemShut {NoStop}%
\bibitem [{\citenamefont {Suchsland}\ \emph {et~al.}(2022)\citenamefont
  {Suchsland}, \citenamefont {Barkoutsos}, \citenamefont {Tavernelli},
  \citenamefont {Fischer},\ and\ \citenamefont {Neupert}}]{Suchsland2022}%
  \BibitemOpen
  \bibfield  {author} {\bibinfo {author} {\bibfnamefont {P.}~\bibnamefont
  {Suchsland}}, \bibinfo {author} {\bibfnamefont {P.~K.}\ \bibnamefont
  {Barkoutsos}}, \bibinfo {author} {\bibfnamefont {I.}~\bibnamefont
  {Tavernelli}}, \bibinfo {author} {\bibfnamefont {M.~H.}\ \bibnamefont
  {Fischer}},\ and\ \bibinfo {author} {\bibfnamefont {T.}~\bibnamefont
  {Neupert}},\ }\bibfield  {title} {\bibinfo {title} {Simulating a ring-like
  {H}ubbard system with a quantum computer},\ }\href
  {https://doi.org/10.1103/PhysRevResearch.4.013165} {\bibfield  {journal}
  {\bibinfo  {journal} {Phys. Rev. Res.}\ }\textbf {\bibinfo {volume} {4}},\
  \bibinfo {pages} {013165} (\bibinfo {year} {2022})}\BibitemShut {NoStop}%
\bibitem [{\citenamefont {Takahashi}(1977)}]{Takahashi1977}%
  \BibitemOpen
  \bibfield  {author} {\bibinfo {author} {\bibfnamefont {M.}~\bibnamefont
  {Takahashi}},\ }\bibfield  {title} {\bibinfo {title} {Half-filled {H}ubbard
  model at low temperature},\ }\href
  {https://doi.org/10.1088/0022-3719/10/8/031} {\bibfield  {journal} {\bibinfo
  {journal} {J. Phys. C Solid State Phys.}\ }\textbf {\bibinfo {volume} {10}},\
  \bibinfo {pages} {1289} (\bibinfo {year} {1977})}\BibitemShut {NoStop}%
\bibitem [{\citenamefont {Hirsch}(1985)}]{Hirsch1985}%
  \BibitemOpen
  \bibfield  {author} {\bibinfo {author} {\bibfnamefont {J.~E.}\ \bibnamefont
  {Hirsch}},\ }\bibfield  {title} {\bibinfo {title} {Attractive interaction and
  pairing in fermion systems with strong on-site repulsion},\ }\href
  {https://doi.org/10.1103/PhysRevLett.54.1317} {\bibfield  {journal} {\bibinfo
   {journal} {Phys. Rev. Lett.}\ }\textbf {\bibinfo {volume} {54}},\ \bibinfo
  {pages} {1317} (\bibinfo {year} {1985})}\BibitemShut {NoStop}%
\bibitem [{\citenamefont {Gros}\ \emph
  {et~al.}(1987{\natexlab{a}})\citenamefont {Gros}, \citenamefont {Joynt},\
  and\ \citenamefont {Rice}}]{Gros1987PRB}%
  \BibitemOpen
  \bibfield  {author} {\bibinfo {author} {\bibfnamefont {C.}~\bibnamefont
  {Gros}}, \bibinfo {author} {\bibfnamefont {R.}~\bibnamefont {Joynt}},\ and\
  \bibinfo {author} {\bibfnamefont {T.~M.}\ \bibnamefont {Rice}},\ }\bibfield
  {title} {\bibinfo {title} {Antiferromagnetic correlations in almost-localized
  {F}ermi liquids},\ }\href {https://doi.org/10.1103/PhysRevB.36.381}
  {\bibfield  {journal} {\bibinfo  {journal} {Phys. Rev. B}\ }\textbf {\bibinfo
  {volume} {36}},\ \bibinfo {pages} {381} (\bibinfo {year}
  {1987}{\natexlab{a}})}\BibitemShut {NoStop}%
\bibitem [{\citenamefont {MacDonald}\ \emph {et~al.}(1988)\citenamefont
  {MacDonald}, \citenamefont {Girvin},\ and\ \citenamefont
  {Yoshioka}}]{MacDonald1988}%
  \BibitemOpen
  \bibfield  {author} {\bibinfo {author} {\bibfnamefont {A.~H.}\ \bibnamefont
  {MacDonald}}, \bibinfo {author} {\bibfnamefont {S.~M.}\ \bibnamefont
  {Girvin}},\ and\ \bibinfo {author} {\bibfnamefont {D.}~\bibnamefont
  {Yoshioka}},\ }\bibfield  {title} {\bibinfo {title} {$\frac{t}{U}$ expansion
  for the {H}ubbard model},\ }\href {https://doi.org/10.1103/PhysRevB.37.9753}
  {\bibfield  {journal} {\bibinfo  {journal} {Phys. Rev. B}\ }\textbf {\bibinfo
  {volume} {37}},\ \bibinfo {pages} {9753} (\bibinfo {year}
  {1988})}\BibitemShut {NoStop}%
\bibitem [{\citenamefont {Gros}\ \emph
  {et~al.}(1987{\natexlab{b}})\citenamefont {Gros}, \citenamefont {Joynt},\
  and\ \citenamefont {Rice}}]{Gros1987}%
  \BibitemOpen
  \bibfield  {author} {\bibinfo {author} {\bibfnamefont {C.}~\bibnamefont
  {Gros}}, \bibinfo {author} {\bibfnamefont {R.}~\bibnamefont {Joynt}},\ and\
  \bibinfo {author} {\bibfnamefont {T.~M.}\ \bibnamefont {Rice}},\ }\bibfield
  {title} {\bibinfo {title} {Superconducting instability in the {large-U} limit
  of the two-dimensional {H}ubbard model},\ }\href
  {https://doi.org/10.1007/BF01471072} {\bibfield  {journal} {\bibinfo
  {journal} {Z. Physik B - Condensed Matter}\ }\textbf {\bibinfo {volume}
  {68}},\ \bibinfo {pages} {425} (\bibinfo {year}
  {1987}{\natexlab{b}})}\BibitemShut {NoStop}%
\bibitem [{\citenamefont {Zhang}\ and\ \citenamefont {Rice}(1988)}]{Zhang1988}%
  \BibitemOpen
  \bibfield  {author} {\bibinfo {author} {\bibfnamefont {F.~C.}\ \bibnamefont
  {Zhang}}\ and\ \bibinfo {author} {\bibfnamefont {T.~M.}\ \bibnamefont
  {Rice}},\ }\bibfield  {title} {\bibinfo {title} {Effective hamiltonian for
  the superconducting cu oxides},\ }\href
  {https://doi.org/10.1103/PhysRevB.37.3759} {\bibfield  {journal} {\bibinfo
  {journal} {Phys. Rev. B}\ }\textbf {\bibinfo {volume} {37}},\ \bibinfo
  {pages} {3759} (\bibinfo {year} {1988})}\BibitemShut {NoStop}%
\bibitem [{\citenamefont {Spa\l{}ek}(1988)}]{Spalek1988}%
  \BibitemOpen
  \bibfield  {author} {\bibinfo {author} {\bibfnamefont {J.}~\bibnamefont
  {Spa\l{}ek}},\ }\bibfield  {title} {\bibinfo {title} {Effect of pair hopping
  and magnitude of intra-atomic interaction on exchange-mediated
  superconductivity},\ }\href {https://doi.org/10.1103/PhysRevB.37.533}
  {\bibfield  {journal} {\bibinfo  {journal} {Phys. Rev. B}\ }\textbf {\bibinfo
  {volume} {37}},\ \bibinfo {pages} {533} (\bibinfo {year} {1988})}\BibitemShut
  {NoStop}%
\bibitem [{\citenamefont {Lee}\ \emph {et~al.}(2006)\citenamefont {Lee},
  \citenamefont {Nagaosa},\ and\ \citenamefont {Wen}}]{Lee2006}%
  \BibitemOpen
  \bibfield  {author} {\bibinfo {author} {\bibfnamefont {P.~A.}\ \bibnamefont
  {Lee}}, \bibinfo {author} {\bibfnamefont {N.}~\bibnamefont {Nagaosa}},\ and\
  \bibinfo {author} {\bibfnamefont {X.-G.}\ \bibnamefont {Wen}},\ }\bibfield
  {title} {\bibinfo {title} {Doping a {M}ott insulator: {P}hysics of
  high-temperature superconductivity},\ }\href
  {https://doi.org/10.1103/RevModPhys.78.17} {\bibfield  {journal} {\bibinfo
  {journal} {Rev. Mod. Phys.}\ }\textbf {\bibinfo {volume} {78}},\ \bibinfo
  {pages} {17} (\bibinfo {year} {2006})}\BibitemShut {NoStop}%
\bibitem [{\citenamefont {Chao}\ \emph {et~al.}(1977)\citenamefont {Chao},
  \citenamefont {Spalek},\ and\ \citenamefont {Oles}}]{Chao1977}%
  \BibitemOpen
  \bibfield  {author} {\bibinfo {author} {\bibfnamefont {K.~A.}\ \bibnamefont
  {Chao}}, \bibinfo {author} {\bibfnamefont {J.}~\bibnamefont {Spalek}},\ and\
  \bibinfo {author} {\bibfnamefont {A.~M.}\ \bibnamefont {Oles}},\ }\bibfield
  {title} {\bibinfo {title} {Kinetic exchange interaction in a narrow
  {S-band}},\ }\href {https://doi.org/10.1088/0022-3719/10/10/002} {\bibfield
  {journal} {\bibinfo  {journal} {J. Phys. C Solid State Phys.}\ }\textbf
  {\bibinfo {volume} {10}},\ \bibinfo {pages} {L271} (\bibinfo {year}
  {1977})}\BibitemShut {NoStop}%
\bibitem [{\citenamefont {Fazekas}(1999)}]{Fazekas1999}%
  \BibitemOpen
  \bibfield  {author} {\bibinfo {author} {\bibfnamefont {P.}~\bibnamefont
  {Fazekas}},\ }\href {https://doi.org/10.1142/2945} {\emph {\bibinfo {title}
  {Lecture Notes on Electron Correlation and Magnetism}}}\ (\bibinfo
  {publisher} {World Scientific},\ \bibinfo {year} {1999})\BibitemShut
  {NoStop}%
\bibitem [{\citenamefont {Auerbach}(1998)}]{Auerbach1998}%
  \BibitemOpen
  \bibfield  {author} {\bibinfo {author} {\bibfnamefont {A.}~\bibnamefont
  {Auerbach}},\ }\href {https://doi.org/10.1007/978-1-4612-0869-3} {\emph
  {\bibinfo {title} {Interacting electrons and quantum magnetism}}}\ (\bibinfo
  {publisher} {Springer},\ \bibinfo {address} {New York},\ \bibinfo {year}
  {1998})\BibitemShut {NoStop}%
\bibitem [{\citenamefont {Childs}\ \emph {et~al.}(2021)\citenamefont {Childs},
  \citenamefont {Su}, \citenamefont {Tran}, \citenamefont {Wiebe},\ and\
  \citenamefont {Zhu}}]{Childs2021}%
  \BibitemOpen
  \bibfield  {author} {\bibinfo {author} {\bibfnamefont {A.~M.}\ \bibnamefont
  {Childs}}, \bibinfo {author} {\bibfnamefont {Y.}~\bibnamefont {Su}}, \bibinfo
  {author} {\bibfnamefont {M.~C.}\ \bibnamefont {Tran}}, \bibinfo {author}
  {\bibfnamefont {N.}~\bibnamefont {Wiebe}},\ and\ \bibinfo {author}
  {\bibfnamefont {S.}~\bibnamefont {Zhu}},\ }\bibfield  {title} {\bibinfo
  {title} {Theory of {Trotter} error with commutator scaling},\ }\href
  {https://doi.org/10.1103/PhysRevX.11.011020} {\bibfield  {journal} {\bibinfo
  {journal} {Phys. Rev. X}\ }\textbf {\bibinfo {volume} {11}},\ \bibinfo
  {pages} {011020} (\bibinfo {year} {2021})}\BibitemShut {NoStop}%
\bibitem [{\citenamefont {Kwon}\ and\ \citenamefont {Park}(2022)}]{Kwon2022}%
  \BibitemOpen
  \bibfield  {author} {\bibinfo {author} {\bibfnamefont {H.}~\bibnamefont
  {Kwon}}\ and\ \bibinfo {author} {\bibfnamefont {K.}~\bibnamefont {Park}},\
  }\bibfield  {title} {\bibinfo {title} {Projected {BCS} theory for the
  unification of antiferromagnetism and strongly correlated
  superconductivity},\ }\href
  {https://doi.org/10.1103/PhysRevResearch.4.013116} {\bibfield  {journal}
  {\bibinfo  {journal} {Phys. Rev. Res.}\ }\textbf {\bibinfo {volume} {4}},\
  \bibinfo {pages} {013116} (\bibinfo {year} {2022})}\BibitemShut {NoStop}%
\bibitem [{\citenamefont {Suzuki}(1985)}]{Suzuki1985}%
  \BibitemOpen
  \bibfield  {author} {\bibinfo {author} {\bibfnamefont {M.}~\bibnamefont
  {Suzuki}},\ }\bibfield  {title} {\bibinfo {title} {Decomposition formulas of
  exponential operators and {L}ie exponentials with some applications to
  quantum mechanics and statistical physics},\ }\href
  {https://doi.org/10.1063/1.526596} {\bibfield  {journal} {\bibinfo  {journal}
  {J. Math. Phys.}\ }\textbf {\bibinfo {volume} {26}},\ \bibinfo {pages} {601}
  (\bibinfo {year} {1985})}\BibitemShut {NoStop}%
\bibitem [{\citenamefont {Hadfield}\ and\ \citenamefont
  {Papageorgiou}(2018)}]{Hadfield2018}%
  \BibitemOpen
  \bibfield  {author} {\bibinfo {author} {\bibfnamefont {S.}~\bibnamefont
  {Hadfield}}\ and\ \bibinfo {author} {\bibfnamefont {A.}~\bibnamefont
  {Papageorgiou}},\ }\bibfield  {title} {\bibinfo {title} {Divide and conquer
  approach to quantum {Hamiltonian} simulation},\ }\href
  {https://doi.org/10.1088/1367-2630/aab1ef} {\bibfield  {journal} {\bibinfo
  {journal} {New J. Phys.}\ }\textbf {\bibinfo {volume} {20}},\ \bibinfo
  {pages} {043003} (\bibinfo {year} {2018})}\BibitemShut {NoStop}%
\bibitem [{\citenamefont {Azzouz}(1993)}]{Azzouz1993}%
  \BibitemOpen
  \bibfield  {author} {\bibinfo {author} {\bibfnamefont {M.}~\bibnamefont
  {Azzouz}},\ }\bibfield  {title} {\bibinfo {title} {Interchain-coupling effect
  on the one-dimensional spin-1/2 antiferromagnetic {H}eisenberg model},\
  }\href {https://doi.org/10.1103/PhysRevB.48.6136} {\bibfield  {journal}
  {\bibinfo  {journal} {Phys. Rev. B}\ }\textbf {\bibinfo {volume} {48}},\
  \bibinfo {pages} {6136} (\bibinfo {year} {1993})}\BibitemShut {NoStop}%
\bibitem [{\citenamefont {Bauer}\ \emph {et~al.}(2011)\citenamefont {Bauer},
  \citenamefont {Carr}, \citenamefont {Evertz}, \citenamefont {Feiguin},
  \citenamefont {Freire}, \citenamefont {Fuchs}, \citenamefont {Gamper},
  \citenamefont {Gukelberger}, \citenamefont {Gull}, \citenamefont {Guertler},
  \citenamefont {Hehn}, \citenamefont {Igarashi}, \citenamefont {Isakov},
  \citenamefont {Koop}, \citenamefont {Ma}, \citenamefont {Mates},
  \citenamefont {Matsuo}, \citenamefont {Parcollet}, \citenamefont
  {Pawłowski}, \citenamefont {Picon}, \citenamefont {Pollet}, \citenamefont
  {Santos}, \citenamefont {Scarola}, \citenamefont {Schollwöck}, \citenamefont
  {Silva}, \citenamefont {Surer}, \citenamefont {Todo}, \citenamefont {Trebst},
  \citenamefont {Troyer}, \citenamefont {Wall}, \citenamefont {Werner},\ and\
  \citenamefont {Wessel}}]{Bauer2011}%
  \BibitemOpen
  \bibfield  {author} {\bibinfo {author} {\bibfnamefont {B.}~\bibnamefont
  {Bauer}}, \bibinfo {author} {\bibfnamefont {L.~D.}\ \bibnamefont {Carr}},
  \bibinfo {author} {\bibfnamefont {H.~G.}\ \bibnamefont {Evertz}}, \bibinfo
  {author} {\bibfnamefont {A.}~\bibnamefont {Feiguin}}, \bibinfo {author}
  {\bibfnamefont {J.}~\bibnamefont {Freire}}, \bibinfo {author} {\bibfnamefont
  {S.}~\bibnamefont {Fuchs}}, \bibinfo {author} {\bibfnamefont
  {L.}~\bibnamefont {Gamper}}, \bibinfo {author} {\bibfnamefont
  {J.}~\bibnamefont {Gukelberger}}, \bibinfo {author} {\bibfnamefont
  {E.}~\bibnamefont {Gull}}, \bibinfo {author} {\bibfnamefont {S.}~\bibnamefont
  {Guertler}}, \bibinfo {author} {\bibfnamefont {A.}~\bibnamefont {Hehn}},
  \bibinfo {author} {\bibfnamefont {R.}~\bibnamefont {Igarashi}}, \bibinfo
  {author} {\bibfnamefont {S.~V.}\ \bibnamefont {Isakov}}, \bibinfo {author}
  {\bibfnamefont {D.}~\bibnamefont {Koop}}, \bibinfo {author} {\bibfnamefont
  {P.~N.}\ \bibnamefont {Ma}}, \bibinfo {author} {\bibfnamefont
  {P.}~\bibnamefont {Mates}}, \bibinfo {author} {\bibfnamefont
  {H.}~\bibnamefont {Matsuo}}, \bibinfo {author} {\bibfnamefont
  {O.}~\bibnamefont {Parcollet}}, \bibinfo {author} {\bibfnamefont
  {G.}~\bibnamefont {Pawłowski}}, \bibinfo {author} {\bibfnamefont {J.~D.}\
  \bibnamefont {Picon}}, \bibinfo {author} {\bibfnamefont {L.}~\bibnamefont
  {Pollet}}, \bibinfo {author} {\bibfnamefont {E.}~\bibnamefont {Santos}},
  \bibinfo {author} {\bibfnamefont {V.~W.}\ \bibnamefont {Scarola}}, \bibinfo
  {author} {\bibfnamefont {U.}~\bibnamefont {Schollwöck}}, \bibinfo {author}
  {\bibfnamefont {C.}~\bibnamefont {Silva}}, \bibinfo {author} {\bibfnamefont
  {B.}~\bibnamefont {Surer}}, \bibinfo {author} {\bibfnamefont
  {S.}~\bibnamefont {Todo}}, \bibinfo {author} {\bibfnamefont {S.}~\bibnamefont
  {Trebst}}, \bibinfo {author} {\bibfnamefont {M.}~\bibnamefont {Troyer}},
  \bibinfo {author} {\bibfnamefont {M.~L.}\ \bibnamefont {Wall}}, \bibinfo
  {author} {\bibfnamefont {P.}~\bibnamefont {Werner}},\ and\ \bibinfo {author}
  {\bibfnamefont {S.}~\bibnamefont {Wessel}},\ }\bibfield  {title} {\bibinfo
  {title} {The {ALPS} project release 2.0: open source software for strongly
  correlated systems},\ }\href
  {https://doi.org/10.1088/1742-5468/2011/05/P05001} {\bibfield  {journal}
  {\bibinfo  {journal} {J. Stat. Mech. Theory Exp.}\ }\textbf {\bibinfo
  {volume} {2011}},\ \bibinfo {pages} {P05001} (\bibinfo {year}
  {2011})}\BibitemShut {NoStop}%
\bibitem [{\citenamefont {Po}(2021)}]{Po2021}%
  \BibitemOpen
  \bibfield  {author} {\bibinfo {author} {\bibfnamefont {H.~C.}\ \bibnamefont
  {Po}},\ }\href@noop {} {\bibinfo {title} {Symmetric {Jordan-Wigner}
  transformation in higher dimensions}} (\bibinfo {year} {2021}),\ \Eprint
  {https://arxiv.org/abs/arXiv:2107.10842v2} {arXiv:2107.10842v2} \BibitemShut
  {NoStop}%
\bibitem [{\citenamefont {Barnes}\ and\ \citenamefont
  {Maekawa}(2001)}]{Barnes2002}%
  \BibitemOpen
  \bibfield  {author} {\bibinfo {author} {\bibfnamefont {S.~E.}\ \bibnamefont
  {Barnes}}\ and\ \bibinfo {author} {\bibfnamefont {S.}~\bibnamefont
  {Maekawa}},\ }\bibfield  {title} {\bibinfo {title} {A {Jordan-Wigner}
  transformation for the {t-J} and {H}ubbard models with holes},\ }\href
  {https://doi.org/10.1088/0953-8984/14/1/103} {\bibfield  {journal} {\bibinfo
  {journal} {J. Phys. Condens. Matter}\ }\textbf {\bibinfo {volume} {14}},\
  \bibinfo {pages} {L19} (\bibinfo {year} {2001})}\BibitemShut {NoStop}%
\bibitem [{\citenamefont {Bravyi}\ and\ \citenamefont
  {Kitaev}(2002)}]{Bravyi2002}%
  \BibitemOpen
  \bibfield  {author} {\bibinfo {author} {\bibfnamefont {S.~B.}\ \bibnamefont
  {Bravyi}}\ and\ \bibinfo {author} {\bibfnamefont {A.~Y.}\ \bibnamefont
  {Kitaev}},\ }\bibfield  {title} {\bibinfo {title} {Fermionic quantum
  computation},\ }\href {https://doi.org/10.1006/aphy.2002.6254} {\bibfield
  {journal} {\bibinfo  {journal} {Ann. Phys.}\ }\textbf {\bibinfo {volume}
  {298}},\ \bibinfo {pages} {210} (\bibinfo {year} {2002})}\BibitemShut
  {NoStop}%
\bibitem [{\citenamefont {Ball}(2005)}]{Ball2005}%
  \BibitemOpen
  \bibfield  {author} {\bibinfo {author} {\bibfnamefont {R.~C.}\ \bibnamefont
  {Ball}},\ }\bibfield  {title} {\bibinfo {title} {Fermions without fermion
  fields},\ }\href {https://doi.org/10.1103/PhysRevLett.95.176407} {\bibfield
  {journal} {\bibinfo  {journal} {Phys. Rev. Lett.}\ }\textbf {\bibinfo
  {volume} {95}},\ \bibinfo {pages} {176407} (\bibinfo {year}
  {2005})}\BibitemShut {NoStop}%
\bibitem [{\citenamefont {Verstraete}\ and\ \citenamefont
  {Cirac}(2005)}]{Verstraete2005}%
  \BibitemOpen
  \bibfield  {author} {\bibinfo {author} {\bibfnamefont {F.}~\bibnamefont
  {Verstraete}}\ and\ \bibinfo {author} {\bibfnamefont {J.~I.}\ \bibnamefont
  {Cirac}},\ }\bibfield  {title} {\bibinfo {title} {Mapping local
  {Hamiltonians} of fermions to local {Hamiltonians} of spins},\ }\href
  {https://doi.org/10.1088/1742-5468/2005/09/P09012} {\bibfield  {journal}
  {\bibinfo  {journal} {J. Stat. Mech. Theory Exp.}\ }\textbf {\bibinfo
  {volume} {2005}},\ \bibinfo {pages} {P09012} (\bibinfo {year}
  {2005})}\BibitemShut {NoStop}%
\bibitem [{\citenamefont {Whitfield}\ \emph {et~al.}(2016)\citenamefont
  {Whitfield}, \citenamefont {Havl\'{\i}\ifmmode~\check{c}\else \v{c}\fi{}ek},\
  and\ \citenamefont {Troyer}}]{Whitfield2016}%
  \BibitemOpen
  \bibfield  {author} {\bibinfo {author} {\bibfnamefont {J.~D.}\ \bibnamefont
  {Whitfield}}, \bibinfo {author} {\bibfnamefont {V.}~\bibnamefont
  {Havl\'{\i}\ifmmode~\check{c}\else \v{c}\fi{}ek}},\ and\ \bibinfo {author}
  {\bibfnamefont {M.}~\bibnamefont {Troyer}},\ }\bibfield  {title} {\bibinfo
  {title} {Local spin operators for fermion simulations},\ }\href
  {https://doi.org/10.1103/PhysRevA.94.030301} {\bibfield  {journal} {\bibinfo
  {journal} {Phys. Rev. A}\ }\textbf {\bibinfo {volume} {94}},\ \bibinfo
  {pages} {030301} (\bibinfo {year} {2016})}\BibitemShut {NoStop}%
\bibitem [{\citenamefont {Havl\'{\i}\ifmmode~\check{c}\else \v{c}\fi{}ek}\
  \emph {et~al.}(2017)\citenamefont {Havl\'{\i}\ifmmode~\check{c}\else
  \v{c}\fi{}ek}, \citenamefont {Troyer},\ and\ \citenamefont
  {Whitfield}}]{Havlicek2017}%
  \BibitemOpen
  \bibfield  {author} {\bibinfo {author} {\bibfnamefont {V.}~\bibnamefont
  {Havl\'{\i}\ifmmode~\check{c}\else \v{c}\fi{}ek}}, \bibinfo {author}
  {\bibfnamefont {M.}~\bibnamefont {Troyer}},\ and\ \bibinfo {author}
  {\bibfnamefont {J.~D.}\ \bibnamefont {Whitfield}},\ }\bibfield  {title}
  {\bibinfo {title} {Operator locality in the quantum simulation of fermionic
  models},\ }\href {https://doi.org/10.1103/PhysRevA.95.032332} {\bibfield
  {journal} {\bibinfo  {journal} {Phys. Rev. A}\ }\textbf {\bibinfo {volume}
  {95}},\ \bibinfo {pages} {032332} (\bibinfo {year} {2017})}\BibitemShut
  {NoStop}%
\bibitem [{\citenamefont {Steudtner}\ and\ \citenamefont
  {Wehner}(2019)}]{Steudtner2019}%
  \BibitemOpen
  \bibfield  {author} {\bibinfo {author} {\bibfnamefont {M.}~\bibnamefont
  {Steudtner}}\ and\ \bibinfo {author} {\bibfnamefont {S.}~\bibnamefont
  {Wehner}},\ }\bibfield  {title} {\bibinfo {title} {Quantum codes for quantum
  simulation of fermions on a square lattice of qubits},\ }\href
  {https://doi.org/10.1103/PhysRevA.99.022308} {\bibfield  {journal} {\bibinfo
  {journal} {Phys. Rev. A}\ }\textbf {\bibinfo {volume} {99}},\ \bibinfo
  {pages} {022308} (\bibinfo {year} {2019})}\BibitemShut {NoStop}%
\bibitem [{\citenamefont {Setia}\ \emph {et~al.}(2019)\citenamefont {Setia},
  \citenamefont {Bravyi}, \citenamefont {Mezzacapo},\ and\ \citenamefont
  {Whitfield}}]{Setia2019}%
  \BibitemOpen
  \bibfield  {author} {\bibinfo {author} {\bibfnamefont {K.}~\bibnamefont
  {Setia}}, \bibinfo {author} {\bibfnamefont {S.}~\bibnamefont {Bravyi}},
  \bibinfo {author} {\bibfnamefont {A.}~\bibnamefont {Mezzacapo}},\ and\
  \bibinfo {author} {\bibfnamefont {J.~D.}\ \bibnamefont {Whitfield}},\
  }\bibfield  {title} {\bibinfo {title} {Superfast encodings for fermionic
  quantum simulation},\ }\href
  {https://doi.org/10.1103/PhysRevResearch.1.033033} {\bibfield  {journal}
  {\bibinfo  {journal} {Phys. Rev. Res.}\ }\textbf {\bibinfo {volume} {1}},\
  \bibinfo {pages} {033033} (\bibinfo {year} {2019})}\BibitemShut {NoStop}%
\bibitem [{\citenamefont {Derby}\ \emph {et~al.}(2021)\citenamefont {Derby},
  \citenamefont {Klassen}, \citenamefont {Bausch},\ and\ \citenamefont
  {Cubitt}}]{Derby2021}%
  \BibitemOpen
  \bibfield  {author} {\bibinfo {author} {\bibfnamefont {C.}~\bibnamefont
  {Derby}}, \bibinfo {author} {\bibfnamefont {J.}~\bibnamefont {Klassen}},
  \bibinfo {author} {\bibfnamefont {J.}~\bibnamefont {Bausch}},\ and\ \bibinfo
  {author} {\bibfnamefont {T.}~\bibnamefont {Cubitt}},\ }\bibfield  {title}
  {\bibinfo {title} {Compact fermion to qubit mappings},\ }\href
  {https://doi.org/10.1103/PhysRevB.104.035118} {\bibfield  {journal} {\bibinfo
   {journal} {Phys. Rev. B}\ }\textbf {\bibinfo {volume} {104}},\ \bibinfo
  {pages} {035118} (\bibinfo {year} {2021})}\BibitemShut {NoStop}%
\bibitem [{\citenamefont {Lyu}\ \emph {et~al.}(2020)\citenamefont {Lyu},
  \citenamefont {Montenegro},\ and\ \citenamefont {Bayat}}]{Lyu2020}%
  \BibitemOpen
  \bibfield  {author} {\bibinfo {author} {\bibfnamefont {C.}~\bibnamefont
  {Lyu}}, \bibinfo {author} {\bibfnamefont {V.}~\bibnamefont {Montenegro}},\
  and\ \bibinfo {author} {\bibfnamefont {A.}~\bibnamefont {Bayat}},\ }\bibfield
   {title} {\bibinfo {title} {Accelerated variational algorithms for digital
  quantum simulation of many-body ground states},\ }\href
  {https://doi.org/10.22331/q-2020-09-16-324} {\bibfield  {journal} {\bibinfo
  {journal} {{Quantum}}\ }\textbf {\bibinfo {volume} {4}},\ \bibinfo {pages}
  {324} (\bibinfo {year} {2020})}\BibitemShut {NoStop}%
\bibitem [{\citenamefont {Keever}\ and\ \citenamefont
  {Lubasch}()}]{Keever2022}%
  \BibitemOpen
  \bibfield  {author} {\bibinfo {author} {\bibfnamefont {C.~M.}\ \bibnamefont
  {Keever}}\ and\ \bibinfo {author} {\bibfnamefont {M.}~\bibnamefont
  {Lubasch}},\ }\href@noop {} {\bibinfo {title} {Classically optimized
  {H}amiltonian simulation}},\ \Eprint {https://arxiv.org/abs/2205.11427}
  {arXiv:2205.11427} \BibitemShut {NoStop}%
\end{thebibliography}%
		
\end{document}